# UV bright red-sequence galaxies: a comparative study between UV upturn and UV weak systems

## Maria Luiza Linhares Dantas

Supervisor: Prof. Paula Rodrigues Teixeira Coelho

Instituto de Astronomia Geofísica e Ciências Atmosféricas

Universidade de São Paulo

Departamento de Astronomia

São Paulo, São Paulo, Brasil.

July 2020

# UV bright red-sequence galaxies: a comparative study between UV upturn and UV weak systems

## Maria Luiza Linhares Dantas

Supervisor: Prof. Paula Rodrigues Teixeira Coelho

Thesis presented to the Department of Astronomy of the Institute of Astronomy, Geophysics, and Atmospheric Sciences at the University of São Paulo, as a partial requisite for obtaining the doctoral degree.

Corrected version. The original version can be found at the Institute.

São Paulo, 2020

If I have seen further it is by standing on the shoulders of Giants.

Isaac Newton

# Acknowledgements

Words fly. Writing remains.

Portuguese proverb

1. I would like to thank first and foremost my advisor Paula Rodrigues Teixeira Coelho, who stood by me since before I became her PhD student. Very few people have the same chance I had to be able to work such wonderful supervisor.

2. I also acknowledge *Coordenação de Aperfeiçoamento de Pessoal de Nível Superior* (CAPES) – including Finance Code 001 – and *Conselho Nacional de Desenvolvimento Científico e Tecnológico* (CNPq) for the grants provided during my doctoral studies.

3. Collaborators from whom I learned enormously: Patricia Sánchez Blázquez, Alberto Krone Martins, Rafael S. de Souza, Thiago Signorini Gonçalves.

4. I thank all the staff in Instituto de Astronomia, Gefísica e Ciências Atmosféricas at Universidade de São Paulo (IAG-USP) for their hard work. Without them it would be impossible to solve the numerous bureaucratic issues, as well as keep the cleanliness of the infrastructure, and so on.

5. I would like to specially acknowledge Marco Antônio Santos and Luis Manrique, the department technicians, who have always welcomed me with my issues – whatever they were – and solved them with the highest standards of professionalism.

6. Friends that stood by me no matter what during my doctoral studies: Marcus Vinícius Costa Duarte, Camile Mendes, Doris Stoppacher, Isabel Rebollido Vázquez, Marina Rodríguez Baras.

7. I thank my parents, Maria Magda Ribeiro Linhares and Josemar Toscano Dantas (in memorian) for always supporting me.

8. My extremely loving kitty, Miúcha, who accompanied me during over half of my doctoral studies; she has been the light of my life.

9. Figs. 1.3, 1.4, and 1.5 have been treated for better visualisation. To that end, this work benefited from https://www.vectorizer.io and GIMP.

10. This work benefited from the following platforms: Overleaf, Github, and Slack.

I thank all those that were there for me, you know who you are.

# Abstract

Ultraviolet (UV) emission from galaxies is associated with hot components, whether from stellar sources or not. It is an important marker for star-formation rate, yet it can also be associated with evolved and rare stellar evolutionary phases. By making use of colour-colour diagrams, early-type galaxies (ETGs) can be classified in terms of their UV emission in mainly three categories: residual star-formation, UV weak, and UV upturn emission. The UV upturn is a phenomenon characterised by an unexpected rise of the fluxes of quiescent ETGs between the Lyman limit and 2500Å. This thesis aims at investigating galaxies presenting UV upturn by comparing them to other systems hosting UV weak emission. This investigation has three fronts: (i) the assessment of the evolution in redshift ($z$) and stellar mass ($\log M_\star$) of the fraction of UV bright red-sequence galaxies (RSGs) that host the UV upturn; (ii) the stratification of the aforementioned study in terms of emission lines; (iii) the comparison of stellar population parameters between UV weak and upturn for retired/passive RSGs. A sample of galaxies has been selected from the Galaxy Mass Assembly (GAMA) aperture-matched with the Sloan Digital Sky Survey (SDSS) and the Galaxy Evolution Explorer (GALEX). To tackle the first front, a Bayesian logistic model was applied. The second front expands on the first, dividing the sample into emission line classes by making use of the WHAN diagram. The final front is focused on the study of stellar population properties of UV weak and UV upturn systems, by making use of value-added catalogues from the GAMA collaboration which provide stellar population properties obtained via spectral energy distribution fitting. To analyse both groups of galaxies, the original samples were balanced in terms of $z$ and $\log M_\star$. With the de-biased samples at hand, three types of comparisons were made: the direct comparison of their stellar population parameters; the estimation of the Spearman correlation rank among these parameters and the differences between both UV classes; and a principal component analysis. The results show that the fraction of UV upturn systems rises up to $z \sim 0.25$, followed by a decline which remains to be confirmed given the thickness of credible intervals. By stratifying the sample into emission line classes, galaxies with star formation have been identified; the galaxies classified as retired/passive – the ones associated with evolved stellar phases – dominate the behaviour with $z$ and $\log M_\star$. Finally, by analysing the stellar populations of both UV weak and upturn systems, some different characteristics emerge such as median ages, metallicities, and time since last burst of star formation. These results seem to indicate that either UV upturn systems evolve more passively, or settled their stellar population at higher $z$ than their UV weak counterparts. Either way, UV upturn systems have narrower star-formation histories, higher metallicities, and slightly older populations.

*Keywords*: Early-type galaxies. Data analysis. Evolution. Ultraviolet.

# Resumo

A emissão ultravioleta (UV) em galáxias está associada a componentes quentes, sejam de fontes estelares ou não. Essa emissão é um importante marcador de formação estelar, mas também pode estar associada a fases estelares evoluídas raras. Fazendo uso de diagramas cor-cor, galáxias do tipo *early-type* (ETGs) podem ser classificadas em três categorias em termos da emissão UV: formação estelar residual, UV fraca e UV *upturn*. O UV *upturn* é um fenômeno caracterizado por uma subida inesperada no fluxo de ETGs quiescentes entre o limite de Lyman e 2500Å. Esta tese tem como objetivo investigar galáxias que apresentam o UV *upturn* ao compar 5-las com outras que abrigam emissão UV fraca. Este estudo tem três partes: (i) avaliar a evolução em *redshift* ($z$) e massa estelar ($\log M_\star$) da fração de galáxias da sequência vermelha que possuem UV *upturn*; (ii) a estratificação da análise anterior em termos de linhas de emissão; (iii) a comparação entre as populações estelares das UV fracas e das UV *upturn* classificadas como aposentadas/passivas. Foi selecionada uma amostra de galáxias do *Galaxy Mass Assembly* (GAMA) combinada com o *Sloan Digital Sky Survey* (SDSS) e o *Galaxy Evolution Explorer* (GALEX). Para a primeira parte do estudo, um modelo logístico Bayesiano foi aplicado. A segunda parte expande a primeira, dividindo a amostra em classes de linhas de emissão por meio do diagrama WHAN. A parte final é focada no estudo das populações estelares de galáxias UV fracas e UV *upturn*, usando catálogos do GAMA que fornecem as propriedades de suas populações estelares. Para analisar ambos os grupos de galáxias, as amostras foram balanceadas em termos de $z$ e $\log M_\star$. Com estas amostras em mãos, três análises foram feitas: a comparação direta entre suas populações estelares; o cálculo do termo de correlação de Spearman entre suas propriedades e as diferenças entre as classes; e uma análise de componentes principais. Os resultados mostram que a fração de sistemas com UV *upturn* cresce até $z \sim 0.25$, seguida por uma aparente descida, que ainda precisa ser confirmada considerando a espessura dos intervalos de credibilidade. Ao estratificar a amostra em classes de linhas de emissão, galáxias com formação estelar foram identificadas; as galáxias classificadas como aposentadas/passivas – associadas às fases estelares evoluídas – são as que dominam o comportamento da fração de UV upturn com $z$ e $\log M_\star$. Enfim, ao analisar as populações estelares de sistemas UV fracos e *upturn*, diferenças nas características gerais aparecem, tais como idades, metalicidades, e período desde o último surto de formação estelar. Estes resultados indicam que as UV *upturn* podem estar evoluindo mais passivamente do que suas contrapartidas UV fracas ou então que suas populações estelares estabilizaram em $z$ maiores. De qualquer forma, os sistemas UV *upturn* possuem histórias se formação estelar mais curtas, maiores metalicidades e populações estelares mais velhas.

*Palavras-chave*: GALÁXIAS ELÍPTICAS. ANÁLISE DE DADOS. EVOLUÇÃO. ULTRAVIOLETA.

# List of Figures

# List of Tables

# Contents

# List of Abbreviations

**Λ-CDM** Λ-Cold Dark Matter. 18, 44

**AAO** Anglo Australian Observatory. 34

**AAT** Anglo Australian Telescope. 34

**AGB** asymptotic giant branch stars. 24, 28

**AGN** active galactic nucleus (sing.) or active galactic nuclei (plur.). 20, 22, 28, 54, 63, 65, 66, 68–70, 75, 76

**BCG** brightest cluster galaxy. 20, 26, 27, 37

**BPT** Baldwin-Phillips-Terlevich diagram. 64–67, 69–71, 75

**CDF** cumulative distribution function. 47

**CMB** cosmic microwave background. 18

**COIN** Cosmostatistics Initiative. 29

**DMU** data management unit. 37, 83, 86

**EHB** extreme horizontal branch stars. 23, 82

**ETG** early-type galaxy. 17, 19, 20, 22, 25, 27, 28, 76, 81, 99

**GALEX** Galaxy Evolution Explorer. 34

**GAMA** Galaxy Mass Assembly. 34

**GLM** generalised linear model. 29, 31

**HB** horizontal branch stars. 23, 82

**HMC** Hamiltonian Markov Chain Monte Carlo. 56

**HOLMES** hot low-mass evolved stars. 28, 66, 76

**HR** Hertzprung-Russell diagram. 22, 24, 106

**S/N** signal-to-noise ratio. 38, 67

**sAGN** strong AGN. 66, 68–70, 75, 76

**SAMI** Sydney/AAO Multi-object Integral-field Spectrograph Survey. 34

**SAURON** Spectroscopic Areal Unit for Research on Optical Nebulae Survey. 25

**SDSS** Sloan Digital Sky Survey. 35

**SED** spectral energy distribution. 22, 24, 36, 43, 80

**SF** star formation/forming. 67, 69–71

**SFH** star formation history. 80, 81, 107

**SFR** star formation rate. 89

**SSP** simple stellar population. 80, 81

**TP-AGB** thermally pulsating asymptotic giant branch stars. 24, 82

**UKIDSS** UKIRT Infrared Deep Sky Survey. 34

**UKIRT** United Kingdom Infra-Red Telescope. 34

**UV** ultraviolet. 17, 21–31, 33, 34, 36–38, 40, 43, 45–47, 49, 51–55, 57, 58, 61–63, 67–72, 74–77, 79–86, 88–90, 92–106, 109–113

**wAGN** weak AGN. 67, 68, 70, 75, 76

**WHAN** $W_{H\alpha} \times W_{N[II]}$ diagram. 64, 66–73, 76, 83

**WMAP** Wilkinson Microwave Anisotropy Probe. 17

The real difference between us and
chimpanzees is the mythical glue that
binds together large numbers of
individuals, families and groups. This
glue has made us the masters of
creation.

Yuval Noah Harari
*Sapiens: A Brief History of
Humankind.*

# 1

# Introduction

This thesis aims at investigating the nature and behaviour of the UV emission in red-sequence galaxies (RSGs), focusing on systems that harbour the so-called UV upturn phenomenon. RSGs are usually classified as early-type galaxies (ETGs), and are typically redder, more massive, and comprised of older stellar populations, when compared to their late-type (LTG) counterparts. In this Introduction, I contextualise this investigation, presenting a general yet current understanding of the Universe.

## 1.1   A quick history of the Universe

The origins of the Universe are somewhat well established. A colossal amount of observational evidence supports the theory in which the Universe has been originated from a singularity, the so-called Big Bang. Since then, it has been expanding with varied rates that depend on the state and density of matter and energy that composes it (Frieman, Turner, & Huterer, 2008). To investigate how this variation of rates evolve, many surveys have been deployed, such as the Wilkinson Microwave Anisotropy Probe (WMAP, e.g. Hinshaw et al., 2013) and Planck (Planck Collaboration et al., 2018), and consequently many of the questions surrounding the origin and evolution of the Universe have been (at least partially) answered. Therefore, our current understanding of the Universe is that it is made of approximately 30% of matter, in which only ~4% is baryonic – i.e. observable – and the other ~26% is dark – not emitting any electromagnetic radiation, but detectable through the gravitational interaction with baryonic matter. The other 70% is even more sinister, it is the so-called *dark energy*, which can only be detected through the accelerated expansion of the Universe (e.g. Peebles & Ratra, 2003). The natures of both dark matter and dark energy are still a mystery.

The currently accepted cosmological model is the so-called Λ-Cold Dark Matter (Λ-CDM), which has been very successful in explaining the expansion of the Universe as well as the anisotropies of the cosmic microwave background (CMB) radiation, and the distribution of primordial elements, among other phenomena. Such anisotropies are 'seeds' of the large scale structure of the Universe: they result in wells of gravitational potential through dark matter haloes, which attract baryonic matter – accreting matter and pulling galaxies – therefore forming galaxy clusters (Blumenthal et al., 1984).

## 1.2   Galaxies: the building blocks of the Universe

Galaxies can be defined as *the building blocks of the observable Universe* as they are responsible for constructing the large-scale structure of the Universe: galaxy clusters. Those systems are a composite of four main ingredients: stars (and their remnants), gas, dust, and dark matter; however, most of their electromagnetic emission comes from their stellar/stellar remnant and dust content (with exceptions, such as the presence of quasars Schmidt, 1963).

As all large collection of objects, galaxies require to be classified in order to simplify the way one studies them. By classifying galaxies (or anything else for that matter), one may group them by general features which they have – or not – in common, therefore allowing us to disentangle a potential shared history behind their formation and evolution. Galaxies can be classified in many different ways (e.g. de Souza et al., 2017); features such as morphology (e.g. Lintott et al., 2008; Dieleman, Willett, & Dambre, 2015), colour (e.g. Strateva et al., 2001; Mateus et al., 2006), mass (e.g. Juneau et al., 2011; Juneau et al., 2014), size, emission lines (e.g. Baldwin, Phillips, & Terlevich, 1981), and so forth, can be used to sort them. With that in mind, Edwin Hubble introduced two broad nomenclatures that are used to this date to sort galaxies – nearly 100 years after his famous work on what was later called *The Hubble Sequence* (Hubble, 1926): early- and late-type galaxies (ETGs and LTGs hereafter). An example of Hubble's Tuning Fork can be observed in Fig. 1.1.

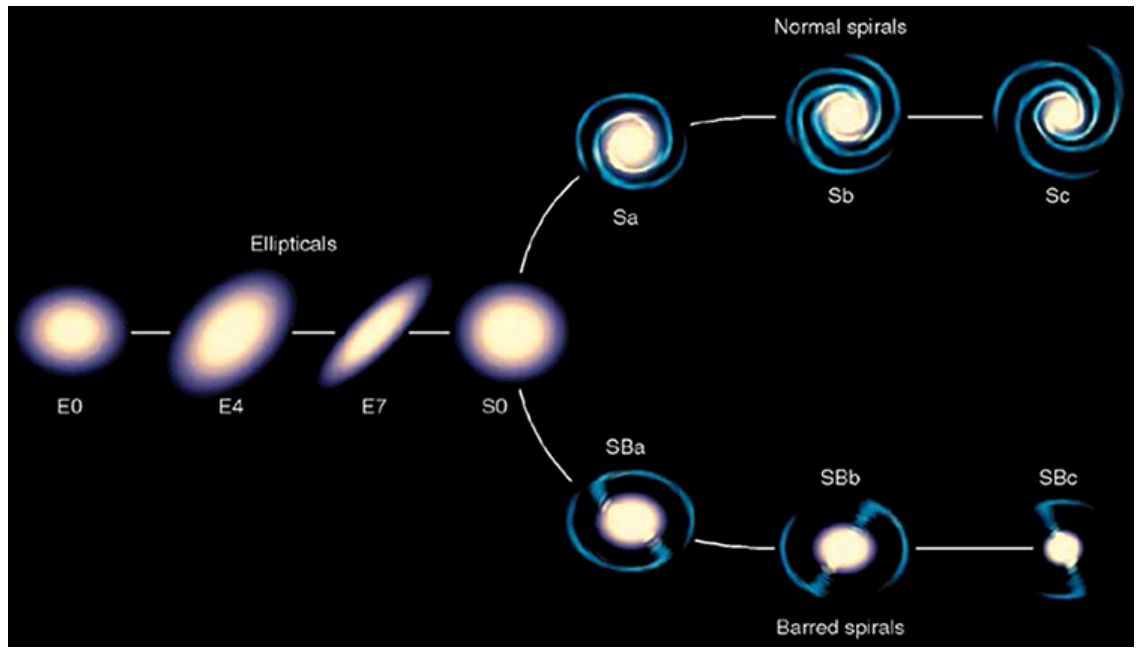

Figure 1.1: Example of Hubble's Tuning Fork. Scheme retrieved from Seigar 2017. Early-type galaxies are depicted in purple whereas late-type galaxies are depicted in blue.

Such nomenclature was introduced as Hubble believed that spiral galaxies had been formed in an earlier time when compared to their elliptical/lenticular counterparts; subsequent studies showed that it was actually the opposite (e.g. Trinh, Balkowski, & Van Tran, 1992). In what follows I deepen the discussion about ETGs and the complexity of such systems.

## 1.2.1   The nature of early-type galaxies

By taking Fig. 1.1 as a visual reference, ETGs are those systems comprised of elliptical and lenticular galaxies, i.e. systems classified between E0 and S0 (with these extremes included). For a long time it was believed that ETGs were simple systems – which was also supported by their visual homogeneity – that once collapsed and were passively evolving. This idea was indeed in accordance with the monolithic collapse theory of galaxy formation (described for the first time by Eggen, Lynden-Bell, & Sandage, 1962). Only with the rise of a second theory, the hierarchical evolution of galaxies (Searle and Zinn, 1978; White and Rees, 1978; see also Cole et al., 2000), was that many other aspects of ETGs started to be fully analysed and understood.

Indeed, ETGs are far from 'simple'; in order for galaxies to evolve into ETGs, they go through various types of dynamical interactions (e.g. Barnes, 1988; Springel, Yoshida, & White, 2001; Springel et al., 2005; De Lucia & Blaizot, 2007; Naab et al., 2014; Schawinski et al., 2014), including mergers that can be wet or dry depending on their original gas reservoirs, which impact the following steps of their evolution (e.g. Sánchez-Blázquez et al., 2009a). Additionally, they are also subject to other types of phenomena that deeply influence their development, such as supernovae and AGN feedback processes (Springel, Di Matteo, & Hernquist, 2005). Such systems can ultimately present a broad range of morphologies, spanning across ellipticals and lenticulars (Sánchez-Blázquez et al., 2009b). Additionally, features such as the initial mass function (IMF, Salpeter, 1955) have been shown to be tricky: they may not be universal (e.g. Kroupa, 2001; Chabrier, 2003; Spiniello et al., 2012; La Barbera et al., 2013; Spiniello, Trager, & Koopmans, 2015; La Barbera et al., 2019).

Additionally, it has been shown that, in fact, the amount of elliptical galaxies has been in the rise, basically doubling since $z \sim 1$, whereas the number of spirals has remained the same (e.g. Bell et al., 2004; Brown et al., 2007; Faber et al., 2007; Ilbert et al., 2013; Sobral et al., 2014). Yet, despite the fact that many advances have been made in order to better understand the formation and evolution of ETGs, their evolutionary mechanisms are yet unclear, including those responsible for the development of the most massive systems known, monstrous ETGs (such as BCGs, see e.g. Arcila-Osejo et al., 2019; Stoppacher et al., 2019).

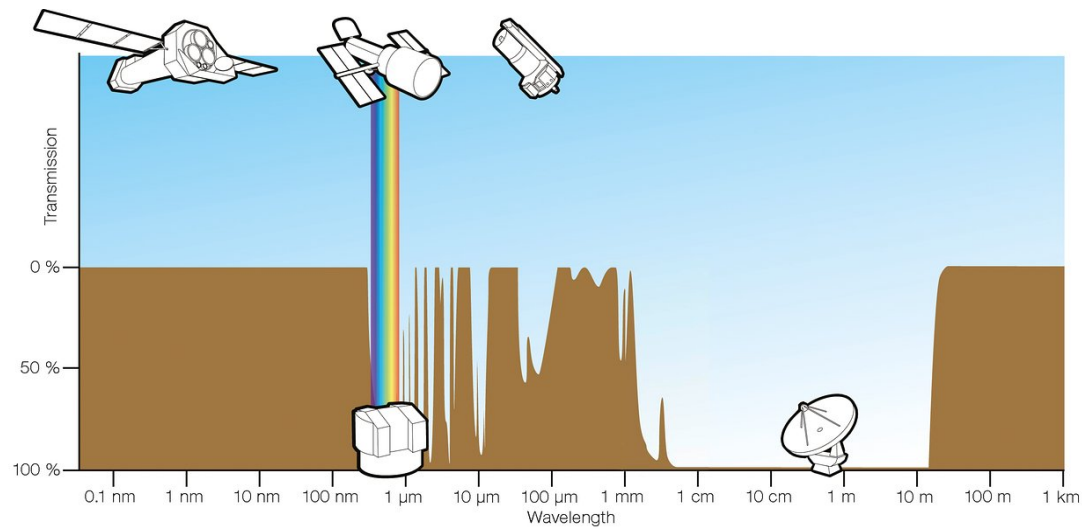

Figure 1.2: Illustration of the transparency of the atmosphere in different wavelengths. Credits: F. Granato (ESA/Hubble).

## 1.3 The Universe seen in the ultraviolet

Observations in the UV range of the electromagnetic spectrum are filled with challenges. The first – and perhaps the most important – obstacle is that the Earth's atmosphere is optically thick to most UV radiation (Rybicki & Lightman, 1991), which forces us to launch UV space telescopes (see Fig. 1.2).

As a consequence, many space telescopes were launched with the goal of grasping the nature of the Universe in shorter wavelengths, not only in the UV, but also in X-rays and $\gamma$-rays (e.g. Madejski, 2005). In the realm of the UV, the observational legacy comes from the Orbiting Astronomical Observatory (OAO, Code, 1969), the International Ultraviolet telescope (IUE, Boggess et al., 1978), the Hopkins Ultraviolet Telescope (HUT, Davidsen et al., 1992; Dixon et al., 2013), and notably the Galaxy Evolution Explorer (GALEX, Martin et al., 2005) as well as the Hubble Space Telescope (HST, e.g. Freedman et al., 2001).

The UV range of the electromagnetic spectrum is associated with hot components, stellar or not. For extragalactic astrophysics, in the local Universe, the simple mainstream assumption is that galaxies that are strong emitters of UV radiation harbour an important amount of young hot stars (e.g. Kennicutt, 1998; Gil de Paz et al., 2007; Salim et al., 2007; Bond et al., 2014; Madau & Dickinson, 2014; de los Reyes & Kennicutt, 2019) or host phenomena such as AGN (e.g. Cid Fernandes, Sodré, and Vieira da Silva, 2000; Chung et al., 2014, Heinis et al., 2016; Padovani et al., 2017). For high-$z$ investigations, the UV has been widely used to inspect Lyman-break galaxies (see Giavalisco, 2002 for a review on the topic; and see Oteo et al., 2014 as an example) by making use of the technique introduced by Steidel, Pettini, and Hamilton (1995). In fact, because of the properties of the atmosphere that block UV radiation, the UV has been historically more used to analyse the characteristics of high-$z$ Universe, which can be assessed by ground-based telescopes.

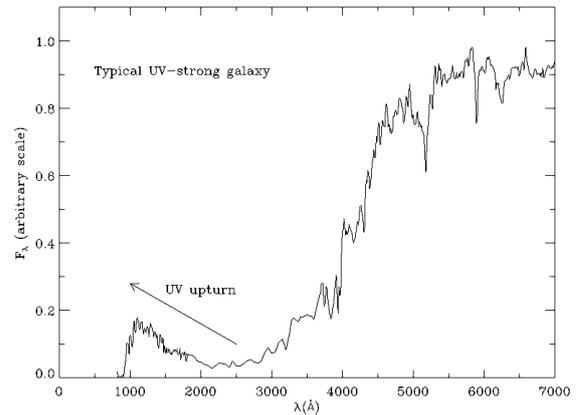

Figure 1.3: Example of an elliptical galaxy presenting UV upturn retrieved from Yi, Demarque, and Oemler, Jr. (1998).

With the launch of OAO group of satellites between 1966 and 1972 (Code, 1969), it was possible to detect a previously unseen phenomenon among ETGs: some of them were actually very bright in the UV (Code & Welch, 1979), including the bulge of M31 – Andromeda, our neighbour spiral galaxy. This is the so-called *UV upturn of elliptical* *galaxies*, their spectral energy distribution (SEDs) show a peculiar up-rise around 1,200–2,500Å (e.g. Bertola, Capaccioli, & Oke, 1982; Dorman, Rood, & O'Connell, 1993; Ferguson & Davidsen, 1993; Dorman, O'Connell, & Rood, 1995; Deharveng, Boselli, & Donas, 2002; Ree et al., 2012) as illustrated in Fig. 1.3.

### 1.3.1 The UV upturn phenomenon

The UV upturn (see O'Connell, 1999, for a review) remained a puzzle for decades, as formerly it was unclear which mechanisms were behind such strong UV radiation; back then it was believed that their stellar populations were purely old and cold.

#### 1.3.1.1 Stellar culprits

In the midst of such debate, some theories started to emerge in order to explain the mystery behind this atypical UV emission. First of all, the idea that such objects could foster some residual star-formation activity gained territory (e.g. Yi et al., 2005; Kaviraj et al., 2007a; Kaviraj et al., 2007b; Pipino et al., 2009; Salim & Rich, 2010; Bettoni et al., 2014; Davis et al., 2015; Haines et al., 2015; Stasińska et al., 2015; Sheen et al., 2016; Vazdekis et al., 2016; Rampazzo et al., 2017; Evans, Parker, & Roberts, 2018; López-Corredoira & Vazdekis, 2018, and references therein).

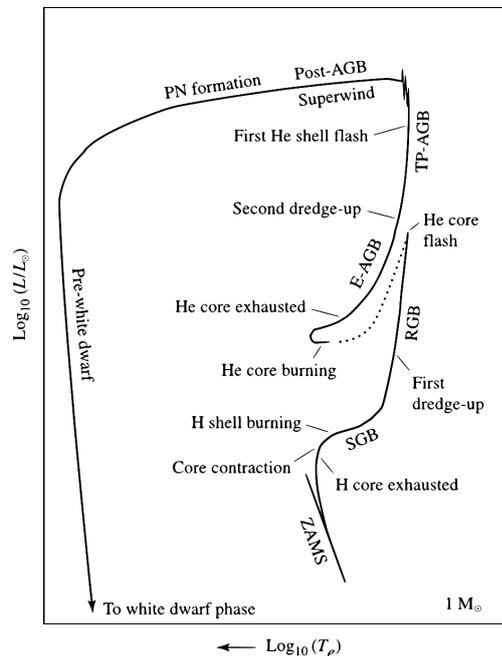

Figure 1.4: HR diagram for a star with $1M_\odot$. Credits: Ostlie and Carroll, 2007, Fig. 13.4 therein.

Yet, with the better understanding of stellar evolutionary phases, other works have shown that stars in some post-main-sequence phases could in fact be efficient UV emitters (e.g. Hills, 1971; Greggio & Renzini, 1990a, 1990b; Brown et al., 1998; Greggio & Renzini, 1999; O'Connell, 1999; Brown et al., 2000; Deharveng, Boselli, & Donas, 2002; Brown, 2004; Lee et al., 2005; Martin et al., 2005; Hernández-Pérez & Bruzual, 2013, 2014). Blue horizontal branch (HB) and extreme horizontal branch (EHB) stars are some of the potential culprits of such phenomenon and the evidence for it has been accumulating (e.g. Yi, Demarque, & Kim, 1997; Brown et al., 1998; Brown et al., 2000; Brown et al., 2003; Yoon et al., 2004; Peng & Nagai, 2009; Donahue et al., 2010; Loubser & Sánchez-Blázquez, 2011; Schombert, 2016; Lonoce et al., 2020). There are mainly two evolutionary paths which explain the blue and extreme HB stars; the first would be that the cores of these helium-burning stars are covered by very thin hydrogen layers, which expose their cores and consequently high inner temperatures, therefore boosting their UV emission. Their cores become exposed due to mass loss which occurs during the red-giant branch (RGB) phase (see Figs. 1.4 and 1.5). The second hypothesis is more modern and is linked to helium-enhancement which is discussed further in this Section.

Other potential culprits include, post-asymptotic giant branch stars (post-AGB) and the 'post-AGB family' of stars: AGB-manqué, post-early-AGB, and so forth (e.g. Greggio & Renzini, 1990a; Brown et al., 1998; Deharveng, Boselli, & Donas, 2002; Donas et al., 2007; Han, Podsiadlowski, & Lynas-Gray, 2007; Chavez & Bertone, 2011, and references therein); even in 'regular' AGB stars, UV excess has been detected (e.g. Ortiz, Guerrero, & Costa, 2019; Guerrero & Ortiz, 2020).

Maraston (2005) and Bruzual (2007) highlight the importance of considering rare evolutionary stellar populations in models, which are frequently overlooked, with special remarks on thermally pulsating-AGB (or TP-AGB) stars (top right regions of Figs. 1.4 and 1.5). For this specific issue, TP-AGB stars are not an issue; they only influence results in stellar populations for those with ages around 2 Gyr. Nonetheless, this issue illustrates the importance of considering all evolutionary steps in order to accurately reproduce the spectral energy distribution (SED) from complex systems, such as galaxies, a quest proposed over three decades ago (Renzini & Buzzoni, 1986).

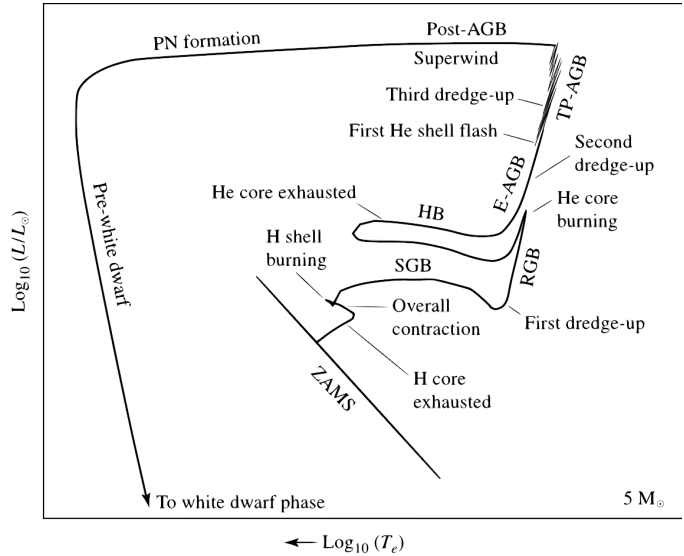

Figure 1.5: HR diagram for a star with $5M_\odot$. Credits: Ostlie and Carroll, 2007, Fig. 13.5 therein.

Also, binary stellar systems should not be forgotten in the quest for the culprits of the UV upturn. Apparently, these may have an important role on the UV emission of quiescent galaxies (Zhang et al., 2005; Han, Podsiadlowski, and Lynas-Gray, 2007, 2010; Hernández-Pérez and Bruzual, 2013, 2014). Not even our own home galaxy, the Milky Way, which is an SBc spiral (e.g. Hodge, 1983; López-Corredoira et al., 2007), is free from old UV bright stars. Smith, Bianchi, and Shiao (2014) have found several amazing features regarding the UV emission of Galactic stars, including solar-like systems:

i. the distribution of (FUV-NUV) for such stars is in fact bimodal;

ii. for G-type stars such colour moves towards bluer values;

iii. and, finally, about 14–18% of F to K-type stars present FUV excesses when compared to their NUV counterpart emission.

The authors attribute this phenomenon to hot dwarf stars and binary systems in interaction, which are some of the potential UV upturn emitters, as aforementioned.

Another aspect explored in the field is the one regarding metallicity features of UV upturn bearers (e.g. Yi, Demarque, & Kim, 1997; Bureau et al., 2011; Jeong et al., 2012; Chung, Yoon, & Lee, 2017). Once again, Burstein et al. (1988) proposed a relation between the UV upturn and Mg$_2$, which was supported by Boselli et al. (2005) and Donas et al. (2007), but not confirmed by Loubser and Sánchez-Blázquez (2011). Bureau et al. (2011), on the other hand, confirmed a positive correlation between Mg$_2$ and the strength of the FUV-*V* colour among 48 ETGs observed by the Spectroscopic Areal Unit for Research on Optical Nebulae (SAURON, Bacon et al., 2001) survey. Another paper by Werle et al. (2020) also finds evidence that ETGs hosting UV upturn are dominated by high metallicity stellar populations. A link between optical metallicity indicators and the strength of the upturn has been explored by previous works: apparently they correlate positively (Yoon et al., 2004; Bureau et al., 2011; Chung, Yoon, & Lee, 2017; Ali et al., 2018c). This behaviour is foreseen when the emission comes primarily from helium-enhanced populations. Also, by making use of absorption lines Le Cras et al. (2016) has shown that stellar populations of different different ages (i.e. young and old) co-exist, reinforcing the need of disentangling the many stellar characters playing a role in the big UV upturn piece.

Additionally, some studies indicate that helium-enhanced populations could explain the UV upturn phenomenon and the UV properties of globular clusters (Faber & Worthey, 1993; D'Antona & Caloi, 2004; Lee et al., 2005; Kaviraj et al., 2007c; Piotto et al., 2007; Peacock et al., 2011; Schiavon et al., 2012; Chung, Yoon, & Lee, 2017; Goudfrooij, 2018; Peacock et al., 2018). Also, it appears that the UV upturn is a common characteristic of a myriad of environments inhabited by old stellar populations, which is the case NGC6791, an old open cluster (Buson et al., 2006; Buzzoni et al., 2012).

### 1.3.1.2   The UV upturn throughout the cosmic timeline

One of the questions that arise when dealing with the UV upturn phenomenon is: *does it evolve in redshift?* To answer this question, many attempts with different approaches have been made, specially considering the strength of the upturn (Brown et al., 1998;

Brown et al., 2000; Brown, 2004; Rich et al., 2005; Ree et al., 2007; Ali et al., 2018a, 2018b).

Rich et al. (2005) explored whether there was a trend in the (FUV-*r*) colour with redshift bins, but found no correlation. On the other hand, Brown (2004) investigated the UV-optical colours of a small sample of galaxies finding a potential evolution up to $z \sim 0.6$. Subsequently, Ree et al. (2007) selected a sample of brightest cluster galaxies (BCGs) located in clusters in $z \sim 0.2$, $z \sim 0.3$, and $z \sim 0.5$ and explored their (FUV - *V*) colours; the authors concluded that there might be a weak evolution in *z*. Recently Ali et al. (2018b) once again tackled this issue and analysed data from four galaxy clusters – one at $z \sim 0.31$, two at $z \sim 0.5$ and one at $z \sim 0.68$; their conclusions indicated that the strength of the phenomenon increases until $z \sim 0.55$, declining subsequently.

Nevertheless, all the aforementioned previous studies have two characteristics in common:

1. their analysis is based on small samples (i.e. under 100 objects – sometimes literally a handful amount of galaxies);

2. their focus is on the strength of the upturn.

In Chapters 3 and 4, this discussion is revisited and a new outlook on the problem is proposed. The evolution of the UV upturn is analysed based on the fraction of UV bright red-sequence galaxies (RSGs) as a function of *z*.

### 1.3.1.3 The many facets of the UV upturn

Besides the potential culprits and the evolution of the phenomenon throughout cosmic time, many other aspects of the UV upturn have been explored by previous studies.

In terms of dynamic characteristics of elliptical galaxies, a seminal work on the topic has been the one that unravelled the 'textbook' Faber-Jackson relation (hereafter FB, Faber & Jackson, 1976) which linked the correlation between $\sigma$ and total luminosity ($L_{\text{total}}$).

Later, a more general relation was introduced comprising FB parameters and the effective radius ($r_e$), which is the so-called Fundamental Plane of Elliptical Galaxies (Djorgovski & Davis, 1987), a phenomenon closely related to the Virial Theorem (Busarello et al., 1997). Concerning the UV upturn, Choi, Goto, and Yoon (2009) have explored its effects on the Fundamental Plane, and have concluded that E+A systems (ellipticals with a minority of A-type stars) behave differently than passively evolving UV dead ETGs (i.e. the values of $\sigma$ are usually smaller for E+A systems than those of their UV dead counterparts).

Burstein et al. (1988) has suggested that the incidence of the UV upturn was correlated to higher velocity dispersion ($\sigma$) values. Nonetheless, such correlation has not been found by Loubser and Sánchez-Blázquez (2011), whereas Yi et al. (2011) suggested that the correlation with mass is weak. This question is in fact revisited in this thesis, specifically in Chapter 3; it is worth briefly mentioning that, in fact, the results reveal that galaxies nesting the UV upturn tend to be more massive than compared to their UV weak counterparts.

From yet another perspective, sedimentation theory predicts that helium should accumulate in the centre of massive galaxy clusters (as described by Peng & Nagai, 2009, and references therein). In this case, the UV flux should be stronger in BCGs when compared to other elliptical galaxies, due to helium-enhancement in their stellar populations (see, for instance, Busso et al., 2007; Catelan, Valcarce, & Sweigart, 2010). That would be an indication that the UV emission from such systems could correlate to their environment. However, such correlation has not been confirmed by other observational papers (for instance, Loubser & Sánchez-Blázquez, 2011; Yi et al., 2011; Ali et al., 2019).

### 1.3.1.4   Clues from emission lines

Emission line diagnostics are a powerful tool to better understand several ionisation processes happening in galaxies (Stasińska, 2007). When investigating the UV upturn phenomenon, one is mainly looking for effects happening in the most 'boring' of systems: passively evolving inactive elliptical red galaxies. Nonetheless, such systems foster a myriad of processes in their insides, which are the complete opposite of boring.

In this context, analyses considering emission lines can be very handful, although not intuitive, as such systems are not expected to foster energetic phenomena that could produce emission lines. Nonetheless, several retired and even passive objects actually show emission lines (Cid Fernandes et al., 2010; Cid Fernandes et al., 2011; Stasińska et al., 2015; Herpich et al., 2018). Emission lines can be detected in ETGs and most of it has been attributed to low-ionisation nuclear emission line regions (LINERs), traditionally considered a type of AGN (Padovani et al., 2017). Another research supporting the fact that UV emission from AGN may not be as important as they seem is the one by Ohl et al. 1998; they argue that AGN has a limited contribution to the far-UV emission in ETGs, which is the case of M87.

A ton of research has been supporting that LINER emissions might not be caused by one sole factor, but that it can actually be a phenomenon produced – at least in part – by hot low-mass evolved stars (HOLMES, Taniguchi, Shioya, and Murayama, 2000; Cid Fernandes et al., 2011; Singh et al., 2013; Belfiore et al., 2016; Percival and James, 2020). In fact, Belfiore et al., 2016 separate LINERs from LIERs (low-ionisation emission line regions – i.e. the 'nuclear' aspect of it is absent) by analysing the spatial emission of LINER-like galaxies through integral field spectroscopy (IFS). In the context of the AGN unification theory (Antonucci, 1993; Urry & Padovani, 1995), the problem concerning the nature of LINERs adds to the pile of issues to be considered (Elitzur & Shlosman, 2006; Netzer, 2015).

Many are the types of stars potentially producing LINER-like emissions, such as Wolf-Rayet stars (e.g. Terlevich & Melnick, 1985), O-type stars (e.g. Filippenko & Terlevich, 1992), and post-AGB stars (e.g. Binette et al., 1994). Therefore, considering that HOLMES – LIERs – have the potential of ionising the interstellar medium of galaxies and that evolved stellar populations (such as post-AGB stars) are also linked to UV emission of quiescent systems (Percival & James, 2020), one may ask: *is there a link between LIERs and the UV upturn?* This is an open question that remains to be properly answered.

## 1.4   Methodological conundrums

Contemporary research in Astronomy is facing an ever-increasing number of observational data, which forces us to innovate in the way we deal with it. It is the so-called *era of big data*. Collecting, processing, storing, and analysing such data require the use of high-end technological tools, as well as statistical/numerical approaches that for a long time stayed overlooked (Zhang & Zhao, 2015).

Astrophysical studies have deep roots in the 'classic' linear (i.e. Gaussian) regression (Isobe et al., 1990; Feigelson & Babu, 1992; Kelly, 2007; Feigelson & Babu, 2012; Sereno, 2016). However, theoretical and observational Astrophysics involve the study of complex issues that often cannot be explained by one sole kind of distribution. In recent years, a few working groups have been exploring novel ways to treat, analyse, and interpret data in Astronomy; notably, the International Astrostatistics Association (IAA)[1] and its most active working group, the Cosmostatistics Initiative (COIN)[2]. As a consequence, many studies arose with the intent of reassessing an enormous amount of astrophysical problems that had been poorly treated before (e.g. Feigelson & Babu, 1992; Krone-Martins, Ishida, & de Souza, 2014; de Souza et al., 2015b; Elliott et al., 2015; de Souza et al., 2016; Sasdelli et al., 2016; Hilbe, de Souza, & Ishida, 2017; Dantas et al., 2020).

In fact, astrophysical research is far from exploring all the techniques and tools made available by statisticians and data experts. An example is the use of generalised linear models (GLMs), which have been used as case of study by the series of papers by de Souza et al. (2015a), Elliott et al. (2015), and de Souza et al. (2015b). GLMs actually are a group of linear models, whereas that astronomers consider only the Gaussian distribution to be linear. Yet, statistics is filled with a myriad of other distributions that are suitable to a given problem and these are usually overlooked (e.g. Hilbe, de Souza, & Ishida, 2017).

---

[1]http://iaa.mi.oa-brera.inaf.it/IAA/home.html
[2]https://cosmostatistics-initiative.org/

Studies involving detection of exoplanet transit have been benefiting from countless techniques previously neglected, specially those involving time series and other similar tools (for instance, Rajpaul, Aigrain, & Roberts, 2016; Jones et al., 2017; Taaki, Kamalabadi, & Kemball, 2020). Other examples of novel techniques are the use of spatial models for IFS studies (e.g. González-Gaitán et al., 2019), hierarchical Bayesian models for inferring physical properties of galaxies (e.g. Sánchez-Gil et al., 2019), to mention a couple.

Additionally, the use of numerical tools such as machine learning (ML, e.g. Hastie, Tibshirani, and Friedman, n.d.) have been increasing with the goal of analysing datasets in different contexts. Some of those are: galaxy emission line classification (e.g. Beck et al., 2016; de Souza et al., 2017; Ucci et al., 2018), galaxy morphological classification (e.g. Dieleman, Willett, & Dambre, 2015; Huertas-Company et al., 2018), supernovae spectral classification (e.g. Sasdelli et al., 2016), photometric redshift estimation (e.g. Krone-Martins, Ishida, & de Souza, 2014; Beck et al., 2017), to mention a few. The meta discussion on the importance of ML to the present and future of Astronomy is further developed by Longo, Merényi, and Tiňo (2019); as well as Feigelson et al. (2020).

In sum, this thesis is inserted in a conjunctural context in which overlooked computational and statistical techniques are on the rise. To extract the most of the data and analyses herein performed, I have tried to make use of the suitable techniques in every facet of this thesis.

## 1.5   About this thesis

Given the context regarding the previous research made about the UV upturn and the methodological patterns of the Astronomical community, the goals and opportunities that arise in this thesis are as follows:

1. to address the evolution of the UV upturn by making use of Bayesian statistical methods, namely GLMs;

2. to understand the role of emission lines in the quest of the UV upturn bearers;

3. to grasp the differences and similarities of the stellar populations inhabiting UV bright RSGs (in other words, RSGs with detectable UV emission) – i.e.  UV weak and UV upturn galaxies.

### 1.5.1   Thesis structure

This thesis is structured as follows. The data is described in Chap. 2, together with some exploratory analysis of the galaxay sample; Chap. 3 provides analyses, discussions, and conclusions on the evolution of the UV upturn with redshift and stellar mass; Chap. 4 tackles the relation between the UV upturn and emission lines; Chap.  5 provides analyses, discussions, and conclusions on the differences and similarities of the stellar populations of UV weak and UV upturn systems. The overall summary and conclusions are displayed in Chap. 6. And, finally, Chap. 7 is an Epilogue which aims at giving some closure to this step in my career as a researcher.  Additionally, I provide an Appendix with relevant additional content to this thesis.

If you torture the data long enough, it
will confess to anything.

Ronald Harry Coase (1910–2013)
1991 Nobel Laureate in Economic
Sciences.

Sometimes even to live is an act of
courage.

Lucius Annaeus Seneca
(54 BCE – 39 CE)

# 2

# Datasets

In this Chapter I describe the galaxy sample used throughout this thesis. Modifications
in the sample herein described are detailed in the Chapters where they are applicable.
Ergo, this Chapter is split in four main Sections:

1. the description of the surveys herein used and the adopted criteria to select the
   observations from their databases are in Sec. 2.1;

2. the description of all the treatment applied to the raw sample in order to accurately
   estimate AB and absolute magnitudes are in Sec. 2.2;

3. the presentation of the dataset according to their UV emission, as well as the complete
   and RSG samples, which are available in Sec. 2.3;

4. exploratory analysis of the complete and RSG samples are provided in 2.4.

## 2.1   Sample selection

In order to build a comprehensive sample of RSGs in terms of wavelength coverage,
quality of observations, and a vast set of data products, the following surveys were used:

   i. the Galaxy Mass Assembly Data Release 3 (GAMA-DR3, Driver et al., 2009; Baldry et al., 2018);

  ii. the Sloan Digital Sky Survey Data Release 7 (SDSS-DR7, York et al., 2000; Abazajian et al., 2009);

 iii. the Galaxy Evolution Explorer Data Release GR6/plus7 (GALEX, Martin et al. 2005).

GAMA-DR3 had been previously aperture-matched with the photometric observations made by SDSS and GALEX. The match procedures can be found in Hill et al. (2011). Details on the characteristics of each respective survey are described in Secs. 2.1.1, 2.1.2, and 2.1.3. The final sets of criteria are displayed in Sec. 2.1.4.

## 2.1.1   GAMA-DR3

The GAMA survey was chosen because of its observing strategy. The strategy behind the observations of GAMA included areas of the sky that had been previously observed by a numerous amount of other surveys, spanning from the UV to the submillimetre/radio wavelengths. Additionally, their public datasets are rich in value-added catalogues, such as emission-line measurements, stellar population parameters, star-formation rate estimations, to enumerate a few. Such characteristics (like the richness data) are of the utmost importance when developing a doctoral thesis, in which one should in all costs avoid research 'bottlenecks'.

GAMA is a spectroscopic and a multiwavelength photometric survey lead by the Anglo Australian Observatory (AAO) and observed by the 3.9m Anglo Australian Telescope (AAT)[1] by making use of the AAOmega spectrograph (Sharp et al., 2006). The survey started in 2008 and has provided three data releases with a total of nearly 220,000 observed objects (Driver et al., 2009; Liske et al., 2015; Baldry et al., 2018). The main focus of GAMA has been to acquire spectroscopic measurements from regions of the sky that have been previously observed by other surveys, specially SDSS and the UKIRT Infrared Deep Sky Survey (UKIDSS, Lawrence et al. 2007) among several others (such description is available in Baldry et al. 2010 and Hill et al. 2011), enabling the acquisition of panchromatic observations for certain sets of data. The AAOmega spectrograph observes in wavelength ranges of 3,700 to 8,800 Å, which is slightly more sensitive in the near-UV and less in the near-IR, when compared to observations from SDSS, which I describe separately in Sec. 2.1.2. Additionally, GAMA spectroscopic observations reach magnitudes up to $\approx 19.8$, whereas SDSS main galaxy sample (MGS) spectroscopy is limited to $\approx 17.7$ (I refer the reader to Table 3 of Driver et al. 2009 and/or Table 1 of Baldry et al. 2018). Thus, I used GAMA-DR3 as a benchmark to select the observations from the following surveys.

---

[1]AAT is home to other important surveys, such as the Sydney/AAO Multi-object Integral-field Spectrograph Survey (SAMI, Bryant et al. 2015) and the WiggleZ Dark Energy Survey (WiggleZ, Blake et al., 2008; Drinkwater et al., 2010), among others. More details can be found at https://www.aao.gov.au/about-us/AAT.

## 2.1.2 SDSS-DR7

SDSS (York et al., 2000) has been a hallmark in observational astrophysics, being one of the first to ever combine a certain threshold of data quality and sky coverage, providing gigantic ever-increasing amounts of data to the community. It represents a major milestone, its data retrieval started in the year 2000 and it is ongoing until the writing of this thesis. SDSS observations are lead by a 2.5m dedicated telescope at the Apache Point Observatory in New Mexico, United States of America. It has been performing observations of the local Universe in both broad-band photometry as well as spectroscopy. The wavelength coverage for the spectroscopic observations has the following range: $3,900 - 9,100$ Å. Photometric observations are retrieved in five bands: *u, g, r, i, z* – their characteristics are summed up in Table 2.1.

SDSS bands cover wavelength ranges that include the near-UV, the entire human eye visible spectrum[2], reaching up to the near-infrared (NIR). SDSS has supplied 16 data releases so far, with the last one documented being DR15 (Aguado et al., 2019). For this study, I made use of SDSS-DR7 (Abazajian et al., 2009) due to its previous cross-match made by the GAMA team (matching details for a combined spectral energy distribution – SED – from different surveys can be found in Hill et al., 2011). For review on the SDSS accomplishments, the reader is referred to Raddick et al. (2014a, 2014b).

Table 2.1: Table featuring the overall characteristics of the SDSS photometric bands; i.e. wavelength coverage of each band, as well as effective wavelength.

| **Characteristics** | *u* | *g* | *r* | *i* | *z* |
|---|---|---|---|---|---|
| Wavelength range (Å) | 2980–4130 | 3630–5805 | 5680–7230 | 6430–8630 | 7730–11230 |
| Effective wavelength (Å) | 3543 | 4770 | 6231 | 7625 | 9134 |

### 2.1.3   GALEX GR6/plus7

GALEX was a 0.5m-space-based telescope (i.e. mounted on a satellite) which gathered observations in both photometry and spectroscopy in the UV range of the electromagnetic spectrum (Martin et al., 2005). GALEX observed approximately 200 million objects and retrieved over 100,000 low resolution spectra. It gathered observations for a little over a decade, starting in the year 2003 and completely decommissioned in 2013, after facing years of technical issues. It operated with both near and far-UV bands (hereafter NUV and FUV) until 2009, when the detector responsible for the FUV measurements failed. From 2009 until 2013 it operated only with NUV observations, which allowed the survey team to observe areas beyond the initial luminosity threshold limit (details on GALEX final observations are available in Bianchi, 2014). Details on the photometry characteristics are displayed in Table 2.2; for further details on calibrations and data products from GALEX, the reader is referred to Morrissey et al. (2007).

---

[2]Which, for comparison purposes, is approximately between 3,800–7,400 Å  (Starr, Evers, & Starr, 2010).

Table 2.2: Table featuring the overall characteristics of the GALEX photometric bands; i.e. wavelength coverage of each band, as well as effective wavelength.

| **Characteristics** | FUV | NUV |
|---|---|---|
| Wavelength range (Å) | 1344–1786 | 1771–2831 |
| Effective wavelength (Å) | 1528 | 2310 |

Table 2.3: The three main GALEX surveys with their respective exposure times ($t_{exp}$) in seconds, sky coverage in square degrees ($deg^2$), and AB magnitude depth for both FUV and NUV bands. Source: Bianchi (2014).

| **GALEX survey** | $t_{exp}$ **(s)** | **coverage ($deg^2$)** | **depth FUV/NUV ($m^{AB}$)** |
|---|---|---|---|
| All Sky Imaging Survey (AIS) | 100 | 26,000 | 20/21 |
| Medium-depth Imaging Survey (MIS) | 1,500 | 1,000 | 22.7 (both) |
| Deep-depth Imaging Survey (DIS) | 30,000 | 80 | 24.8/24.4 |

GALEX performed three main surveys, the All Sky Imaging (AIS), Medium-depth Imaging (MIS) and Deep-depth Imaging (DIS). An overall description of these surveys can be seen in Table 2.3. For this thesis, I made use of the MIS sample, which provides the best compromise in terms of sky coverage and image depth (see Table 2.3), a similar approach taken by previous UV upturn studies making use of GALEX data (for instance, Ree et al. 2007 made use of both MIS and DIS observations to create a sample of brightest cluster galaxies, BCGs). A detailed discussion on the use of the different GALEX surveys to study the UV upturn phenomenon can be found in Yi (2008).

## 2.1.4 GAMA, SDSS, & GALEX selection criteria

The dataset was selected by making use of data management units (DMUs) from the GAMA-DR3 survey[3]. The main DMUs used were: `ApMatchedCat`, `InputCatA`, `SpecLinesSFR`, and `GalexMain`. Other DMUs have been used and will be mentioned in the applicable Chapters. The criteria used to build the first sample of galaxies is described as follows.

---

[3]Which can be accessed at http://www.gama-survey.org/dr3/http://www.gama-survey.org/dr3/

   i. The GAMA-DR3 catalogue identification number – i.e. `CATAID` – was used across all data selection steps, including the retrieval of data products in DMUs mentioned in further Chapters;

  ii. the dataset was 'filtered' in order to select objects identified as galaxies (`TYPE=3`) in `InputCatA`;

 iii. only objects detected in all five SDSS and both GALEX bands were considered, excluding objects missing data: -9999.0;

  iv. objects with multiple match results (for instance one observation in the visible, but two in ultraviolet) were not considered (`NMATCHUV=1` and `NMATCHOPT=1`);

   v. flags that indicated potential issues with UV observations were taken into account: `FUVFLAG=0` and `NUVFLAG=0`;

  vi. to ensure good quality of $z$ measurements, the recommendations by Baldry et al. (2018) were adopted:

      a. the probability of $z$ being correctly estimated: `PROB>0.8`;

      b. the normalised quality parameter for $z$, in order to select systems with acceptable quality: `NQ>2` – as described by Liske et al. (2015, Sec. 2.3.4);

 vii. regarding the spectral signal-to-noise ratio (S/N, no restriction was imposed as it could impact the number of objects in the sample, as discussed in Cid Fernandes et al. (2010). Given that the usual cut for emission-line detection is S/N>3 (see Kauffmann et al., 2003a)[4], it is important to note that only about 0.4% of the final sample (described in Sec. 2.3.1) is impacted with objects with S/N<3;

---

[4]This will be important in Chapter 4.

As a consequence of the aforementioned criteria, the minimum and maximum $z$ were naturally limited to $z_{min} = 0.06$ and $z_{max} = 0.4$, which is our final range of $z$.

## 2.2 Treatment

In this Section I describe the treatment applied to the aforementioned dataset in order to estimate AB magnitudes and, consequently, absolute magnitudes. Such treatment includes extinction, k-corrections, and instrumental offsets when applicable.

### 2.2.1 AB magnitudes

The magnitudes used in this work are in the AB system (as defined by Oke & Gunn, 1983). Equation 2.1 displays the corrections made onto the observed magnitudes in order to estimate the 'corrected' magnitudes of our sources:

$$m_i^{AB} = m_i^{obs} - e_i \pm k_i \pm o_i, \tag{2.1}$$

in which:

  i. $m_i^{AB}$ is the corrected magnitude for the $i^{th}$ band;

 ii. $m_i^{obs}$ is the observed magnitude (or apparent magnitude) for the $i^{th}$ band;

iii. $e_i$ is the galactic extinction for the $i^{th}$ band;

 iv. $k_i$ is the k-correction for the $i^{th}$ band;

  v. $o_i$ is the offset for the $i^{th}$ band (only applicable to SDSS observations due to equipment degradation, see Doi et al., 2010).

In what follows – i.e. Secs. 2.2.1.1 to 2.2.1.4 – I discuss each item in Eq. 2.1 and show how they were estimated.

#### 2.2.1.1   Apparent (or observed) magnitude

Apparent magnitude or observed magnitude is a measure of the brightness of an object as seen from the Earth. The first magnitude system was created by the (ancient) Greek astronomer Hipparchus (160-125 BCE) and formalised in modern times by Pogson (1856). A major review of astronomical magnitude systems and their calibration is given by Bessell (2005).

In the AB system adopted throughout this thesis (Oke & Gunn, 1983), magnitudes are defined in terms of flux in erg s$^{-1}$ Hz$^{-1}$ cm$^{-2}$ (Eq. 2.2), thus relating the magnitude measurement to physically interpretable values:

$$m_{AB} = -2.5 \log f_\nu + 48.6 \qquad (2.2)$$

#### 2.2.1.2   Foreground extinction

Extinction can be briefly defined as the light that is scattered and/or absorbed by a certain medium; that leaves us with basically two kinds of extinction: foreground and internal. The first takes into account the 'loss' of radiation due to the scattering and absorption that happens in our observable foreground sky (in other words, the effects due to the gas and dust close to us, that is inside our own galaxy – the Milky Way); such effects must be taken into account for all sorts of observations independently of being point-like (e.g. stars) or extended sources (e.g. galaxies). On the other hand, when observing extended objects, such as galaxies, these also suffer from effects of their own gas and dust, which causes their 'internal extinction'. In this Sec. I discuss only the foreground extinction, i.e. the one caused by the Milky Way.

Most extinction laws are in fair agreement in the visible region of the electromagnetic spectrum, but the UV range is subject to a vast discussion (e.g. Kong et al., 2004; Conroy, 2010; Peek & Schiminovich, 2013; Gordon et al., 2016; Narayanan et al., 2018; Werle et al., 2019) and it is mostly divergent among such laws (such as Seaton, 1979; Cardelli,

Clayton, & Mathis, 1989; Fitzpatrick, 1999; Calzetti et al., 2000). This problem arises from the fact that absorption and scattering by gas and dust is still a matter of debate, when it comes to the UV. Examples of different extinction laws are displayed in Fig. 2.1

To estimate the foreground extinction, I made use of the extinction law described by Fitzpatrick (1999), using the maps of Schlegel, Finkbeiner, and Davis (1998), and implemented it by making use of the PYTHON package PYNEB (Luridiana, Morisset, & Shaw, 2015). The extinction law proposed by Fitzpatrick (1999) is still one of the best choices when dealing with UV observations, as it was developed after several works on UV extinction curves retrieved from the International Ultraviolet Explorer (IUE, Boggess et al. 1978): Fitzpatrick and Massa (1986, 1988, 1990).

**Extinction in a nutshell**    In what follows, I present how excess colour is estimated and how it gives us the parameters needed to explore the curves for the Milky Way (Metchev, 2013).

$$E(B - V) = A_B - A_V = (B - V) - (B - V)_0, \qquad (2.3a)$$

in which:

i. B and V are the standard bands for colour excess mapping;

ii. E(B-V) is the colour excess of the colour (B-V) as provided by Schlegel, Finkbeiner, and Davis (1998);

iii. $A_B$ and $A_V$ are the photometric extinctions for B and V bands.

That can also be written as:

$$E(B - V) = A_B - A_V = \left( \frac{A_B}{A_V} - 1 \right) A_V. \qquad (2.3b)$$

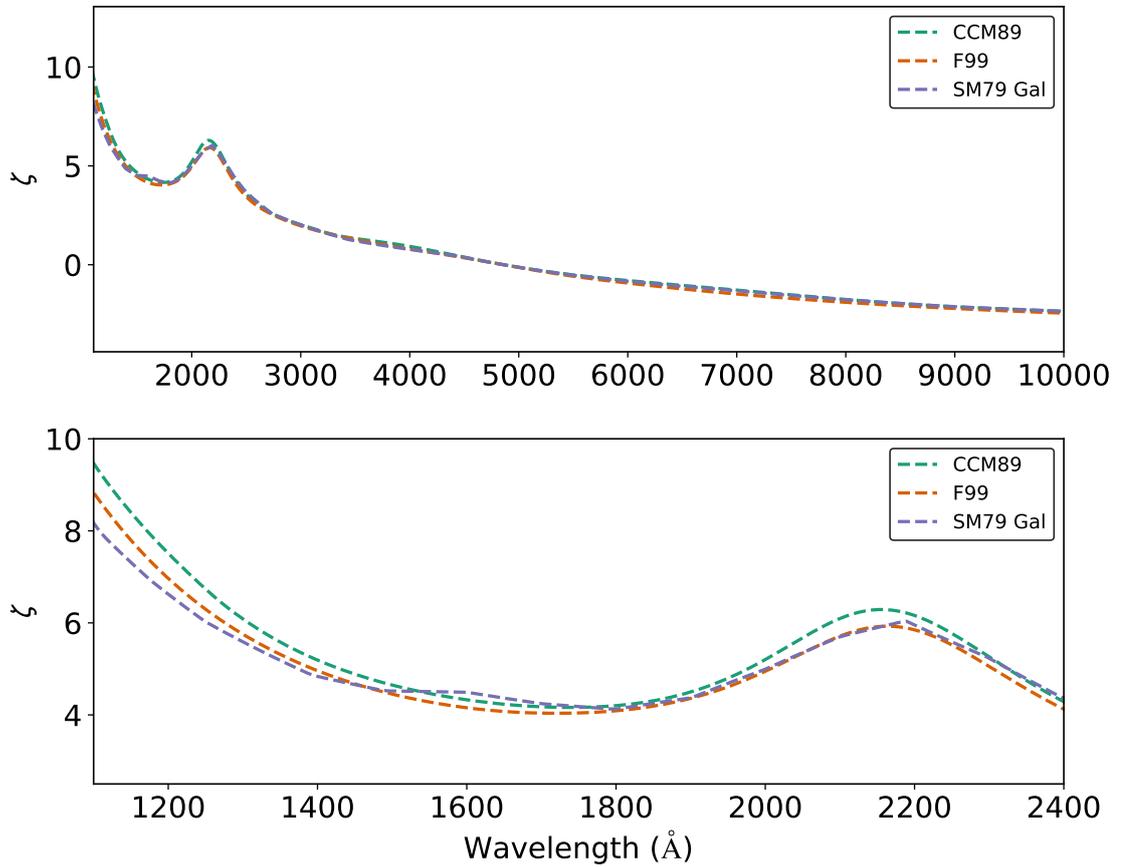

Figure 2.1: Examples of three extinction laws: Seaton (1979) represented by 'SM79 Gal',
Cardelli, Clayton, and Mathis (1989) represented by 'CCM89', and Fitzpatrick (1999) represented
'F99'. The x and y axes, respectively, represent the wavelength (Å) and $\zeta$ is a function that
represents the excess colour in terms of wavelength ($\lambda$): $\zeta = E(\lambda - V)/E(B - V)$. The upper
panel displays the entire curve of extinction between 1,000–10,000 Å, while the lower panel is
zoomed into 1,100–2,400 Å. This figure was created based on the implementation of the curves
in the following PYTHON package: PYNEB (Luridiana, Morisset, & Shaw, 2015).

Since, $\frac{A_B}{A_V} \approx 1.324$ for the Milky Way, we have that

$$E(B - V)) = 0.324 A_V. \tag{2.3c}$$

From Eq. 2.3c, we have that E(B-V) is measured throughout the sky area, 0.324 is a constant associated with the Milky Way, therefore:

$$\therefore A_V \simeq 3.1 E(B - V) \tag{2.3d}$$

in which 3.1 is the so-called $R_V$ value (see Rieke & Lebofsky, 1985; Cardelli, Clayton, & Mathis, 1989; Fitzpatrick, 1999). It is the $R_\lambda$ that changes for each extragalactic object in each wavelength $\lambda$, which gives us:

$$\therefore R_\lambda = \frac{A_\lambda}{E(B - \lambda)} \tag{2.3e}$$

in which $A_\lambda$ is the extinction at a given wavelength $\lambda$.

### 2.2.1.3 K-correction

K-corrections are necessary when dealing with photometry of sources that are at different $z$. In other words, K-correction is a tool used to 're-frame' these objects at a common $z$ (it can be $z = 0$ or $z = 0.1$, for instance) with the purpose of making comparisons among them.

For extragalactic objects, one must estimate the K-corrections for each galaxy and bandpass. This is usually done by making use of SED fitting (e.g. Blanton & Roweis, 2007), but other methods have also been proposed in the literature (e.g. Chilingarian, Melchior, & Zolotukhin, 2010; O'Mill et al., 2011).

In this work, K-corrections have been previously estimated by the GAMA team which are available on the DMU kCorrections; they used the package K_CORRECT (Blanton

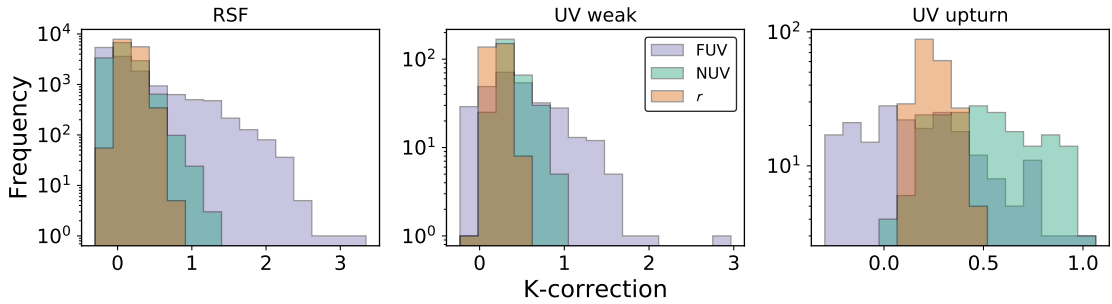

Figure 2.2: K-corrections the three main bands used in this study: SDSS *u*-band and both GALEX bands (FUV and NUV). The sub-plots depict the different classes according to the paradigm of Yi et al. (2011), which are: residual star-formation (RSF), UV weak, and UV upturn systems. These are described in Sec. 2.3.

Table 2.4: SDSS-DR7 apparent magnitude offsets as described by Doi et al. (2010).

| **SDSS-DR7 bands** | *u* | *g* | *r* | *i* | *z* |
|---|---|---|---|---|---|
| **Offset correction** | -0.04 | +0.01 | +0.01 | +0.01 | +0.02 |

& Roweis, 2007) and provided K-corrections for $z = 0.0$ and $z = 0.1$ – I chose the first option. It is worth mentioning that K_CORRECT makes use of Bruzual and Charlot (2003) stellar population models and Chabrier (2003) initial mass function (IMF). The distributions of the K-corrections for the most important bands (SDSS *r*-band, and both GALEX bands) used in this work are displayed in Fig. 2.2. Further details on the UV classes displayed on it are described in Sec. 2.3.

### 2.2.1.4    Offsets

Offsets are particularly applicable to measurements of SDSS-DR7 bands, specially the *u*-band that is susceptible to near-UV radiation. Due to deterioration of parts of SDSS instrumentation – which was further corrected – offset corrections in the observed magnitudes were made necessary for DR7 (Doi et al., 2010). Such offsets are displayed in Table 2.4.

### 2.2.2 Absolute magnitudes

Absolute magnitudes are estimated according to the equation 2.4.

$$M_i = m_i^{\mathrm{AB}} - 5\log(D_i^L),\qquad(2.4)$$

in which:

i. $M_i$ is the absolute magnitude for the $i^{\mathrm{th}}$ band;

ii. $D_i^L$ is the luminosity distance for the $i^{\mathrm{th}}$ band in Mpc;

iii. other variables have been previously defined.

In order to estimate $D_i^L$, I made use of the PYTHON package ASTROPY (Astropy Collaboration et al., 2013). To that end, a set of cosmological parameters have been adopted as follows.

**Adopted cosmology**   Throughout this work, I have considered the standard Λ-(Λ-CDM) cosmological model with the parameters retrieved from Taylor et al. (2011), as follows:

i  Hubble constant, $H_0 = 70 \mathrm{km\,s^{-1}\,Mpc^{-1}}$;

ii  mass contribution (baryonic and dark): $\Omega_M = 0.3$;

iii  dark energy contribution: $\Omega_\Lambda = 0.7$.

## 2.3 UV characterisation of the galaxy sample

In order to classify the sample of galaxies herein used, I have made use of the prescription made by Yi et al. (2011). In this paper, the authors define three criteria in order to segregate galaxies with different UV emission characteristics. Their norms are as follows.

i. (NUV-$r$) > 5.4: this is an attempt to limit the contamination of young stellar populations. It is marked by the long vertical line in Fig. 2.3, in which bluer objects on the left side of the chart are characterised as those presenting *residual star formation* (RSF) therein marked in green circles. Those on the right side (represented by two shades of orange) are the ones we refer in this thesis as *UV bright red-sequence galaxies* (RSGs);

ii. (FUV-NUV) < 0.9: this criterion measures the rising slope for lower wavelengths, which is an attempt to measure the *strength* of the upturn – in Fig. 2.3 this is marked by the horizontal black line on the right side of the chart;

iii. (FUV-$r$) < 6.6: this criterion is a measure of the strength of the FUV flux. These objects are not directly represented in Fig. 2.3, as it is not one of the dimensions in which it is displayed. However, objects that are within these characteristics are displayed in dark orange diamond-shaped markers and are those we refer to as *UV upturn*. Systems that respect criteria (i) and (ii), but not this one are the so-called *UV weak* systems, which are displayed in light orange squares.

### 2.3.1   The complete and the RSG samples

The complete sample is comprised of 14,331 objects, with most of them being classified as RSF – this sample is revisited in Chapter 4. Nonetheless, the main focus of the investigation presented in this thesis is the UV bright RSGs, which exclude the RSF class.

Therefore, by making use of the criteria described in Sec. 2.1, treating the dataset as described in Sec. 2.2, and classifying the objects according to their UV class by making use of the prescription suggested by Yi et al. (2011), the final sample of UV bright RSGs is as follows: **it is constituted of 506 objects, of which 296 are classified as UV weak and 210 as UV upturn.**

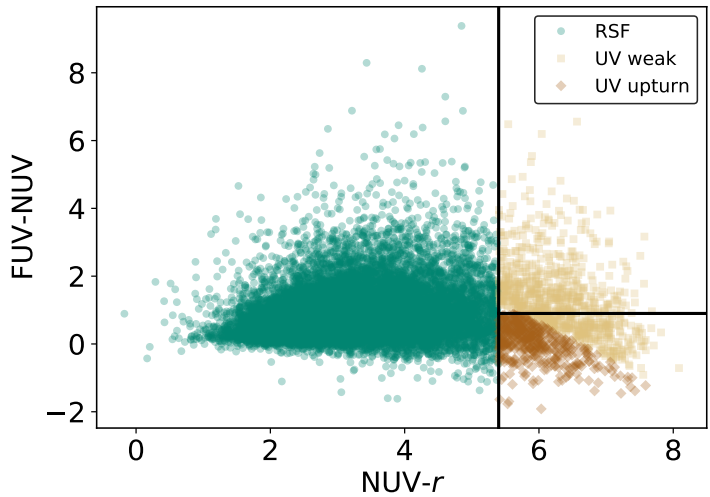

Figure 2.3: Colour-colour diagram with the UV classes according to Yi et al. (2011) featuring UV (y-axis: FUV-NUV) and UV-optical (x-axis: NUV-$r$) colours. The vertical line at (NUV-$r$) = 5.4 and the horizontal line at (FUV-NUV) < 0.9 expose two out of the three criteria to used to select UV weak and upturn systems.

## 2.4 Exploratory analysis

In this Section I briefly explore some characteristics of the sample of UV bright RSGs, namely the distributions of stellar masses (log $M_\star$, Fig. 2.4)[5], and colour-magnitude diagrams (Figs. 2.5 and 2.6).

Fig. 2.4 depicts the cumulative distribution function (CDF) for log $M_\star$. It shows that UV upturn systems tend to be slightly

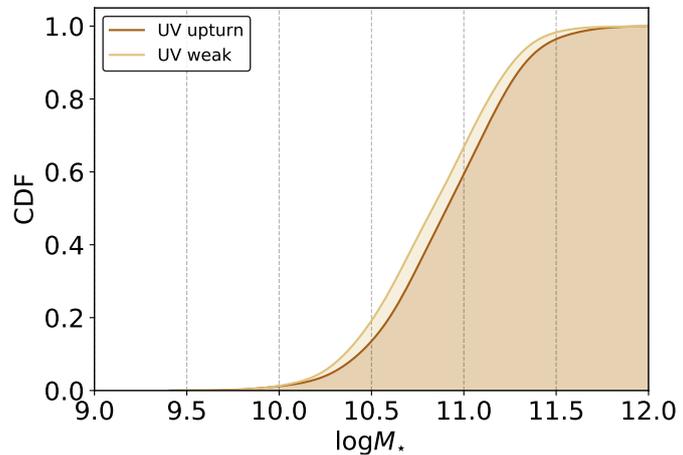

Figure 2.4: Cumulative distribution function (CDF) for log $M_\star$ for UV bright RSGs.

[5]In solar masses.

more massive when compared to the distribution of their UV weak counterparts.

In terms of colour-magnitude diagrams, Fig. 2.5 shows that UV upturn galaxies are slightly redder in the optical when compared to their UV weak counterparts (see $g - r$ against $M_r$ therein, as well as $g - r$ against $M_{FUV}$ and $M_{NUV}$ in Fig. 2.6). The effects of the UV classification criteria described by Yi et al. (2011) can be seen in the distributions of FUV-NUV, NUV-$r$, and FUV-$r$ colours against $M_r$, $M_{FUV}$ and $M_{NUV}$ (respectively in Figs. 2.5 and 2.6).

An interesting difference that can be seen in Fig. 2.6 is that the distributions of $M_{FUV}$ for each UV class are relatively separated (in the sense that UV upturn systems peak at higher luminosities than UV weak galaxies), whereas for $M_{NUV}$ they mostly overlap. This is an effect of the classification proposed by Yi et al. (2011), since UV weak and UV upturn galaxies are mainly classified using two colours dependent on FUV (FUV-NUV and FUV-$r$), whereas NUV-$r$ is only used to eliminate RSF systems. Therefore, the distributions featuring NUV-$r$ and $g - r$ are far more blended than the others.

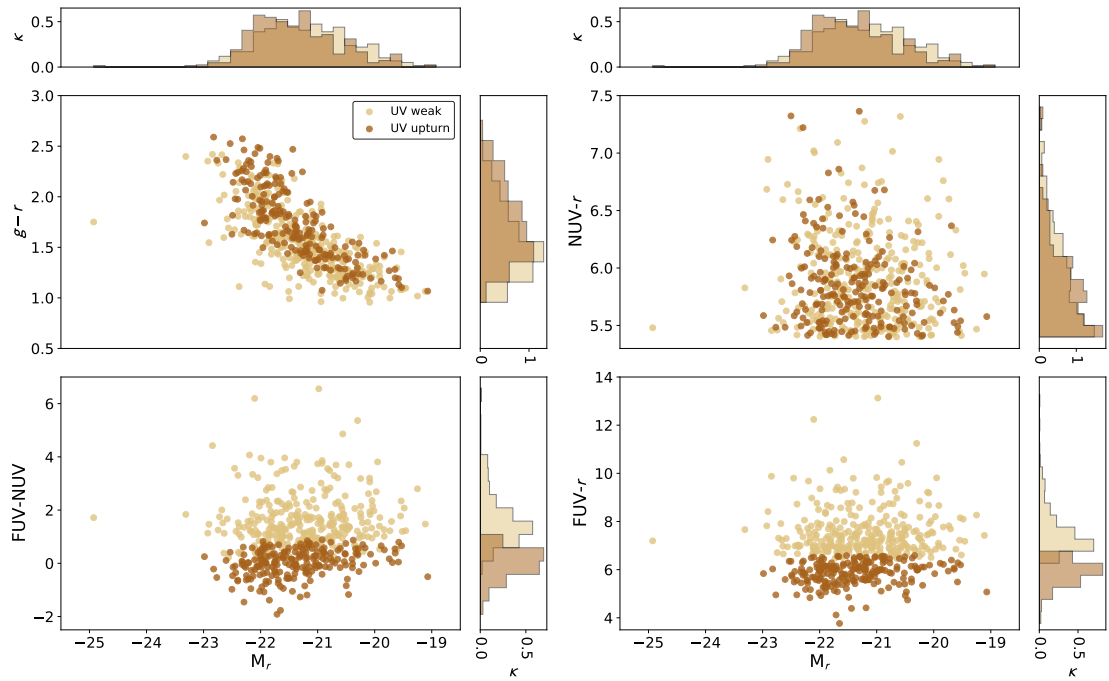

Figure 2.5: Colour-magnitude diagrams for four different colours with their respective normalised distributions ($\kappa$) on the adjacent histograms. From top to bottom: (first row) $g - r$ and NUV-$r$ against absolute magnitude in the $r$ band ($M_r$), (second row) FUV-NUV and FUV-$r$ also against $M_r$.

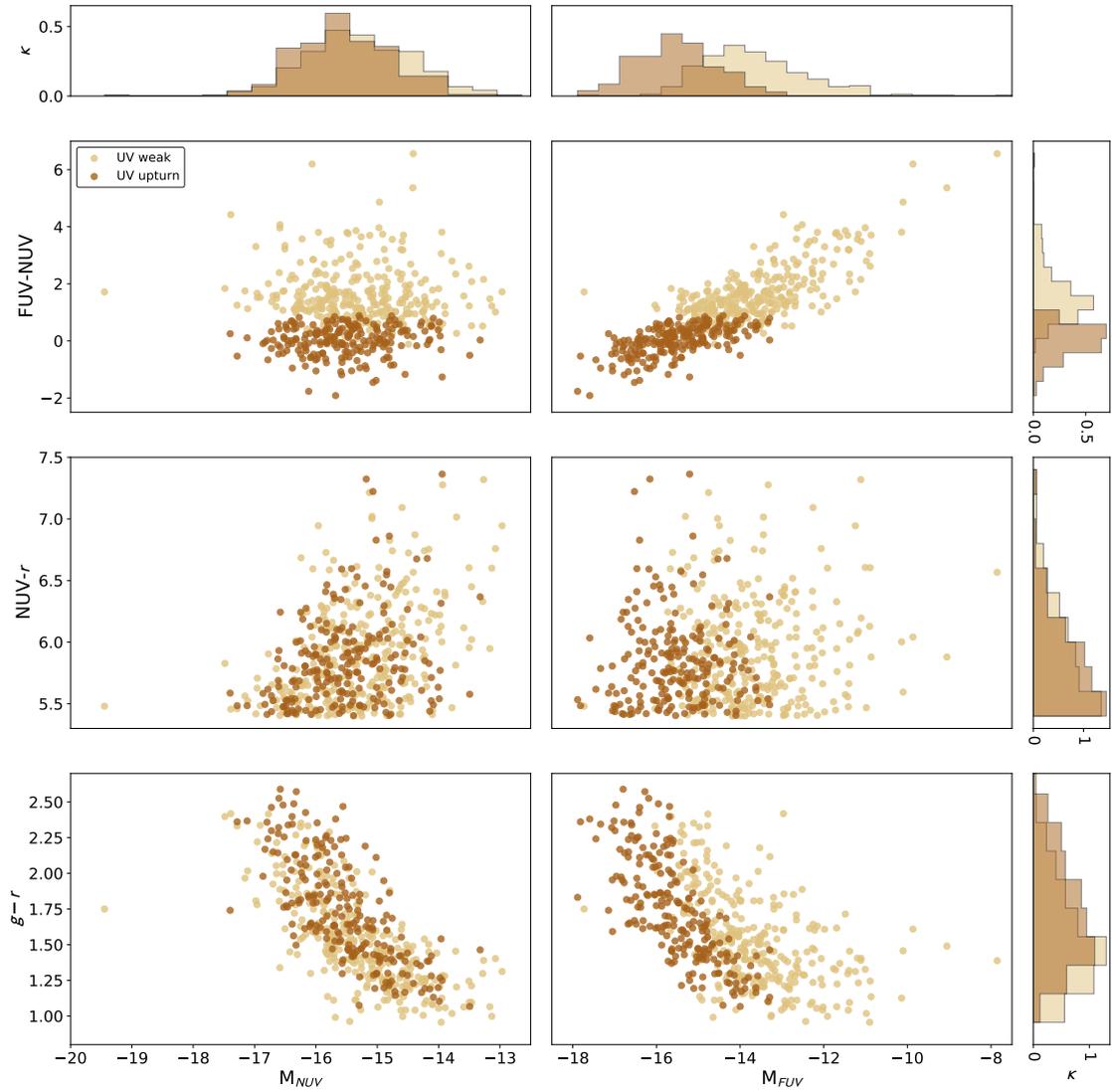

Figure 2.6:  Colour-magnitude diagrams for three different UV/UV-optical colours and two UV absolute magnitudes.  The first row displays FUV-NUV according $M_{NUV}$ and $M_{FUV}$ respectively; the second and third, NUV-$r$ and $g - r$ respectively according to the same absolute magnitudes.  Their respective normalised distributions can be seen in the adjacent histograms represented by $\kappa$.

If you're not growing, you are dying;
and if you're dying, then, by
definition, you're not living a
meaningful life.

---

Joshua Fields Millburn &
Ryan Nicodemus
*Minimalism: Live a Meaningful Life.*

# 3

# Evolution

---

**Based on:**

Dantas, Coelho, de Souza, and Gonçalves (2020);

de Souza, Dantas, Krone-Martins, Cameron, Coelho, Hattab, de Val-Borro, Hilbe, Elliott, Hagen, and COIN Collaboration (2016).

---

The question of whether the UV upturn evolves – that is, changes throughout cosmic time – has been deeply investigated throughout several previous works, as described in Sec. 1.3.1.2. In this Chapter, I propose a new outlook on the way the evolution of the UV upturn could be addressed; instead of taking into account the changes in the strength of the upturn throughout $z$ (such as what has been made by Brown et al., 1998; Rich et al., 2005; Ree et al., 2007; Ali et al., 2018b), one can investigate whether the fraction of UV upturn evolves when compared to the entire UV bright RSG population.

With the goal of mitigating some of the observational biases, such as the Malmquist bias (Malmquist, 1922; Sandage, 2000), $\log M_\star$ has also been directly embedded in the analysis. This approach has a second benefit, as it also provides results for the incidence of the UV upturn phenomenon in terms of $\log M_\star$, which will be discussed throughout this Chapter. To grasp potential effects of the Malmquist bias, the distribution of $M_r$ against bins of $z$ (with width of 0.05) is displayed for UV weak and UV upturn galaxies in Fig. 3.1. It is possible to see that the distributions of UV weak and upturn systems change monotonically with bins of $z$. Also, as supported by the distribution patterns seen in Fig. 3.2, the fraction is suitable to damp selection biases, as both groups of galaxies are potentially subject to the same effects.

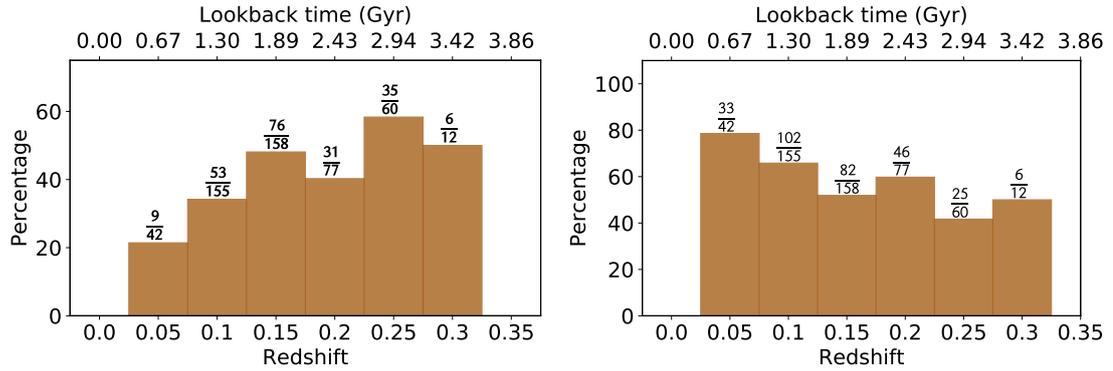

Figure 3.2: Proportion of UV upturn (left panel) and weak (right panel) systems in different bins of $z$ (lower x-axis) and corresponding look-back time (upper x-axis).

With this in mind, in this Chapter I describe the methodology used and the results for the evolution of the UV upturn in terms of $z$ and $\log M_\star$.

## 3.1   Probing the evolution: methodology

In this Section, the methodology applied in order to probe whether the UV upturn evolves or not in $z$ is presented.

### 3.1.1   Clues of evolution

In order to check whether there is a potential evolution with $z$ of the UV upturn, exploratory bar-plots depicting the fraction of UV upturn and UV weak among the entire UV bright RSG population in bins of $z$ were made and are available in Fig. 3.2. It is possible to see a growth in the fraction of UV upturn galaxies (left panel) up to $z \sim 0.25$ (look-back time around 3 Gyr). Conversely, the fraction of UV weak RSGs appears to grow with decreasing $z$ (right panel).

Indeed, such approach can only be interpreted with a lot of caveats, since bar-plots – just as histograms – are 'fragile' in what concerns data interpretation: they are subject to effects of binning, and consequently they do not convey the same physical interpretability as regression models.

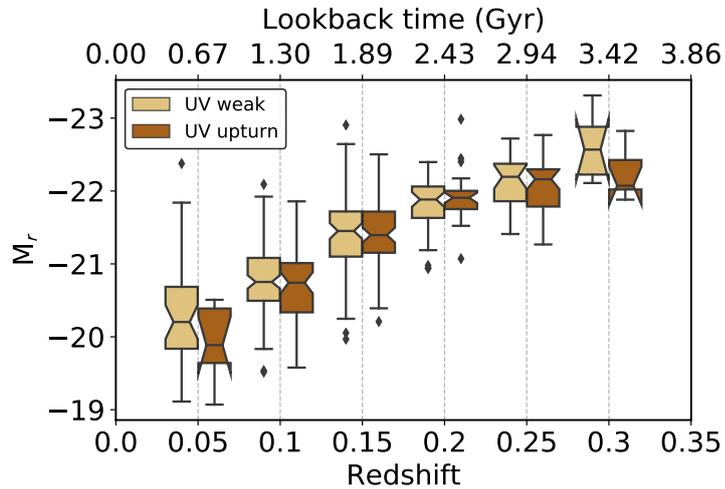

Figure 3.1: Distribution of absolute SDSS $r$-band magnitudes ($M_r$) in $z$ bins represented by boxplots. The UV upturn systems are represented by the dark brown ones, whereas the UV weak are represented by the light brown boxplots. The interquartile ranges are displayed by the coloured regions; the bars display approximately $3\sigma$ of the distribution; the notch represent 96% of confidence interval around the median (horizontal line); the outliers are displayed as diamond-shaped markers (fliers). More details on the information synthesised in boxplots can be found in Tukey (1977), and more recently in Hofmann, Wickham, and Kafadar (2017).

Another important caveat is the differences of luminosity of each group of galaxy; UV upturn systems are brighter in the UV when compared to their UV weak counterparts. Because of that, by moving towards higher $z$, the detection of galaxies becomes harder, specially for UV weak systems, creating an impression of higher amounts of UV upturn galaxies in higher $z$. The model deals with this issue in a way that a simple bar-plot cannot, by verifying the distribution of all the systems throughout $z$, not only their bins.

Nonetheless, this exploratory step certainly entices a deeper investigation into this subject.

### 3.1.2   Binomial models

Binomial models are extremely valuable to analyse a myriad of astrophysical problems, although they are still rarely used in the area.  An example of the use of binomial models is described in the work by de Souza et al. (2016), who have analysed the fraction of galaxies hosting Seyfert activity in terms of their cluster/group centric distance and morphological class.  In such, the model was crucial to show that the fraction of ellipticals harbouring AGN rises with increasing distance, whereas spirals are not affected by the environment.

The binomial regression is useful to model problems containing binary variables; for instance the presence or not of AGN in a galaxy, the existence or not of a habitable planet around a star, and so on.  Hence, a binomial distribution describes a sequence of independent trials (or experiments), which possess two possible results: 0 or 1, just like the flipping of a coin.  The variable in question then assumes integer values of 1 or 0, to indicate such 'yes' or 'no' condition.  Intermediate values are, therefore, not considered in model, as well as values lower than 0 and higher than 1.  The binomial function is expressed by Eq. 3.1:

$$f(y; p, m) = \binom{m}{y} p^y (1 - p)^{m-y},  \tag{3.1}$$

in which

$$\binom{m}{y} = \frac{m!}{y!(m-y)!};  \tag{3.2}$$

$y$ is the response variable, which is the one assuming the values of 0 or 1; $p$ is the probability; and $m$ is the number of trials.

For the problem herein tackled, UV upturn systems receive the response variable $y = 1$, while UV weak galaxies receive $y = 0$.  In the current application, the Bernoulli

model – a particular case of the Binomial distribution – is more suitable to this problem; it is further described in Sec. 3.1.2.1.

### 3.1.2.1 Bernoulli logistic model

The Bernoulli distribution is a particular case of the binomial distribution, in which $m = 1$, simplifying Eq. 3.1:

$$f(y; p) = p^y (1 - p)^{1-y}.$$ (3.3)

In order to model the fraction of UV upturn galaxies among all UV bright systems in terms of $\log M_\star$ and $z$, a customised Bayesian logistic regression (see Hilbe, de Souza, & Ishida, 2017, section 5.3 for an overview) was used. To that end, I made use of the STAN (Gelman, Lee, & Guo, 2015), a programming language focused on statistical modelling, through PYSTAN (Riddell et al., 2018), a PYTHON package that serves as an interface with STAN. Equations 3.4–3.6 describe the logistic model applied to the problem.

$$y_i = \text{Bern}(p_i)$$ (3.4)

$$\eta_i \equiv \log\left(\frac{p_i}{1 - p_i}\right)$$ (3.5a)

$$\eta_i = X_i \beta_i,$$ (3.5b)

which in this case can be re-written as:

$$\eta = \beta_0 + \beta_1 \log M_\star + \beta_2 \log M_\star^2 + \beta_3 z + \beta_4 z^2.$$ (3.6)

Additionally, the log-like likelihood for the Bernoulli model is:

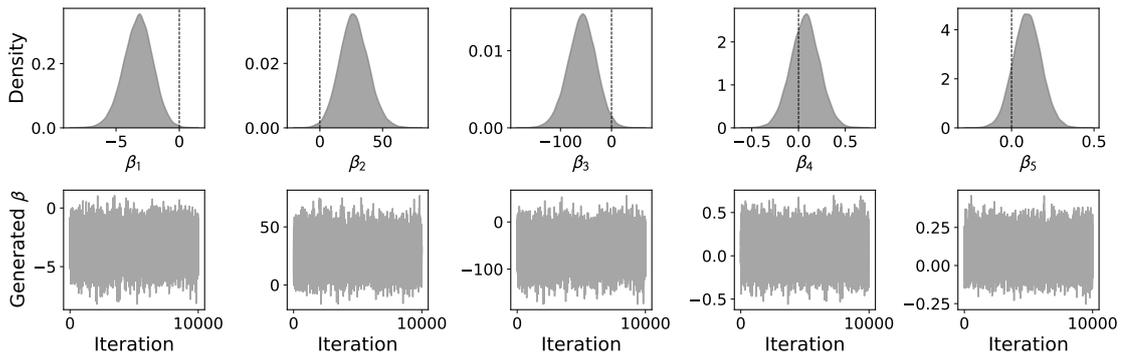

Figure 3.3:  Posteriors represented by $\beta_1$ to $\beta_5$.  The first row is simply the kernel density of each $\beta$, with a vertical dashed line at $x = 0$ for reference.  The second row is filled with the correspondent trace-plots, which enable us to verify whether the HMC chains converged or not.

$$\mathscr{L}(p, y) = \sum_{i=1}^{n} \left\{ y_i \ln \frac{p_i}{1 - p_i} + \ln(1 - p_i) \right\} \tag{3.7}$$

for further information, the reader is referred to Chap. 8 in Hilbe, de Souza, and Ishida, 2017.

Hamiltonian Markov chains (HMC, Brooks et al., 2011; Carpenter et al., 2017) sampler was initiated by making use of different initial values.  The total number of iterations was set at 15000 with 5000 initial burn-in phases – in other words, the 5000 initial steps were discarded.  The priors are weakly informative for $\beta_i$: the mean is set to zero and standard deviation at 10, assuming a Normal distribution.  These parameters are sufficient to ensure the convergence of the model according to the Gelman-Rubin statistics (Gelman & Rubin, 1992).  The STAN code for reproducing this model is available in Appendix A.

### 3.1.2.2   Posteriors

In this brief Section, I describe the results for the posteriors of the logistic model (depicted in Fig. 3.3).  The posteriors are the $\beta$ values described in Eqs. 3.5 and 3.6.

Fig.  3.3 displays the distributions of $\beta_i$, as described in Eq.  3.6 (top row).  The corresponding trace-plots are displayed on the bottom row. Dashed lines are drawn at $x = 0$, serving as visual aid; the further away $\beta$ is from zero, the more important the term is to the regression. The values of $\beta_i$ highlight how important each term is to the regression.

The trace-plots depicted in the second row of Fig. 3.3 indicate how well the chains converged. In other words, if they are spread as they are – and showing no patterns such as 'knots' –, it means that the chains converged successfully, which is the case.

## 3.2 Results: dependence on redshift and stellar mass

In this Section we present and discuss the results for the model described in Sec. 3.1. Figs. 3.4 to 3.7 depict the main results for the evolution of the fraction of the UV upturn ($f_{\text{upturn}}$). It can also be interpreted as the probability of a UV bright RSG of harbouring the UV upturn phenomenon.

Fig. 3.4 depicts a slice of the regression in $z$ (with $\log M_\star$=11, left image) and another in $\log M_\star$ (with $z$=0.261, right image). In both images 50% and 95% of credible intervals are displayed in the blue shaded areas.

In Fig. 3.5 such results are displayed in a 3-dimensional space, with $f_{\text{upturn}}$, $\log M_\star$, and $z$ simultaneously; the credible intervals are not displayed to ease the visualisation. Also, these results can be seen from different angles, exposing the horse saddle shape of the regression.

The results for $\log M_\star$ show a decrease in $f_{\text{upturn}}$ for $\log M_\star$ up to $\sim 11$ followed by an accentuated increase. One must have in mind that galaxies with $\log M_\star$<10 are nearly absent (see Fig. 2.4) and, therefore the results for $\log M_\star$<10 are not very reliable. In Fig. 3.6 a dashed line is added at $\log M_\star$=10 and the area below it is tinted with light grey.

For what concerns $z$, $f_{\text{upturn}}$ seems to increase until $z \sim 0.25$ followed by an in-fall.

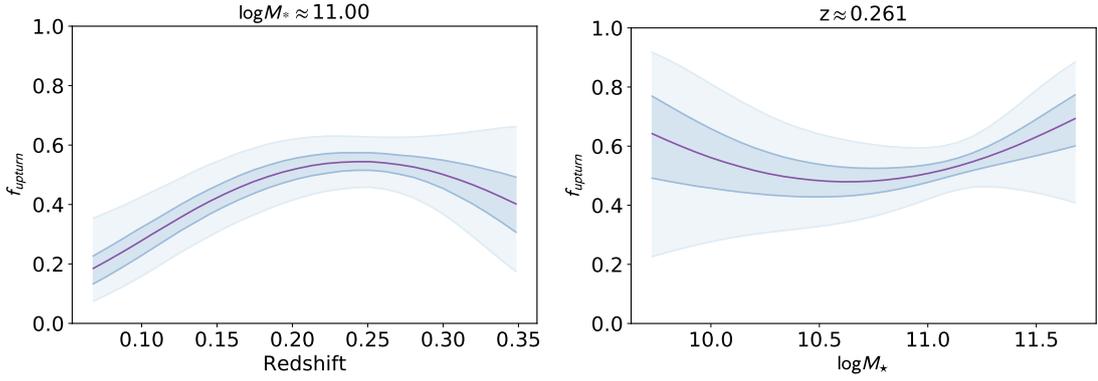

Figure 3.4:  Fraction of UV upturn systems in terms of $z$ and log $M_\star$, respectively.  On the left panel, the fraction of UV upturn galaxies are displayed in a slice of log $M_\star \approx 11$ in terms of $z$; the results indicate that the fraction of UV upturn galaxies rises up until approximately $z \sim 0.25$ with a subsequent in-fall up to $z \sim 0.35$.  On the right panel, the fraction of UV upturn systems in a slice of $z$ – in this case, the $z$ median, $z \approx 0.261$ – in terms of log $M_\star$; the results indicate that the fraction of UV upturn galaxies rises up with log $M_\star$ from log $M_\star \approx 10.5$.  The blue shaded areas depict intervals with 50% and 95% of credibility.

However, with increasing $z$, the credible intervals tend to widen up – specially due to the decreasing number of systems – and the actual trend after at $z > 0.25$ is unknown.  The credible intervals indicate a higher chance of it decreasing at $z > 0.25$, but it is possible that the trend plateaus or even continues to increase.  For log $M_\star$ around 11, the credible intervals are quite tight, showing a nice behaviour in the rise of $f_{\text{upturn}}$.  To properly answer what happens when $z > 0.25$, better data for intermediate-to-high $z$ is necessary. These results are in agreement with the previous suggestions of Boissier et al. (2018), who mentioned a higher number of UV upturn systems around $z \sim 0.25$.

All in all, the results point to an evolution of $f_{\text{upturn}}$ in terms of $z$, as well as log $M_\star$.

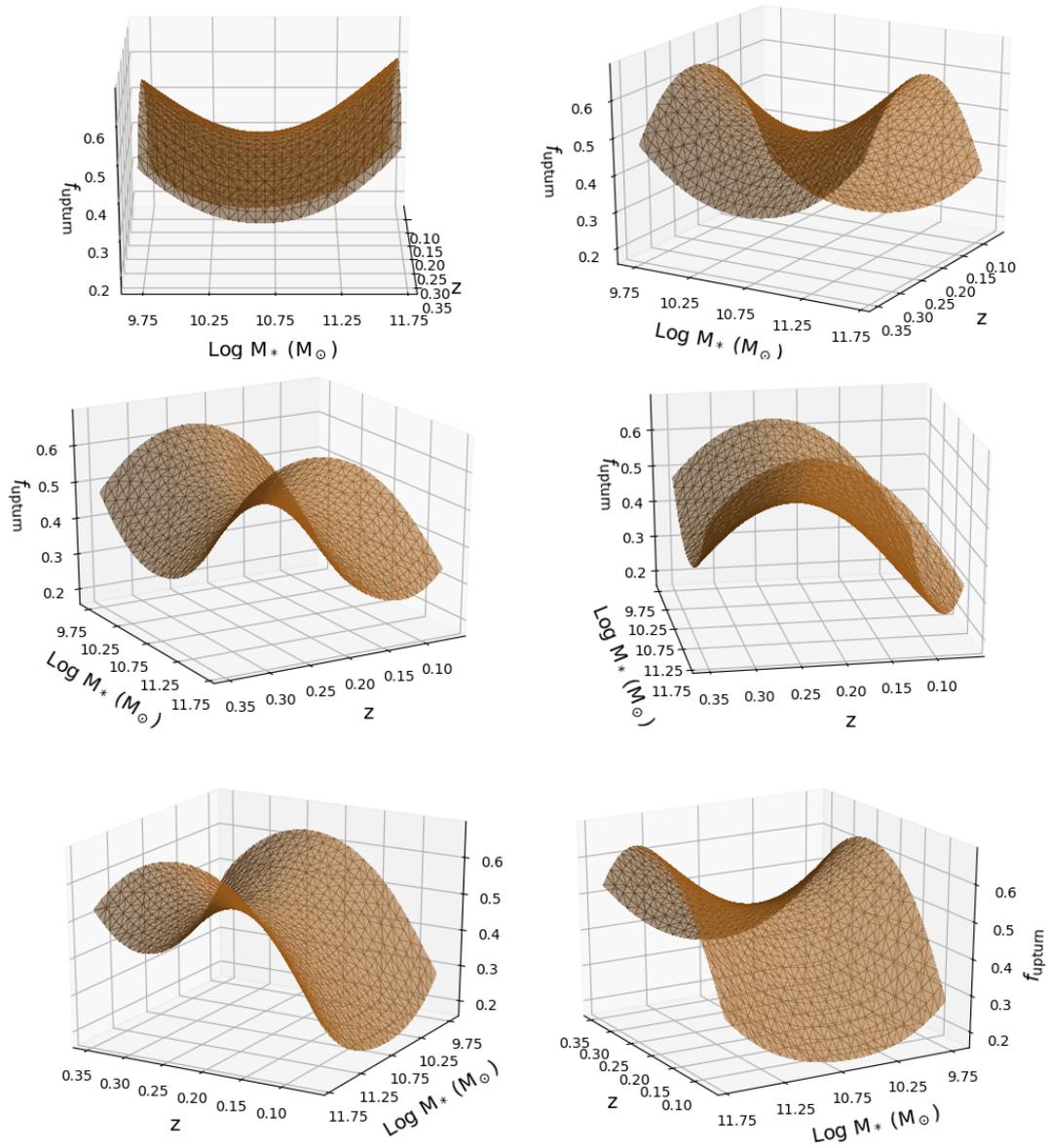

Figure 3.5: 3D regression results for $\log M_\star$ and $z$ seen through 6 different angles. The results depict $f_{upturn}$ which turns out to be in the shape of a horse saddle.

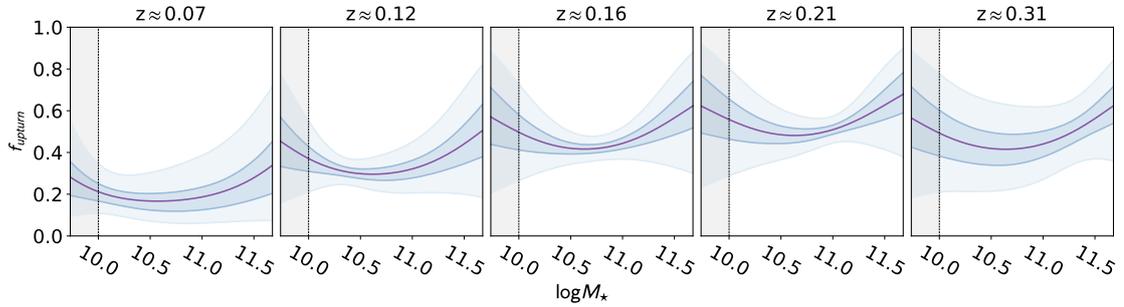

Figure 3.6: Results depicting the dependence of the $f_{\text{upturn}}$ with $\log M_\star$ in five redshift slices: 0.07, 0.12, 0.16, 0.23, 0.31. The blue shaded areas depict intervals with 50% and 95% of credibility. The dashed line in $\log M_\star = 10$ with the shaded area below is a reminder for the reader regarding the almost absent values of $\log M_\star < 10$ – see Fig. 2.4.

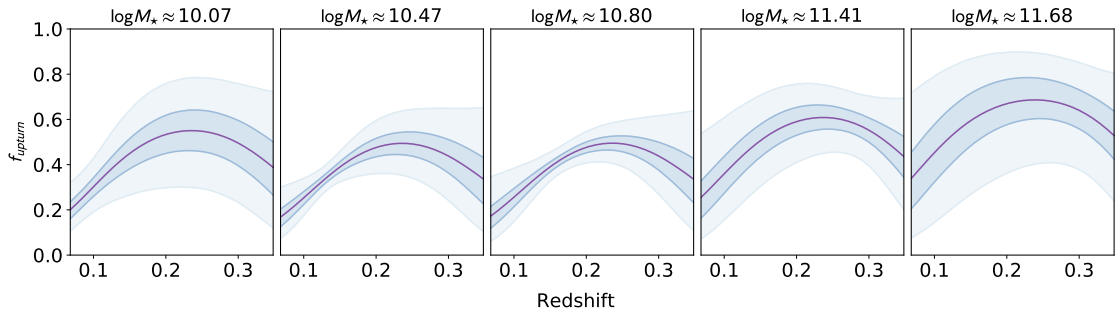

Figure 3.7: Results depicting the dependence of the $f_{\text{upturn}}$ with $z$ in five $\log M_\star$ slices: 10.07, 10.47, 10.80, 11.41, 11.68. The blue shaded areas depict intervals with 50% and 95% of credibility.

## 3.3 Volume-limited validation

To validate the results presented in Sec. 3.2, a complementary analysis is made. In this step, a volume-limited sub-sample has been adopted: $M_r \leq -22$ and $0.03 \leq z \leq 0.35$, which can be seen in Fig. 3.8. By applying this volume restriction, the number of objects declines drastically when compared to the previous analysis; with such limitation, the sample is composed of 91

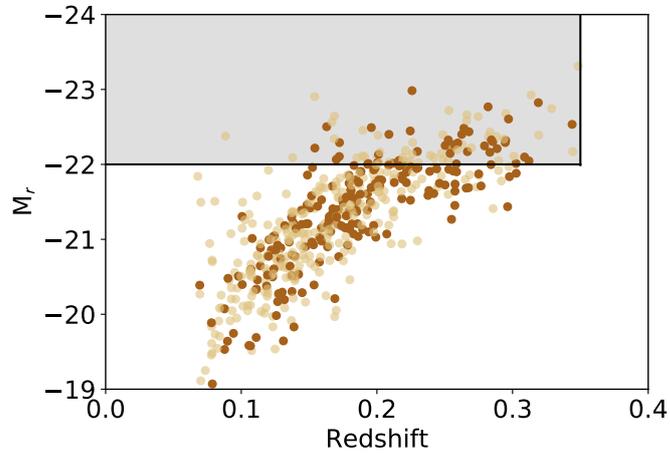

Figure 3.8: $M_r$ against $z$ chart. The grey area depicts the sub-sample used in the analysis of the volume-limited validation. UV weak and upturn galaxies are represented by the light and dark shades of brown respectively.

RSGs: 50 UV weak and 41 UV upturn systems. The model was run again with the same parameters described in Sec. 3.1.2.1.

This complementary approach is made to validate that the analysis made throughout this Chapter is physically robust. The use of $\log M_\star$ embedded in the analysis as well as the fraction of UV upturn galaxies can be considered alternative measures for dealing with the Malmquist bias. Indeed, the results presented in Figs. 3.9 and 3.10 are very similar to those for the complete sample. It is worth noting, though, that the values of $\log M_\star$ have been reset to higher lower values; with that in mind, the results for $\log M_\star > 10.75$ for both complete and volume-limited samples are equivalent.

All in all, the results from the complete sample and its validation (using a volume-limited sub-sample) point to a clear evolution of the fraction of the UV upturn among UV bright RSGs and the dependence with $\log M_\star$.

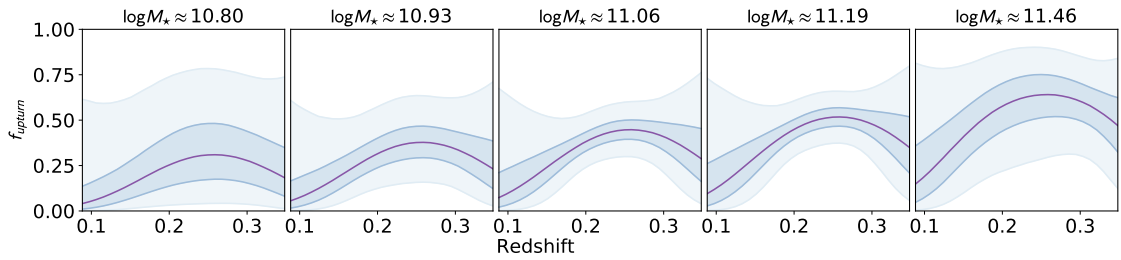

Figure 3.9: Dependence of the fraction of UV upturn systems among all the UV bright sample with $z$, in different slices of $\log M_\star$. These images depict the results for the model considering a volume-limited sub-sample.

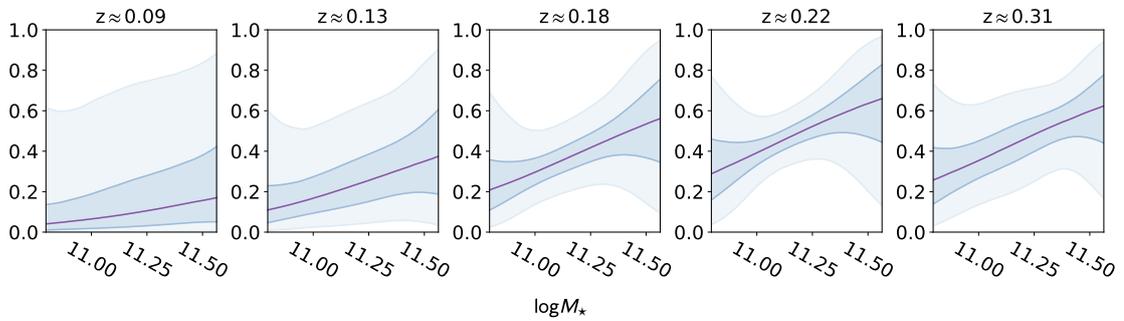

Figure 3.10: Dependence of the fraction of UV upturn systems among all the UV bright sample with $\log M_\star$, in different slices of $z$. These images depict the results for the model considering a volume-limited sub-sample.

Knock on the sky and listen to the
sound.

―――――――――――――――

Zen Buddhist proverb

# 4

# Emission Lines

**Based on:**

In Chapter 3 the evolution in $z$ of the fraction of UV upturn systems over the total of
UV bright RSGs has been presented, as well as its dependence with mass. As we have
seen, the criteria used to classify the UV upturn galaxies is photometric, based in only
three colours. Given this scenario, many questions may arise, such as:

- what about their spectral features? Are these systems hosting the classical understanding of the UV upturn (in the sense of not being contaminated by star-formation)?

- AGN have been pointed to influence the UV emission of galaxies, could those be meddling in the sample (see e.g. Adams et al., 2020)?

Therefore, in this Chapter I discuss the importance of emission lines in characterising the sample, as well as some of the issues involving the use of diagnostic diagrams.

## 4.1    An overview of diagnostic diagrams

Emission lines can give us many clues about the photo-ionisation processes happening basically everywhere in the Universe (Stasińska, 2007), consequently providing essential information about the chemical composition and physical conditions of a certain system – in our case, galaxies. It is remarkable that they arise in all ranges of wavelength, from submillimetre/radio to $\gamma$-rays (Lobanov & Zensus, 2007; Stasińska, 2007).

When investigating galaxies, emission lines are useful to separate systems according to their main ionisation source, revealing crucial information about the nebular emission from the ionised gas inhabiting them (e.g. Baldwin, Phillips, & Terlevich, 1981; Veilleux & Osterbrock, 1987; Rola, Terlevich, & Terlevich, 1997; Kewley et al., 2001; Kauffmann et al., 2003a; Stasińska et al., 2006; Schawinski et al., 2007; Cid Fernandes et al., 2011; Juneau et al., 2011; Yan et al., 2011; Juneau et al., 2014; Feltre, Charlot, & Gutkin, 2016; de Souza et al., 2017). Notorious works on emission lines from galaxies include the so-called *seagull* diagram (or simpy, the BPT, Baldwin, Phillips, & Terlevich, 1981), the WHAN diagram (Cid Fernandes et al., 2010; Cid Fernandes et al., 2011) the colour-excitation diagram (Yan et al., 2011), the Trouille–Barger–Tremonti diagram (Trouille, Barger, & Tremonti, 2011), the mass-excitation diagram (Juneau et al., 2014), to mention a few.

Among the aforementioned diagnostic diagrams, two stand out for their simplicity, yet richness of information: the BPT and WHAN. The first diagram is probably the most used so far in the literature; the latter stands out for its similarity with the BPT, however encompassing a larger number of galaxies that are not covered by the BPT, namely retired and passive systems; this makes the WHAN diagram suitable for this study. Details about such diagrams are discussed in the following sections.

### 4.1.1 Baldwin-Phillips-Terlevich diagnostic diagram

The technique prescribed by the BPT to classify emission line galaxies is one of the most successful in extragalactic astrophysics[1]. There are three versions of the BPT chart, which is composed of two line ratios; the most used version being: $\log([\text{OIII}]/\text{H}\beta)$ – depicted in the y-axis, making use of [OIII] $\lambda5007$ – and $\log([\text{NII}]/\text{H}\alpha)$ – in the x-axis, making use of [NII] $\lambda6583$. Both other versions include $\log([\text{OIII}]/\text{H}\beta)$, but vary the x-axis with $\log([\text{OI}]/\text{H}\alpha)$ and $\log([\text{SII}]/\text{H}\alpha)$. The distribution of galaxies in this diagram is in the shape of a *seagull*, with two main wings, namely the star-forming and the AGN wings. Many have attempted to classify the objects in each of the different wings of the BPT: Kewley et al. (2001) proposed a theoretical starburst line, whereas Kauffmann et al. (2003a) recommended the use of an empirical line; and finally, Stasińska et al. (2006) prescribed a hybrid solution which makes use of theoretical and empirical precepts to separate AGN and star-forming systems. Equations 4.1 to 4.3 display the lines that separate AGN from star-forming systems from the three aforementioned studies.

$$\log\left(\frac{[\text{OIII}]}{\text{H}\beta}\right) = \frac{0.61}{\log([\text{NII}]/\text{H}\alpha) - 0.47} + 1.19 \qquad \text{(Kewley et al., 2001)} \qquad (4.1)$$

$$\log\left(\frac{[\text{OIII}]}{\text{H}\beta}\right) = \frac{0.61}{\log([\text{NII}]/\text{H}\alpha) - 0.05} + 1.3 \qquad \text{(Kauffmann et al., 2003a)} \quad (4.2)$$

$$\log\left(\frac{[\text{OIII}]}{\text{H}\beta}\right) = (-30.787 + 1.1358x + 0.27297x^2)\cdot \qquad (4.3)$$

$$\tanh(5.7409x) - 31.093 \qquad \text{(Stasińska et al., 2006)}$$

---

[1] As of the writing of this thesis, this paper had whopping 3282 citations in the Astrophysics Data System (https://ui.adsabs.harvard.edu/abs/1981PASP...93....5B/abstract) and 4117 according to Google Scholar, and it continues to grow in a daily basis.

in which $x = \log([\text{NII}]/\text{H}\alpha)$.

With so many options to separate galaxies ionised by star-forming from those by AGN, the region between these lines (e.g. the region between the line of Kewley et al. and the one by Stasińska et al.; or, mostly used, the one by Kewley et al. and the line by Kauffmann et al.) is considered a transitional area called *composite*. This region serves as an 'uncertainty' region regarding ionising sources in those systems, in which both phenomena are likely to co-exist.

In the AGN region of the BPT diagram, it is also possible to distinguish between Seyfert and LINERS. For example, Schawinski et al. (2007) proposes an empirical Seyfert-LINER separation, illustrated by the following equation:

$$\log\left(\frac{[\text{OIII}]}{\text{H}\beta}\right) = 1.05\log\left(\frac{[\text{NII}]}{\text{H}\alpha}\right) + 0.45. \qquad (4.4)$$

An example of the BPT diagram can be seen in the top panel of Fig. 4.1. The data therein presented will be discussed further in this Chapter.

### 4.1.2   WHAN diagnostic diagram

The WHAN diagram receives its name because of its use of the equivalent width of H$\alpha$, i.e. its notation $\text{W}_{\text{H}\alpha}$ and $\text{W}_{[\text{NII}]}$ in Cid Fernandes et al. (2011). The main difference between the BPT and the WHAN diagrams is that the latter allows us to spot different sources of ionisation that usually are blind to the BPT. These sources are usually in the retired/passive region and are associated with HOLMES. Of course, still retired/passive systems can be split in other two groups of systems: completely lineless or 'liny' (Herpich et al., 2018) – the completely lineless cannot be detected in neither diagram. Cid Fernandes et al. (2011) tag retired/passive lineless galaxies that are not detected as 'undetected'.

The criteria for the classes therein described are as follows.

i. log([NII]/Hα)<-0.4 and log EW(Hα)>3Å: star-forming galaxies (SF);

ii. log([NII]/Hα)>-0.4 and log EW(Hα)>6Å: strong AGN (sAGN);

iii. log([NII]/Hα)>-0.4 and 3Å≤log EW(Hα)≤6Å: weak AGN (wAGN);

iv. log EW(Hα)<3Å: retired galaxies (RGs);

v. log EW(Hα)  and log EW([NII])<0.5Å: passive.

In the case of this work, even the undetected systems may not be entirely comprised of retired/passive lineless objects; some of them can be objects that failed to be accurately measured for unknown reasons, such as potential low S/N. For the purposes of this work, all of the galaxies that could not be detected – the 'undetected' systems by Cid Fernandes et al. (2011) or the ones without measured lines due to the other aforementioned reasons – are tagged as 'unclassified'. An example of WHAN diagram can be seen in the bottom panel of Fig. 4.1 and it will be discussed throughout this Chapter.

## 4.2 UV upturn and emission lines

First of all, it is necessary to clarify a potential vocabulary mix-up. In this Chapter, the sample used for scientific purposes is the RSG sample described in Sec. 2.3.1, i.e. the same used in Chap. 3. However, for illustration purposes (Fig. 4.1 and left panel of Fig. 4.2), the entire sample is included (i.e. also RSF described by Yi et al. – see also Sec. 2.3.1). In terms of specific vocabulary, RSF are always the systems described as such by Yi et al. (2011); that is, based on photometric criteria. In terms of emission line classification, SF are the systems occupying the star-formation regions of the BPT and/or WHAN diagram; that is, based on emission line criteria, unless otherwise specified.

As discussed in Sec. 1.3.1.4, emission lines can be a tool to investigate the nature of the UV upturn. In fact, the complete sample described in Chap. 2 was analysed according to their emission lines (or lack thereof). To that end, I made use of the BPT (Baldwin, Phillips, & Terlevich, 1981) and WHAN (Cid Fernandes et al., 2010; Cid Fernandes et al., 2011) diagrams. Fig. 4.1 displays the complete sample projected in the BPT (top panel) and WHAN (bottom panel); also, the sample is stratified by UV class, according to Fig. 2.3. In terms of numbers and classification of objects, they can be assessed in Tables 4.2 and 4.1. Additionally, Fig. 4.2 left panel works as a visual aid, by displaying these same diagnostic diagrams in subplots according to their UV class.

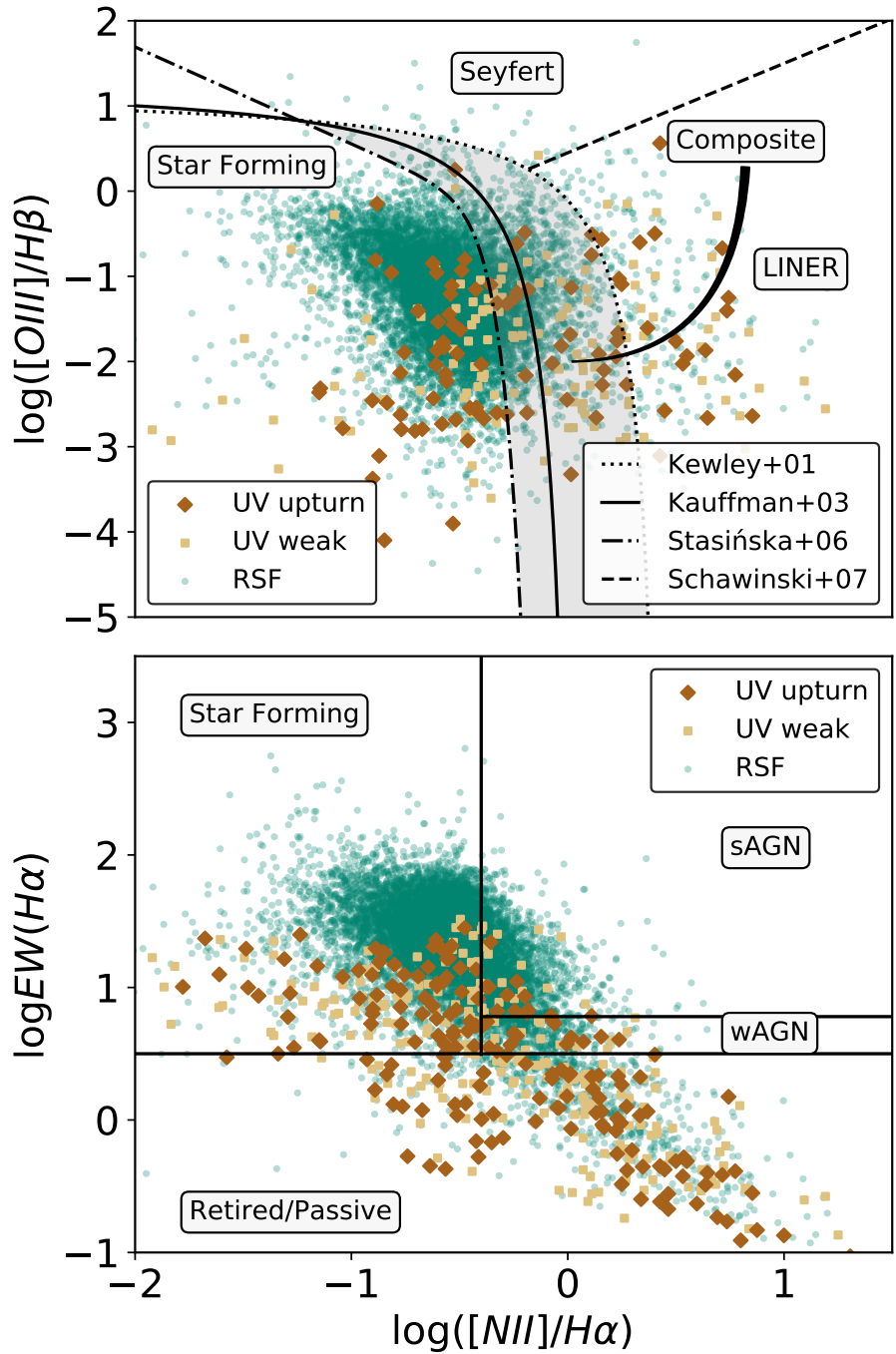

Figure 4.1:  BPT (top figure, Baldwin, Phillips, & Terlevich, 1981) and WHAN (bottom figure, Cid Fernandes et al., 2011) diagrams for the entire sample. The sample is colour-coded according to their UV class: green circles depict RSF galaxies, light brown squares the UV weak, and the dark brown diamonds the UV upturn. **Upper panel**:  the solid line represents the division by Kauffmann et al. (2003a); the dotted line, the one by Kewley et al. (2001); and the dot-dashed line, the one by Stasińska et al. (2006). **Lower panel**:  the divisions represented in the WHAN diagram are those proposed by Cid Fernandes et al. (2011); i.e.  star-forming (SF), weak and strong AGN (wAGN and sAGN respectively), and retired/passive (R/P) galaxies.

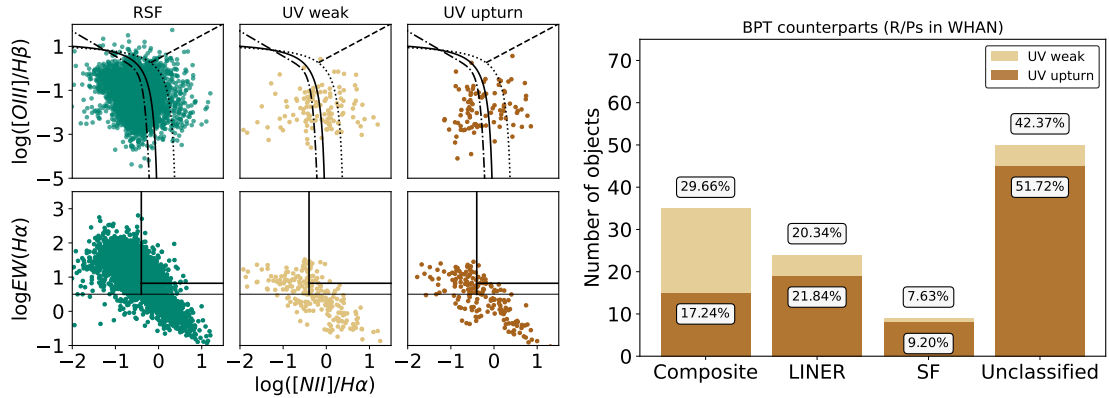

Figure 4.2: Left panel: BPT and WHAN diagrams (similar to those of Fig. 4.1), although stratified by UV class for better visualisation. Right panel: Bar-plot depicting the retired/passive systems in the WHAN diagram and the cross-classification from the BPT according to their UV class. Composite, LINER, SF, and unclassified correspond to approximately 29.66%, 20.34%, 7.63% and 42.37% respectively for UV weak galaxies; and 17.24%, 21.84%, 9.20%, and 51.72% for UV upturn systems.

It is noteworthy how the sample occupies the entire BPT and WHAN *loci*; RSF galaxies occupy both wings of the BPT diagram, but they are, in fact, mostly located in the SF wing and composite regions. The same behaviour is seen in the WHAN diagram, RSF systems are spread all over the diagram, but mainly located at the star-forming region. Yet, no UV bright RSGs are in the extreme of the SF wing of the BPT, the starburst region.

Also, UV weak and upturn galaxies do not occupy the Seyfert region of the BPT diagram and they are mainly located at the lower regions of the sAGN *locus* of the

WHAN chart, where no 'official' intermediary regions exist (yet, since they are so close
to boundary areas, these objects should be treated more carefully de Souza et al., 2017).
In fact, strong AGN (such as Seyferts) are predominant in green-valley (e.g. Smolčić,
2009) systems and/or in spiral galaxies (e.g. Orban de Xivry et al., 2011).

Indeed, an important result from this analysis is that the photometric criteria proposed
by Yi et al. (2011) is robust against strong AGN (such as Seyferts) and starburst activity,
but it is not enough to entirely eliminate UV upturn systems from which the UV emission
is originated from newly born stars. For example, as illustrated in Table 4.2, 210 objects
are classified as UV upturn according to Yi et al. (2011), but 49 occupy the SF *locus* of
the WHAN and 68 of the BPT; which corresponds to a contamination of 23% and 32%
respectively.

It is possible to see differences in the number of objects tagged as 'unclassified' in
both BPT and WHAN diagrams (see Tables 4.2 and 4.1). This is due to the fact that the
WHAN chart makes use of EW(H$\alpha$), which succeeds in detecting more objects than the
BPT; these systems are mostly classified as retired/passive. Additionally, Fig. 4.2 right
panel displays the cross-classification of WHAN retired/passive systems seen in the BPT
diagram (for both UV classes); composite, LINER, SF, and unclassified correspond to
approximately 29.66%, 20.34%, 7.63% and 42.37% respectively for UV weak galaxies,
whereas 17.24%, 21.84%, 9.20%, and 51.72% for UV upturn systems. Both UV weak
and UV upturn have similar percentages of LINER and SF galaxies therein, and the
highest differences occur in the composite region and unclassified systems. This is an
important indicator of the lack of consensus in terms of emission line galaxy classification
(this is specially discussed in de Souza et al., 2017, who explore this issue in detail).

Table 4.1: Number of galaxies in each UV class in the WHAN diagram; it includes all the 13,050 with measured emission lines plus 1,281 remaining systems with no measurable emission lines (unclassified depicted as 'unc.'). The last two rows, in bold, consist on the UV bright RSGs.

| UV class | **WHAN classification** | | | | | |
|---|---|---|---|---|---|---|
| | SF | sAGN | wAGN | R/P | unc. | **total** |
| RSF | 9,253 | 2,156 | 557 | 669 | 1,190 | 13,825 |
| **weak** | **78** | **19** | **19** | **118** | **62** | **296** |
| **upturn** | **68** | **9** | **17** | **87** | **29** | **210** |

Table 4.2: Number of galaxies in each UV class in the BPT diagram; it includes all the 11,647 with measured emission lines plus the remaining objects, analogously to Table 4.1. Seyferts are displayed as 'Sy', composites as 'comp.' and unclassified as 'unc.'. The last two rows, in bold, consist on the UV bright RSGs.

| UV class | **BPT classification** | | | | | |
|---|---|---|---|---|---|---|
| | SF | Sy | LINER | comp. | unc. | **total** |
| RSF | 9,223 | 37 | 197 | 1,975 | 2,402 | 13,834 |
| **weak** | **50** | **1** | **27** | **50** | **168** | **296** |
| **upturn** | **49** | **0** | **20** | **27** | **114** | **210** |

## 4.3 Model tailoring

To tackle the dependence of $z$, $\log M_\star$, and WHAN classes, the model described in Sec. 3.1.2.1 was modified to take into account the four classes of the WHAN diagram plus unclassified systems (in practical terms, five classes). Such modification can be seen in Eq. 4.5 – and in more detail in Eq. 4.6 –, in which the $i$ is the linked to the $i^{\text{th}}$ term of the regression and $k$ is the $k^{\text{th}}$ emission line class.

$$\eta_{i[k]} = X_{i[k]}\beta_{i[k]} \qquad (4.5)$$

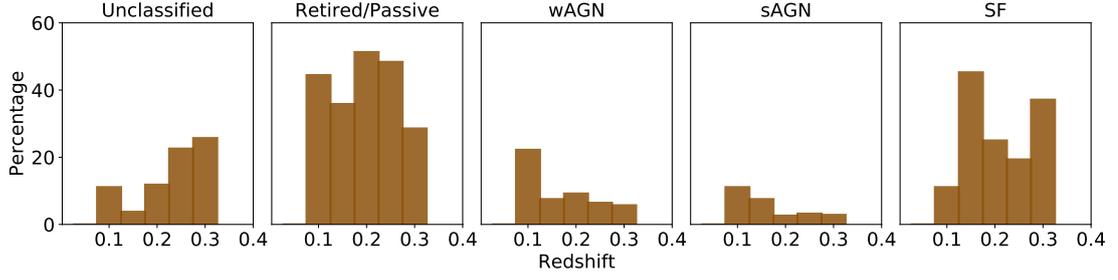

Figure 4.3:  Bar-plots similar to those of Fig. 3.2 stratified by WHAN classes.

$$
\begin{bmatrix}
\eta_{11} & \dots & \eta_{1k} \\
\eta_{21} & \dots & \eta_{2k} \\
\vdots & \ddots & \vdots \\
\eta_{n1} & \dots & \eta_{nk}
\end{bmatrix}
=
\begin{bmatrix}
1 & \log M_{\star 1} & \log M_{\star 1}^2 & z_1 & z_1^2 \\
1 & \log M_{\star 2} & \log M_{\star 2}^2 & z_2 & z_2^2 \\
\vdots & & \ddots & & \vdots \\
1 & \log M_{\star n} & \log M_{\star n}^2 & z_n & z_n^2
\end{bmatrix}
\begin{bmatrix}
\beta_{11} & \dots & \beta_{1k} \\
\beta_{21} & \dots & \beta_{2k} \\
\vdots & \ddots & \vdots \\
\beta_{j1} & \dots & \beta_{jk}
\end{bmatrix}
\tag{4.6}
$$

Because of the five emission line classes, the model generates 25 posteriors. Analogously to the model described in Chapter 3, the sampler is initiated at random values, with 3000 burn-in steps and 7000 iteration steps; the priors are weakly informative for $\beta_{i[k]}$, with the mean set to zero and standard deviation at 10, assuming a Normal distribution.

Similarly to Fig. 3.2, new bar-plots are displayed in Fig. 4.3 in which the systems are stratified by WHAN classes.

### 4.3.1   Posteriors

Analogously to Sec. 3.1.2.2, the posteriors of the model herein used are displayed in Fig. 4.4. In this case, instead of dealing with only five posteriors, the model produces 25 of them, that is 5 for each WHAN class, which are detailed in the y-axis of each row of Fig. 4.4. In this case, the trace-plots are not represented, as they converge just as the ones displayed in Fig. 3.3, not adding any new relevant information.

## 4.4 Results: dependence on redshift and stellar mass

In what follows, I discuss the main results for $z$ and $\log M_\star$ dependence according to the WHAN classes of UV upturn systems.

### 4.4.1 Unclassified and retired/passive

Unclassified and R/P systems display a very similar behaviour in terms of $z$ in Fig. 4.5, with a rising probability up to $z \approx 0.25$ followed by an in-fall. On the other hand, despite de large credible intervals, it appears that the dependence with $\log M_\star$ is in fact different among them. For unclassified systems, the probability of nesting the UV upturn phenomenon appears to be constant across the range of $\log M_\star$, with large credible intervals in the extremes; whereas for R/P galaxies, the trend appears to be slightly declining for $\log M_\star < 10.5$ with a subsequent strong rise. Unclassified and R/P galaxies seem to be the most influencing classes when looking at the overall model, specially in terms of $z$ (see Fig. 3.7).

These results are the most important ones for this Chapter, as they indicate that the trend seen in Chapter 3 is supported by retired/passive as well as unclassified objects (heavily populated by lineless retired and passive systems). These systems are the UV upturn bearers: their UV emission is not related to star-formation and as such they are strong candidates to host the rare stellar evolutionary phases mentioned in Sec. 1.3.1.1.

### 4.4.2 wAGN and sAGN

Analysing AGN classes in this context involves discussing the following issues: how their classification is made, the unified AGN model (Antonucci, 1993; Urry & Padovani, 1995), and its issues (see for instance Elitzur & Shlosman, 2006; Netzer, 2015, for a review on the unifed AGN model and its controversies) – the reader is referred to Sec. 1.3.1.4.

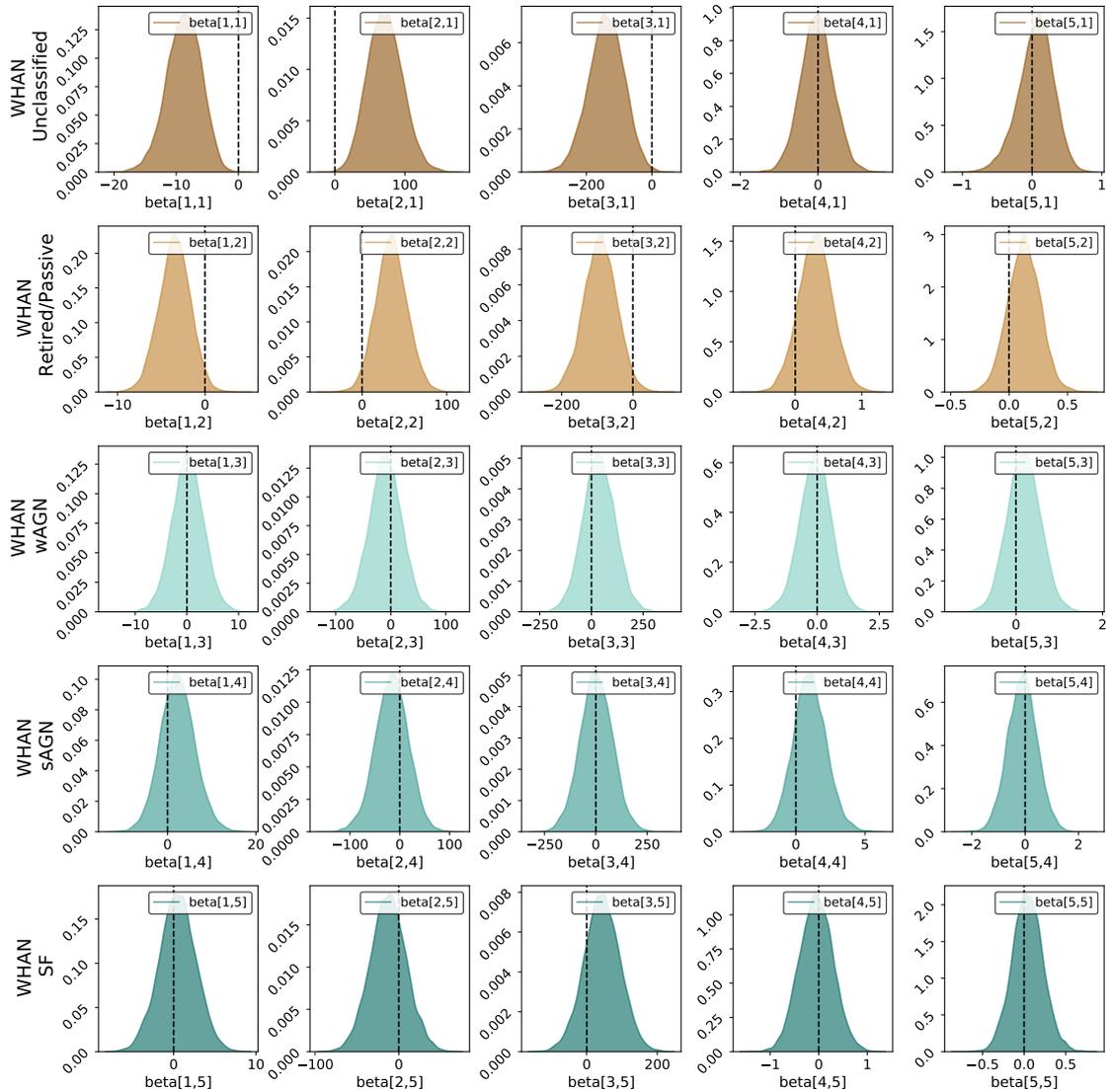

Figure 4.4: Results of the 25 posteriors. Each row represents the five posteriors for each WHAN class (Cid Fernandes et al., 2010). The interpretation is analogous to the one presented in Fig. 3.3.

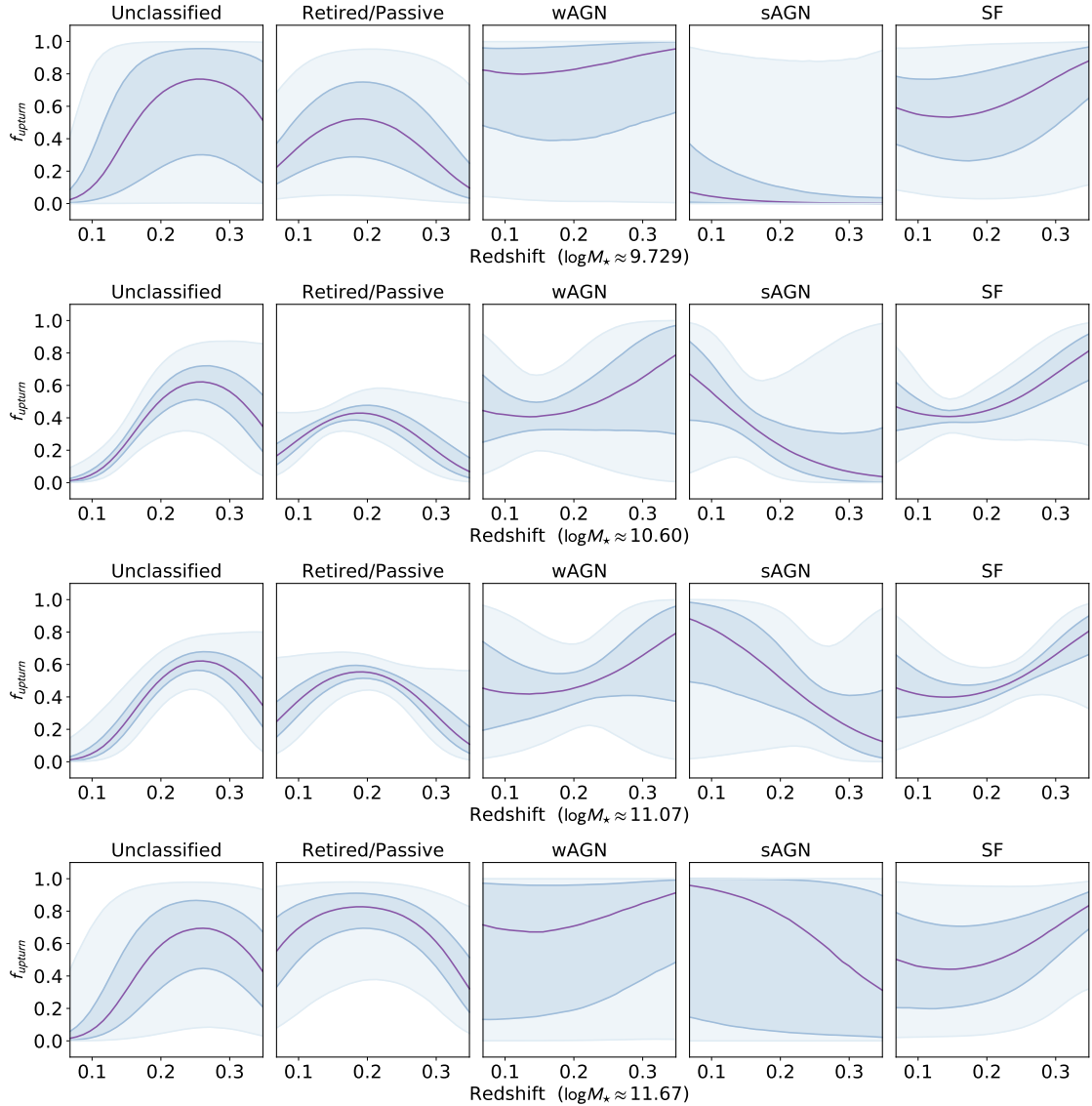

Figure 4.5: Regression displaying the fraction of UV bright RSG galaxies that host the UV upturn phenomenon given their emission line classification (WHAN diagram) in terms of $z$. The results herein displayed are valid for systems with $\log M_\star$ between 9.7 and 11.7

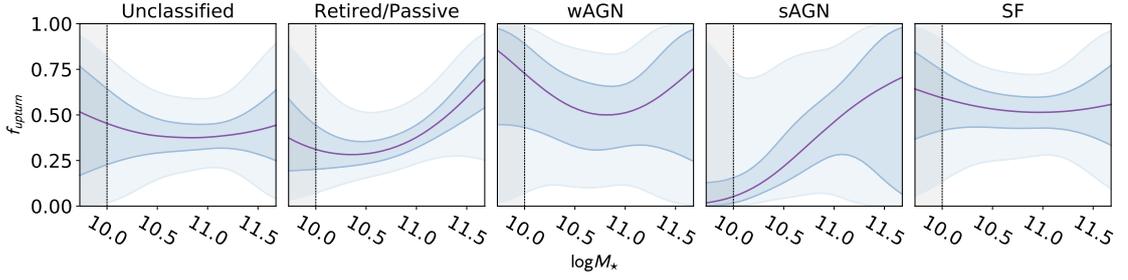

Figure 4.6:  Regression displaying the fraction of UV bright RSG galaxies that host the UV upturn phenomenon given their emission line classification (WHAN diagram).  The results herein displayed are valid for systems with $z \approx 0.261$, the median $z$ of the sample.

Although the BPT is not being actively used for this analysis, it is still useful:  no Seyferts are detected in the RSG sample, which means that there are no 'real' strong AGN. By checking the WHAN diagram, those classified as sAGN are very close to the boundaries – the lack of transitioning areas makes it hard to guarantee their classification (see, for instance, the discussion about this issue presented by de Souza et al., 2017).

For the wAGN group, the idea behind it would be to check any potential links between LINER galaxies and the UV upturn.  It is worth mentioning that the classification of LINERs is subject to controversies involving the AGN unified model (see Sec. 1.3.1.4). However,LINER galaxies not only occupy the wAGN area in the WHAN chart, as they can spread towards the retired/passive area (the *loci* where HOLMES reside, as seen in Fig.  1 of Cid Fernandes et al., 2011, and further discussed by de Souza et al.). Nonetheless, the actual numbers for wAGN are very low, which thickens credibility areas (interquartile ranges), making it very hard to confirm any potential trends for $f_{\text{upturn}}$.

Finally, given the small number of UV bright RSGs classified as sAGN and wAGN (see Table 4.1), the uncertainties regarding boundary areas, and the thickness of credible intervals, it is not possible at the moment to withdraw any conclusions for sAGN or wAGN.

### 4.4.3 Star-forming

Galaxies presenting some level of star-formation should not be overlooked in the quest for UV upturn bearers. In fact, many studies have shown that ETGs can go through several rejuvenating processes that lead to nesting younger stellar populations. For instance, Bettoni et al. (2014) discuss the effects of rejuvenation among counter-rotating ETGs; they argue that these galaxies have recently gone through dynamic interactions, such as minor mergers, and that approximately 50% of them show strong far-UV emission.

This result goes hand-in-hand with the findings of Werle et al. (2020), who have shown two kinds of UV upturn galaxies: those influenced by rare stellar populations and those suffering from somewhat recent (~1 Gyr) episodes of star-formation. In the latter case, they seem to have accreted 'pristine' (metal-poor) gas from the intergalactic medium, resulting in an average lower metallicity than their star-formation free counterparts.

In the sample of RSGs with star-formation identified here, $f_{upturn}$ rises with increasing $z$, which is an expected result. This is due to the fact that the UV emission from these galaxies come from rapidly evolving massive stars. These stars are bluer and more massive with increasing $z$ (e.g. Madau et al., 1996; Vink, 2018).

We are made of star stuff. We are a
way for the Cosmos to know itself.

Carl Sagan
*Cosmos TV series*

# 5

# Stellar Populations

**Based on:**

Dantas, Coelho, and Sánchez-Blázquez (2020, in press as of the release of this thesis);

de Souza, Dantas, Krone-Martins, Cameron, Coelho, Hattab, de Val-Borro, Hilbe,
Elliott, Hagen, and COIN Collaboration (2016).

As seen in Chapters 3 and 4, the fraction of UV upturn systems evolves in $z$ and is
also dependent on $\log M_\star$, specially in galaxies classified as retired/passive according to
the WHAN diagram. Such evolution indicates a peak of UV upturn galaxies at $z \sim 0.25$.
Given this scenario, many questions arise such as:

- what about their stellar populations? Would those change as well?

- What differences and similarities can stellar population analyses reveal about UV
  weak and upturn RSGs?

- How can we make sure to control $z$ and $\log M_\star$ in this scenario (given that *we know*
  these are important variables)?

To answer these questions, it is necessary to make use of samples which are comparable in terms of $\log M_\star$ and $z$. To that end, two sub-samples are drawn based on the RSGs classified as retired/passive (Chap. 4), by making use of *propensity score matching* (PSM). It is a technique that mitigates biases between two sets of data – in this case, galaxies classified as UV weak or UV upturn.

In this Chapter I discuss the properties of stellar populations among passive/retired UV bright red-sequence galaxies and highlight the differences and similarities among each galaxy UV class.

## 5.1   A brief overview on stellar populations and SED fitting history

In this brief Section I contextualise the overall approach that is the basis of this Chapter (analysis of stellar populations by making use of SED fitting technique[1]).

The SED of a galaxy conceals a remarkable amount of information, be it on their stellar populations or their dynamics and kinematics. In other words, by looking at the SEDs of galaxies, one looks at the integrated light from all their the inhabiting objects. Through evolutionary stellar population synthesis, which was introduced by Tinsley (1968; 1972; and notably the seminal article by Tinsley and Gunn, 1976), one has the potential to decipher the inner baryonic content of galaxies, i.e. stars, gas, dust; as well as to retrieve parameters such as star formation history (SFH), metallicity ($\langle Z/Z_\odot \rangle$), velocity dispersion ($\sigma$), IMF shape, and so forth (see both Walcher et al., 2011; Conroy, 2013, for reviews on the topic).

Throughout the years, several simple stellar population (SSPs) libraries have been developed to help decipher the information encrypted in galaxy SEDs (e.g. Bruzual A., 1983; Buzzoni, 1989; Worthey, 1994; Fioc & Rocca-Volmerange, 1997; Leitherer et al., 1999; Bruzual & Charlot, 2003; Pietrinferni et al., 2004; Maraston, 2005; Martins et al., 2005; Coelho et al., 2007; Cordier et al., 2007; Lee et al., 2009; Percival et al., 2009; Walcher et al., 2009; Vazdekis et al., 2010; Chung et al., 2013; Vazdekis et al., 2015; Vazdekis et al., 2016; Chung, Yoon, & Lee, 2017; Villaume et al., 2017), as well as a myriad of codes to fit such templates (e.g. Bolzonella, Miralles, & Pelló, 2000; Heavens, Jimenez, & Lahav, 2000; Cappellari & Emsellem, 2004; Le Borgne et al., 2004; Cid Fernandes et al., 2005; Ocvirk et al., 2006; Tojeiro et al., 2007; da Cunha, Charlot, & Elbaz, 2008; Koleva et al., 2009; Kotulla et al., 2009; Noll et al., 2009; Chevallard & Charlot, 2016; Cappellari, 2017; Wilkinson et al., 2017; Carnall et al., 2018; Robotham et al., 2020).

---

[1]In this context, I refer to SED fitting as a comprehensive term for the fitting of both photometry and spectroscopy.

The reason why so many templates and fitting codes exist is that the goal behind each of them is different. In terms of fitting code, for instance, one can be non-parametric (such as STARLIGHT, Cid Fernandes et al., 2005; PPXF, Cappellari and Emsellem, 2004 and Cappellari, 2017) which assumes nothing about the SFH or chemical enrichment histories; whereas another one can be parametric (such as CIGALE, Noll et al., 2009; BEAGLE, Chevallard and Charlot, 2016). Additionally, one can be aimed at fitting spectra (e.g. Cappellari & Emsellem, 2004; Cid Fernandes et al., 2005), whereas other at fitting photometrical data (e.g. da Cunha, Charlot, & Elbaz, 2008; Robotham et al., 2020). Some codes were also built to take into account problems potential problems caused by overfitting noisy spectra, such as PPXF, which penalises the noise in the final fit solution. Other codes have adopted fitting techniques other than the traditional $\chi^2$, such as Markov Chain Monte Carlo (MCMC, e.g. Acquaviva, Gawiser, and Guaita, 2011). The use of one or another code is a choice of the user, who knows (or should know) the requirements behind their research purposes.

In terms of model templates, as it is very difficult to access all the potential idiosyncrasies of galaxies and therefore different SSP libraries seek to cover certain wavelength ranges or different issues. In other words, each stellar population library is usually aimed at solving a certain *class* of galaxy. For instance, STARBURST99 (Leitherer et al., 1999) was developed to decode the aspects of galaxies with very high rates of star-formation, mainly composed of young hot stars; whereas MILES (Vazdekis et al., 2010) and its UV extension, E-MILES (Vazdekis et al., 2016), were developed with the goal to decipher properties of ETGs, mostly inhabited by old and evolved stellar populations. Additionally, the need for considering accurate $\alpha$-enhanced models is on the rise, which has been leading to an improvement in stellar libraries towards a chemical fine tuning, with the aim of solving (or at least mitigating) this issue (first explored in spectral models by Coelho et al., 2007). Another choice to be made is the conundrum involving the adoption of empirical *versus* theoretical libraries; their usage has not reached a consensus among the community (a deeper discussion can be found in Coelho, Bruzual, & Charlot, 2020). Additionally, some libraries take into account rare stellar populations, that are often overlooked; e.g., Maraston (2005) developed a library that takes into account TP-AGB stars, Hernández-Pérez and Bruzual (2013) propose another one rich in binary systems; and Percival and Salaris (2011) developed a library rich in HB and EHB stars with the aim of solving issues such as the UV upturn. However, such rare stellar populations still remain to be properly studied and considered, specially to better quantify phenomena such as the UV upturn (Chavez & Bertone, 2011).

Finally, it is worth mentioning other uses of SED fitting that are not directly linked to stellar population analysis, but that highlight its usefulness. It is suitable for estimating K-corrections (e.g. Blanton & Roweis, 2007), retrieving photometric redshifts (e.g. Benítez, 2000; Battisti et al., 2019), fitting emission-lines for AGN studies (e.g. Calistro Rivera et al., 2016), to mention a few.

## 5.2   A controlled sample

One of the main problems when dealing with data analysis – which is essentially what is done in Astrophysics research – are all the biases that permeate observations. The problems involving biases (whichever they are and in any research area) are old and have been tackled by numerous articles (e.g. Cochran, 1968; Rosenbaum & Rubin, 1983; Ho et al., 2007), and this is not different in Astronomy (e.g. de Souza et al., 2016; Trevisan, Mamon, & Khosroshahi, 2017). Nevertheless, it is difficult to assess all the potential biases to which one can be subject of; therefore I address a particular type of bias in this Section: the one involving the comparison of two sets of data (i.e. UV weak and UV upturn).

To deal with this kind of bias, PSM is used to select a sub-sample of UV weak galaxies that are the closest to their UV upturn counterparts. PSM is a statistical technique largely used (specially in Medicine) that aims at reducing biases among two (or more) sets of data, by selecting one of those as the benchmark set (in this case, the UV upturn systems). The other sets are then matched according to the chosen parameters.

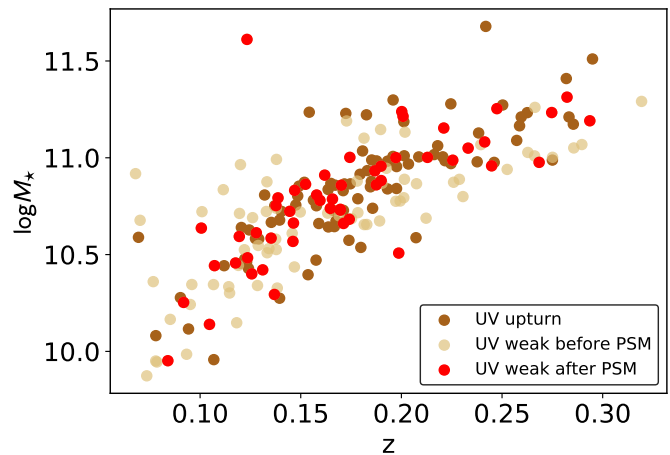

Figure 5.1: Scatter-plot featuring UV weak and upturn systems before and after propensity score matching (PSM; $z$ in x-axis and $\log M_\star$ in y-axis). As UV upturn systems (marked in dark brown) are the reference objects, UV weak galaxies were matched using PSM. UV weak systems before and after PSM are marked in light brown and red respectively.

Consequently, it is possible to compare groups of galaxies that do not necessarily share the same parameter distribution otherwise (the reader is referred to de Souza et al., 2016, in which the two sets of data are originally very different). This highly mitigates biases such as comparing systems with very different mass distributions or evolving in different $z$.

Table 5.1: Number of UV bright systems before and after PSM. It is noticeable that UV upturn galaxies remain the same, as they are the reference sample in order to select UV weak counterparts.

| UV class | Before PSM | After PSM |
|----------|-----------|-----------|
| UV upturn | 87 | 87 |
| UV weak | 118 | 51 |

In the context of this thesis, the goal is to select UV weak galaxies that have the most similar distribution of $z$ and $\log M_\star$ to the UV upturn sub-sample; it is comprised of the UV upturn RSGs classified as retired/passive in the WHAN diagram as defined in Chapter 4. It is important to remind that the $\log M_\star$ used is retrieved from the `StellarMasses` DMU from GAMA-DR3, which is explained in further detail in Taylor et al. (2011). More specifically, I have made use of nearest neighbour algorithm available in SCIKIT-LEARN, a PYTHON package built with machine learning tools (Pedregosa et al., 2011). The confounding variables, $X_c$, herein used are as aforementioned:

$$X_c = \{\log M_\star, z\}. \tag{5.1}$$

Considering the UV upturn systems as the ones of reference, it is possible to select a 'twin' UV weak counterpart with the closest values of $X_c$. In this specific case, some of the UV weak counterparts were detected as the twins of more than one UV upturn system; therefore, these UV weak objects have been considered only one time, which caused a shrinkage in the number of UV weak systems after PSM. The result is a 'twin' or 'mirror' distribution of UV weak galaxies when compared to their upturn counterparts. Such numbers are available in Table 5.1.

To visualise how PSM works, Fig. 5.1 displays a scatter-plot in which the x and y axes represent $z$ and $\log M_{\star}$ respectively. The UV weak post-PSM systems appear in red markers, whereas the original UV weak and upturn are respectively displayed in light and dark shades of brown. Additionally, Fig. 5.2 illustrates the distributions of the parameters used in the two samples for $X_c$ before and after PSM, in the shape of violinplots. It is remarkable the fine tuning of the median (dashed lines) and percentiles (25th and 75th marked in dotted lines) after the PSM, as well as the shape of the distribution. It is also noticeable that the shape of the distribution of $\log M_{\star}$ for the UV weak population is more similar to the upturn counterparts after the PSM (Fig. 5.2): the UV upturn systems show a 'bump' for $\log M_{\star} \sim 11.2$, which was not present UV weak before the PSM, but has been mimicked after it.

In the following Sections I present an analysis and results fof the SED fitting results obtained for the galaxies herein described. These were results retrieved from the `MagPhys` DMU from the GAMA-DR3 repository.

# 5.3 SED fitting results from value-added catalogues in the GAMA database

In this Section, I present and discuss the results for stellar population properties retrieved from the `MagPhys` DMU made available by the GAMA team (details can be found in Baldry et al., 2018; Driver et al., 2018). This DMU contains the SED fitting results obtained with MAGPHYS code (da Cunha, Charlot, & Elbaz, 2008).

MAGPHYS is a parametric[2] code that makes use of photometric data for SED fitting. To run MAGPHYS, the GAMA collaboration made use of Bruzual and Charlot (2003) synthetic library, IMF of Chabrier (2003), and dust models by Charlot and Fall (2000). To obtain the results herein studied, they have used the Panchromatic Data Release (Driver et al.,

---

[2]Further in this chapter, one of the parameters used is $\langle \gamma \rangle$ (its description is available in Table 5.2), which comes from the parametric characteristic of MAGPHYS.

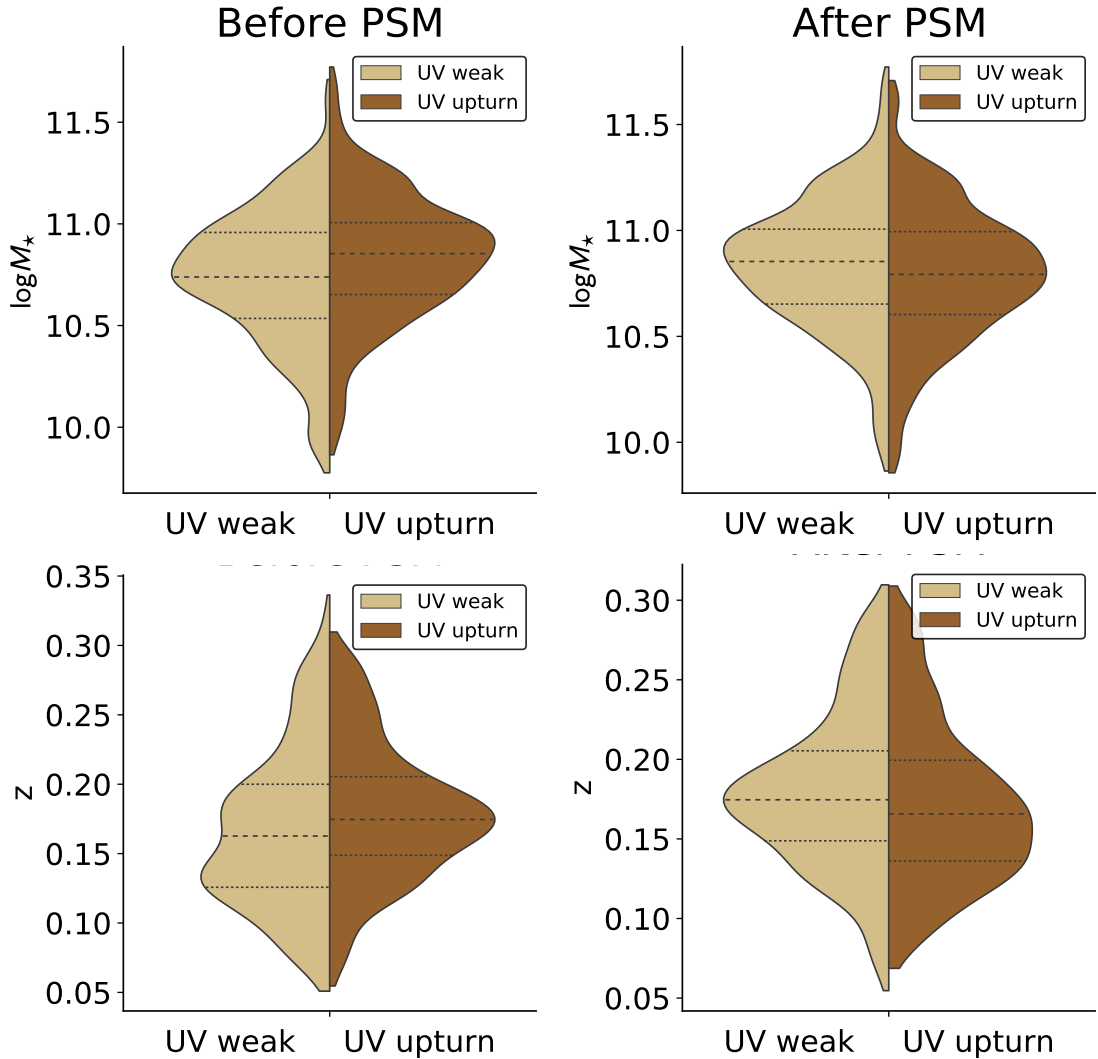

Figure 5.2: Violinplots featuring the distribution of $\log M_\star$ and $z$ among UV bright systems before and after PSM (left and right panels respectively). The median, $25^{th}$, and $75^{th}$ quantiles are respectively displayed through the dashed and dotted lines. As expected, after the PSM, the quantiles and the shape of the distribution for UV weak systems approximate to those of UV upturn galaxies.

2016) treated with the LAMBDAR code (Wright et al., 2016), which makes use of 21 band-passes (FUV, NUV, ugriz, ZYJHK, W1234, PACS100/160, SPIRE 250/350/500, see Sec. 2.1 of Driver et al., 2018) spanning from FUV (GALEX) to far-infrared/submillimetre (Herschell[3] telescope, Pilbratt et al., 2010). Their approach is extremely convenient, as it considers the UV magnitudes to generate their SED fitting results. Additionally, the choice of code is convenient as it has also been proven to be very robust (see Hayward & Smith, 2014). The use of this DMU is advantageous for two main reasons:

i. the use of UV photometry in the fit causes the stellar population ages to be better constrained, in particular those around $10^7$–$10^8$ yr (Werle et al., 2019).

ii. by making use of Charlot and Fall (2000) dust model, MAGPHYS carefully takes into account dust effects in SED fitting, by computing the absorbed luminosity from birth stellar clouds and its re-emission in the infrared (IR). (da Cunha, Charlot, & Elbaz, 2008).

In other words, the way the code was built and how it was used by the GAMA team makes it suitable for dealing with the problems herein addressed (da Cunha, Charlot, & Elbaz, 2008). The parameters used in this analysis are listed in Tab. 5.2.

## 5.3.1 Dust impact in UV bright RSGs

For this study, I have not attempted to correct the colours of the sample for internal extinction. Hence, it is possible that some contamination of green-valley systems is present; therefore some systems may be misclassified in the Yi et al. (2011) diagram (for additional reading on the extinction degeneracy, the reader is referred to Worthey, 1994; de Meulenaer et al., 2013; Sodré, Ribeiro da Silva, & Santos, 2013, and references therein).

---

[3]Named after sir William Herschel, who discovered the infrared region of the spectrum.

Table 5.2: Description of the parameters retrieved from MAGPHYS SED fitting.

| Parameter | Description |
| --- | --- |
| $\log M_\star$ | stellar mass in terms of solar masses in logarithmic scale (best fit) |
| $D_n 4000$ | 4000Å break |
| $\langle \log t \rangle_r$ | median light-weighted age ($r$-band) |
| $\langle \log t \rangle_m$ | median mass-weighted age |
| $\langle Z/Z_\odot \rangle$ | metallicity in solar units |
| $\langle \text{SFR} \rangle$ | median star formation rate at 0.1 Gyr |
| $\langle \text{sSFR} \rangle$ | median specific star formation rate at 0.1 Gyr |
| $\langle t_{\text{form}} \rangle$ | median age of oldest stars in the galaxy |
| $\langle t_{\text{last}} \rangle$ | median time since last burst of star formation ended |
| $\langle f_{\text{burst}} \rangle$ | median fraction of stellar mass formed in the corresponding time-scale |
| $\left\langle f_{\text{burst}}^{\text{2Gyr}} \right\rangle$ | median fraction of stellar mass formed in the last 2 Gyr |
| $\langle \gamma \rangle$ | median star formation timescale |

To analyse to what extent the dust would impact the galaxies herein studied, the work of Davies et al. (2019a) has been used as a benchmark. The authors explore several issues that arise from dust in a series of papers of the so-called 'Dustpedia', and in this particular paper the connection between morphology and dust-to-stellar mass ratio, among other relations. Their work predicts that, in the absence of external sources of dust (such as mergers), the minimum value for $\log M_\star/M_{\text{dust}}$ for E/S0 systems is 2.5. In the post-PSM sample used in this thesis, only 2 objects could be considered heavily obscured by dust: one UV weak and one for UV upturn. The values of $\log M_\star/M_{\text{dust}}$ for each of these systems are 2.21 and 2.32 for UV upturn and UV weak galaxies respectively; these values are very close to the lower limit, indicating that these samples are not heavily impacted by dust. The distributions of $\log M_\star$ and $\log M_\star/M_{\text{dust}}$ are available in Fig. 5.3. The left panel depicts the distributions for $\log M_\star$ estimated by Taylor et al. (2011, used in the previous Chapters) and da Cunha, Charlot, and Elbaz (2008, used in this Chapter). The right panel depicts the distributions for $\log M_\star/M_{\text{dust}}$ (estimated by MAGPHYS, da Cunha, Charlot, and Elbaz, 2008) for both UV weak and UV upturn systems.

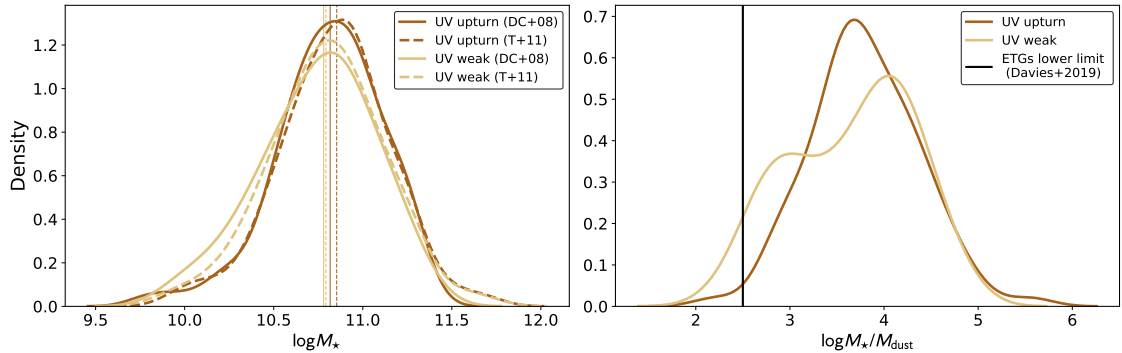

Figure 5.3: The left panel displays the comparison between the distributions of $\log M_\star$ retrieved from MAGPHYS (da Cunha, Charlot, & Elbaz, 2008) and the estimates of Taylor et al. (2011, with the medians represented by the corresponding dashed lines) – which have been previously used in Chapters 3 and 4 – for UV weak and upturn systems. This image aims at showing that both distributions are in good agreement before exploring mass ratios exclusively retrieved from da Cunha, Charlot, and Elbaz (2008). The right panel displays the distribution of $\log M_\star/M_{\rm dust}$ for MAGPHYS; the black straight line is the threshold of 2.5 established by Davies et al. (2019a).

### 5.3.2 Direct comparison of stellar population properties

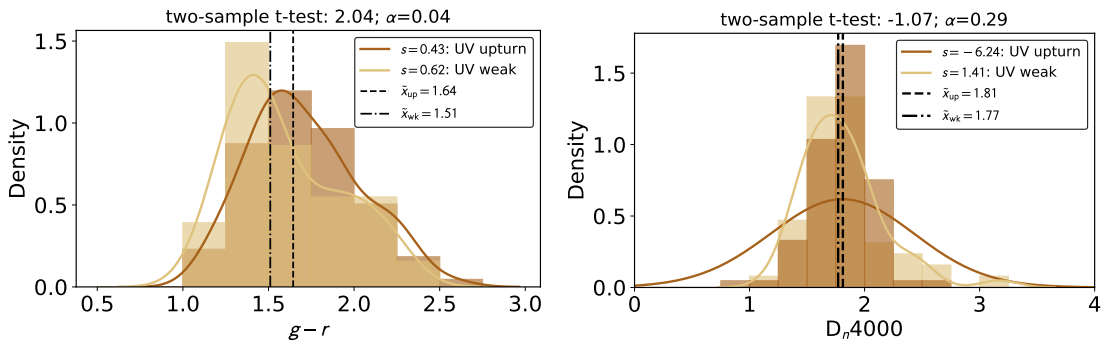

Figure 5.4: Distributions of $(g - r)$ and $\mathrm{D}_n4000$ respectively for UV weak and UV upturn systems (in light and dark shades of brown). A two-sample t-test and p-value are displayed at the top of each subfigure.

As first step for analysing the differences and similarities between UV weak and upturn systems, a direct comparison of the distributions of several parameters has been made. Fig. 5.4 displays the distributions of the $(g - r)$ colour and $\mathrm{D}_n4000$ (Bruzual A., 1983). These two parameters are very important; both are correlated with age and the latter also with metallicity (e.g. Strateva et al., 2001; Mateus et al., 2006). It is possible to see that $(g − r)$ is in fact higher for UV upturn systems, even considering that both types of galaxies have the same range of $\log M_\star$ (i.e. masses are correlated to $(g − r)$; for instance, see Kauffmann et al., 2003b). Additionally, $D_n 4000$ seem to be higher for UV upturn systems, yet this difference appears to be less important. These results indicate a potential gap in ages and metallicites, which will be discussed later in this Chapter.

In all Figs. (i.e. Figs. 5.4, as well as 5.5 and 5.6) a t-test (Student, 1908) and p-value ($\alpha$) are displayed, as some may find them useful. However, given the difficulties in interpreting these values and the controversies involved in the use of p-value (see, for instance, Lin, Lucas, & Shmueli, 2013; Nuzzo, 2014; Halsey et al., 2015; Wasserstein & Lazar, 2016), the discussion around these results is not developed throughout the text. Moreover, the medians ($\tilde{x}$) and skewness ($s$) of each distributions are displayed in these Figs. Regarding $s$, it is a measure of asymmetry of the distribution; the distribution is skewed to the left for negative $s$, right otherwise, and for symmetrical cases it is null (for more details, see Groeneveld & Meeden, 1984).

Subsequent results can be seen in Fig. 5.5, which displays the estimates for $\langle \log t \rangle_r$, $\langle \log t \rangle_m$, $\langle t_{last} \rangle$, $\langle t_{form} \rangle$, $\langle sSFR \rangle$, $\langle SFR \rangle$, $\langle Z/Z_\odot \rangle$, and $\langle \gamma \rangle$. The SFRs and corresponding fraction of stellar mass formed in the last $10^6$–$2 \times 10^9$ yr are available in Fig. 5.6.

In the first row of Fig. 5.5, it is possible to notice that the overall distribution of the mean ages – in both cases, weighted by mass or light – are similar, with a slight shift towards higher ages in the UV upturn systems. Yet, considering the error bars (see Fig. B.1 available in Appendix B), these ages can be considered equivalent.

The second row depicts $\langle t_{last} \rangle$ and $\langle t_{form} \rangle$. The first displays a gap of nearly 0.44 Gyr[4] between UV weak and UV upturn systems. This might indicate that UV upturn galaxies finished their last burst of star-formation earlier than UV weak systems, although the difference is small compared to our current precision for measuring ages (of the order of 1–2 Gyr, see e.g. Chaboyer, 2008; Dotter, Sarajedini, & Anderson, 2011). The second parameter, $\langle t_{\text{form}} \rangle$, is very similar for both UV groups, showing nearly no difference among the median values displayed by the dashed lines.

_______________

[4]The median $\langle t_{form} \rangle$ is 9.402 and 9.472 for UV weak and upturn galaxies respectively. Transforming this to linear results, the difference is of ~0.44 Gyr.

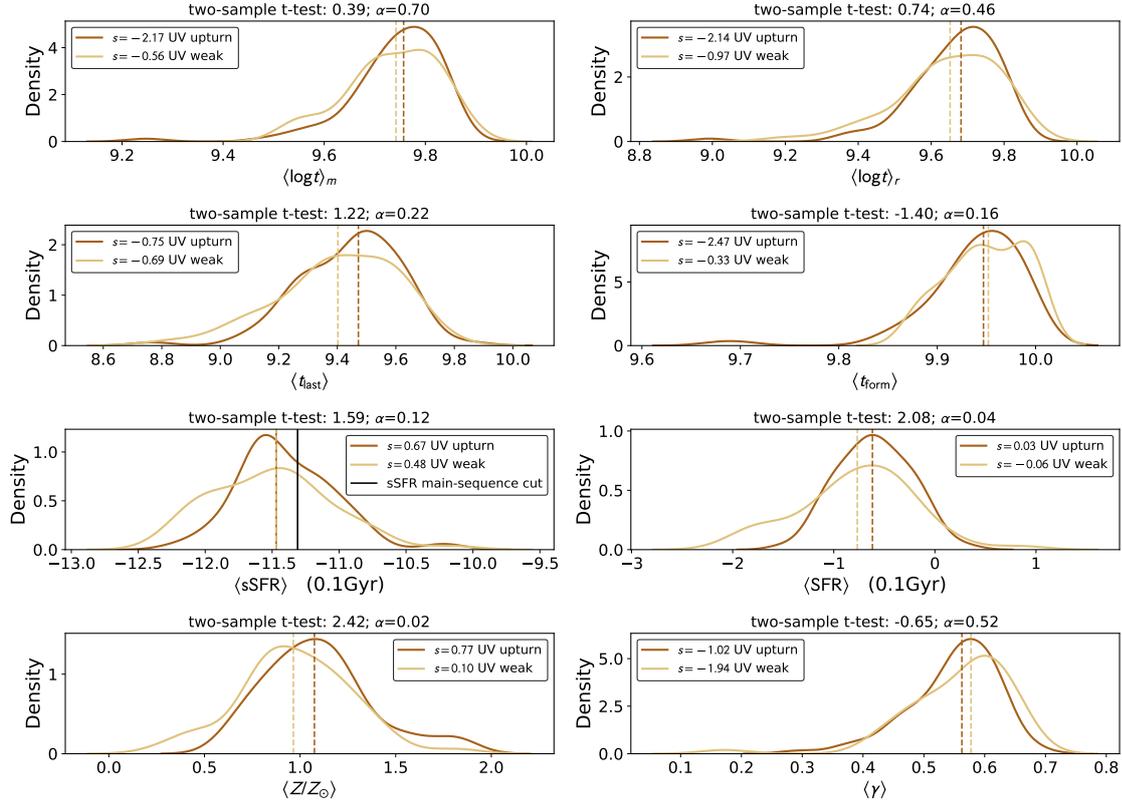

Figure 5.5: Distributions of several parameters which resulted from SED fitting by MAGPHYS (da Cunha, Charlot, & Elbaz, 2008). Left panel (from top to bottom): mass-weighted median stellar age ($\langle \log t \rangle_m$); median time since last burst of star-formation ($\langle t_{\text{last}} \rangle$); median specific star-formation over the past 0.1 Gyr ($\langle \text{sSFR} \rangle$); median metallicity ($\langle Z/Z_\odot \rangle$). Right panel (from top do bottom): light-weighted (based on the $r$-band) median stellar age ($\langle \log t \rangle_r$); median age of the oldest stars in the galaxy ($\langle t_{\text{form}} \rangle$); median star-formation rate over the past 0.1 Gyr ($\langle \text{SFR} \rangle$); median star-formation timescale ($\langle \gamma \rangle$). The median values for each distribution is displayed in dashed lines, except for $\langle \text{sSFR} \rangle$, which the medians are exactly the same; therefore, the median of UV upturn systems is displayed in a continuous line to allow the reader to see the overlapped values. Also, the star-formation main sequence cut is displayed via a black straight line for $\langle \text{sSFR} \rangle$ at $-11.31$ (see, for instance, Davies et al., 2019b). A two-sample t-test and p-value are displayed at the top of each subfigure.

The third row of Fig. 5.5 shows $\langle \text{sSFR} \rangle$ and $\langle \text{SFR} \rangle$ (both in the last 0.1 Gyr). The values are very low (consistent with zero) which is expected regarding the types of systems herein studied. UV weak systems present a longer tail towards lower values of both parameters. The medians for $\langle sSFR \rangle$ are exactly the same for both UV classes (at -11.47), but for $\langle SFR \rangle$ UV upturn systems seem to have a higher median. To check this effect, Fig. 5.6 shows the values of $\langle SFR \rangle$ and $f_{\mathrm{burst}}$ in different timescales, which are also available in Tab. 5.3. The median values of $f_{\mathrm{burst}}$ in all timescales show that, in fact, UV weak systems have had higher bursts of star-formation when compared to their UV upturn counterparts. This result is in fact more important when looking at $\left\langle f_{\mathrm{burst}}^{\mathrm{2Gyr}} \right\rangle$. Hence, $\left\langle f_{\mathrm{burst}}^{\mathrm{2Gyr}} \right\rangle$ is used heretofore in order to complement the analysis.

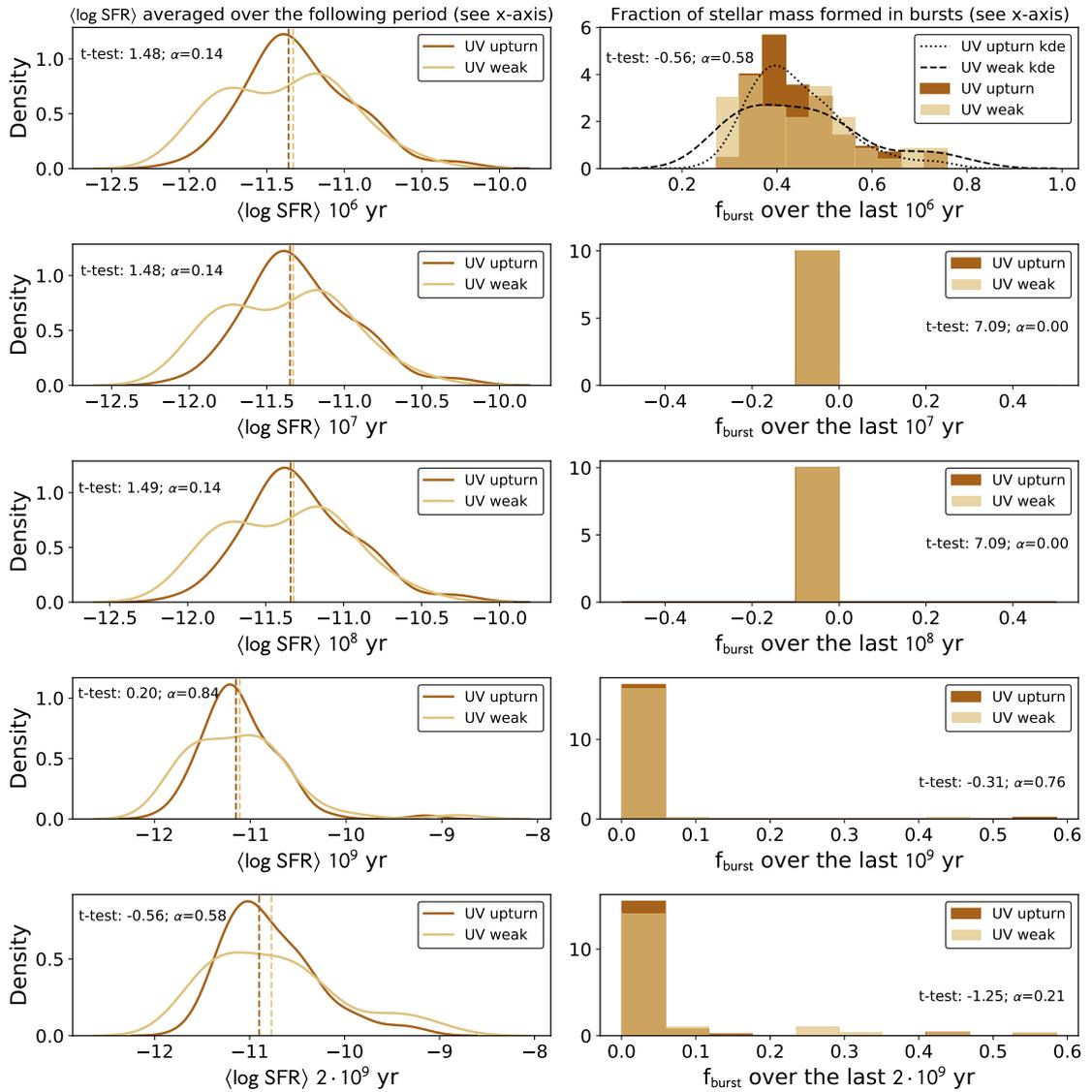

Figure 5.6: Results from SED fitting by MAGPHYS (da Cunha, Charlot, & Elbaz, 2008). Left kernel density plots: median star-formation rate (SFR) respectively over the last $10^6$, $10^7$, $10^8$, $10^9$, and $2 \cdot 10^9$ years. Right histogram plots: fraction of stellar mass formed over the last $10^6$, $10^7$, $10^8$, $10^9$, and $2 \cdot 10^9$ years. Results show that UV upturn systems have systematically lower $\langle SFR \rangle$ in all time ranges, and the gap is the widest at $2 \cdot 10^9$ yr. A two-sample t-test and p-value are displayed at the top of each subfigure.

Additionally, the values for $\langle sSFR \rangle$ are much higher than expected – see black straight line at $-11.31$ that marks the main sequence for star-forming systems (e.g. Davies et al., 2019b), indicating that ~40% of the sample is consistent with star-forming activity. This result is not in accordance with all the measures taken to mitigate star-formation activity (i.e. the cuts made according to the paradigms of Yi et al. 2011 and Cid Fernandes et al. 2010; Cid Fernandes et al. 2011). Therefore, this appears to be an effect caused by the overfitting of young stellar populations in Bruzual and Charlot (2003) and its lack post-main-sequence stellar evolutionary phases. These two parameters seem to be the ones most affected by the use of the models by Bruzual and Charlot (2003). Other important effects of young setllar populations seem to be mitigated due to the use of IR bands in the SED fitting process. In sum, the estimates of $\langle SFR \rangle$ and $\langle sSFR \rangle$ seem to be highly influenced by the young stellar populations that are filling the gaps in the UV range of the SED and, therefore, they should be evaluated with caution.

The fourth and final row depicts $\langle Z/Z_{\odot} \rangle$ and $\langle \gamma \rangle$. In the case of $\langle Z/Z_{\odot} \rangle$, the gap between UV weak and upturn galaxies is wider. Also, the shapes are slightly different, with the UV weak systems spanning over lower values of $\langle Z/Z_{\odot} \rangle$. In other words, these results indicate that UV upturn galaxies may be more rich in metals than their UV weak counterparts, which is in agreement with the recent paper by Werle et al. (2020).

The parameter $\langle\gamma\rangle$ for the star formation timescale[5] is the result of the parametric customisation of MAGPHYS. The higher $\langle\gamma\rangle$, the shorter is the star-formation timescale. The distribution of this parameter is similar for both UV weak and upturn systems, although the median shows slightly lower values for UV upturn systems when compared to their weak counterparts.

### 5.3.3 Correlations between the parameters from MAGPHYS

Another useful approach to analyse the results from MAGPHYS fitting is to check the correlations (or lack thereof) among the output parameters. To that end, I made use of visualisation tools such as heatmaps and clustermaps which are very useful to explore high-dimensional sets of data (see de Souza & Ciardi, 2015, for a reference in visualisation tools in Astronomy). Figs. 5.7 and 5.8 provide a straightforward visualisation of such correlations. For this thesis, the Spearman correlation rank ($\rho$) is used (Spearman, 1904). To guide the reader, two tables with the respective values of $\rho$ are displayed in Tables 5.4 and 5.5.

Fig. 5.7 displays two heatmaps, in which the left panel depicts the UV upturn systems, whereas the right panel shows the UV weak counterparts; the order of the variables are the same among both groups of galaxies, which serves as a visual aid in order to compare the level of correlation among the same parameters. In Fig. 5.8, the same correlations are shown, however their order is shifted to depict groups of variables that are clustered according to their level or correlation; such clustering is easily verified by the depicted upper and lateral dendrograms. Therefore, the stronger the correlation, the closer $\rho$ is to 1 (darker brown); and the stronger the anticorrelation, the closer $\rho$ is to -1 (darker cyan); the weaker the correlation/anticorrelation, the closer $\rho$ is to 0 (very light shades of brown/cyan). Additionally, details about such correlations/anti-correlations can be found in Secs. 5.3.3.1 to 5.3.3.4.

Table 5.3: Table depicting the fraction of stellar mass formed over several timescales (from $10^6$ to $2 \times 10^9$ yr) for both UV weak and UV upturn galaxies. The descriptive statistics for each period is available below. Whenever the results are marked as null values, they correspond to nominal values of $5 \times 10^{-4}$ (these are results from fluctuations during the fit.

| $\langle f_{\text{burst}} \rangle$ | UV weak | UV upturn |
|---|---|---|
| $10^6$ yr | min: 0.26 | min: 0.27 |
| | max: 0.80 | max: 0.76 |
| | mean: 0.46 | mean: 0.44 |
| $10^7$ yr | min: null | min: null |
| | max: null | max: null |
| | mean: null | mean: null |
| $10^8$ yr | min: null | min: null |
| | max: null | max: null |
| | mean: null | mean: null |
| $10^9$ yr | min: null | min: null |
| | max: 0.46 | max: 0.59 |
| | mean: 0.01 | mean: 0.01 |
| $2 \times 10^9$ yr | min: null | min: null |
| | max: 0.53 | max: 0.59 |
| | mean: 0.05 | mean: 0.03 |

_______________

[5]Considering that SFR is estimated by the following expression: $\text{SFR}(t) = \exp^{-\gamma t}$.

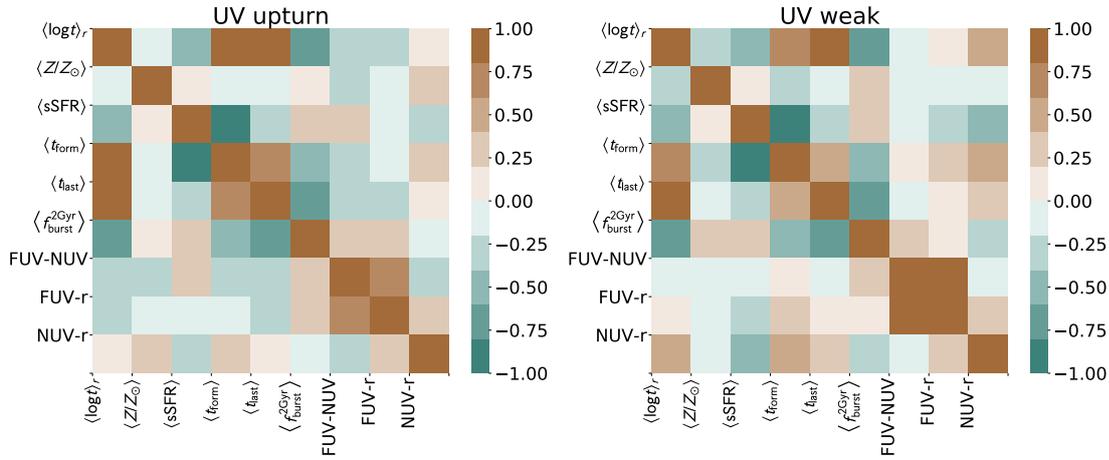

Figure 5.7: Heatmaps featuring the correlations among the parameters from MAGPHYS. The UV upturn systems are in the left while the UV weak are in the right panel. Each heatmap displays its colormap on right side with graduations from -1 (dark cyan) to 1 (dark brown). These graduations are Spearman's rank correlation indices ($\rho$, Spearman, 1904): the closer to 1 or -1, the higher the correlation or anticorrealtion, respectively; whereas the closer to 0, the shallower the correlation/anticorrelation.

First of all, it is necessary to provide a disclaimer on correlations that are equal to 1. These cases happen only when the $\rho$ is calculated between the same parameters (e.g. $\langle \log t \rangle_r$ vs. $\langle \log t \rangle_r$) – see the diagonal colours of heatmaps and clustermaps: the darkest shade of brown. Also, some parameters have been excluded from this step: $\langle \log t \rangle_m$ for being a duplicated age estimation; $\langle SFR \rangle$ is also a redundant parameter since $\langle sSFR \rangle$ is being considered; and $\log M_\star$, since it has been used as a PSM parameter.

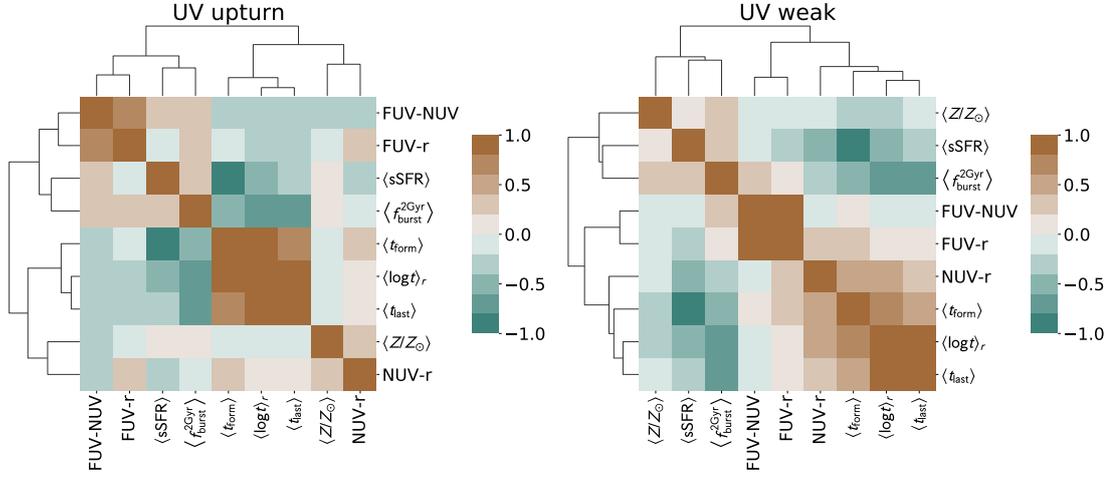

Figure 5.8: Clustermaps (heatmaps with dendrograms) featuring eaturing the correlations among the parameters from MAGPHYS. The UV upturn systems are in the left while the UV weak are in the right panel. Each clustermap displays its colormap on right side with graduations from -1 (dark cyan) to 1 (dark brown). These graduations are Spearman's rank correlation indices ($\rho$, Spearman, 1904): the closer to 1 or -1, the higher the correlation or anticorrealtion, respectively; whereas the closer to 0, the shallower the correlation/anticorrelation. The dendrogram groups parameters that are closely linked in terms of correlations or anti-correlations.

Table 5.4: Correlations table for the UV upturn systems.

|  | $\langle \log t \rangle_r$ | $\langle Z/Z_\odot \rangle$ | $\langle sSFR \rangle$ | $\langle t_{form} \rangle$ | $\langle t_{last} \rangle$ | $\left\langle f_{burst}^{2Gyr} \right\rangle$ | FUV-NUV | FUV-$r$ | NUV-$r$ |
|---|---|---|---|---|---|---|---|---|---|
| $\langle \log t \rangle_r$ | 1.00 | -0.08 | -0.54 | 0.81 | 0.93 | -0.72 | -0.32 | -0.29 | 0.11 |
| $\langle Z/Z_\odot \rangle$ | -0.08 | 1.00 | 0.08 | -0.06 | -0.00 | 0.03 | -0.24 | -0.05 | 0.30 |
| $\langle sSFR \rangle$ | -0.54 | 0.08 | 1.00 | -0.83 | -0.31 | 0.40 | 0.22 | -0.02 | -0.37 |
| $\langle t_{form} \rangle$ | 0.81 | -0.06 | -0.83 | 1.00 | 0.63 | -0.54 | -0.27 | -0.12 | 0.27 |
| $\langle t_{last} \rangle$ | 0.93 | -0.00 | -0.31 | 0.63 | 1.00 | -0.74 | -0.32 | -0.33 | 0.05 |
| $\left\langle f_{burst}^{2Gyr} \right\rangle$ | -0.72 | 0.03 | 0.40 | -0.54 | -0.74 | 1.00 | 0.32 | 0.24 | -0.12 |
| FUV-NUV | -0.32 | -0.24 | 0.22 | -0.27 | -0.32 | 0.32 | 1.00 | 0.75 | -0.34 |
| FUV-$r$ | -0.29 | -0.05 | -0.02 | -0.12 | -0.33 | 0.24 | 0.75 | 1.00 | 0.29 |
| NUV-$r$ | 0.11 | 0.30 | -0.37 | 0.27 | 0.05 | -0.12 | -0.34 | 0.29 | 1.00 |

Table 5.5: Correlations table for the UV weak systems.

| | $\langle \log t \rangle_r$ | $\langle Z/Z_\odot \rangle$ | $\langle \text{sSFR} \rangle$ | $\langle t_{\text{form}} \rangle$ | $\langle t_{\text{last}} \rangle$ | $\left\langle f_{\text{burst}}^{\text{2Gyr}} \right\rangle$ | FUV-NUV | FUV-$r$ | NUV-$r$ |
|---|---|---|---|---|---|---|---|---|---|
| $\langle \log t \rangle_r$ | 1.00 | -0.36 | -0.57 | 0.73 | 0.82 | -0.74 | -0.01 | 0.14 | 0.42 |
| $\langle Z/Z_\odot \rangle$ | -0.36 | 1.00 | 0.12 | -0.24 | -0.10 | 0.21 | -0.09 | -0.15 | -0.16 |
| $\langle \text{sSFR} \rangle$ | -0.57 | 0.12 | 1.00 | -0.87 | -0.25 | 0.28 | -0.00 | -0.21 | -0.42 |
| $\langle t_{\text{form}} \rangle$ | 0.73 | -0.24 | -0.87 | 1.00 | 0.42 | -0.40 | 0.01 | 0.20 | 0.42 |
| $\langle t_{\text{last}} \rangle$ | 0.82 | -0.10 | -0.25 | 0.42 | 1.00 | -0.76 | -0.06 | 0.08 | 0.36 |
| $\left\langle f_{\text{burst}}^{\text{2Gyr}} \right\rangle$ | -0.74 | 0.21 | 0.28 | -0.40 | -0.76 | 1.00 | 0.32 | 0.16 | -0.32 |
| FUV-NUV | -0.01 | -0.09 | -0.00 | 0.01 | -0.06 | 0.32 | 1.00 | 0.82 | -0.14 |
| FUV-$r$ | 0.14 | -0.15 | -0.21 | 0.20 | 0.08 | 0.16 | 0.82 | 1.00 | 0.38 |
| NUV-$r$ | 0.42 | -0.16 | -0.42 | 0.42 | 0.36 | -0.32 | -0.14 | 0.38 | 1.00 |

By using correlation maps, we gain access to 81 values of $\rho$, of which 72 are effective ones (when discarding those calculated between the same parameters – see diagonals); the remaining $\rho$ appear twice, giving us a total of 36 unique values of $\rho$. The discussion is focused on the most remarkable similarities and differences among UV weak and upturn systems, focusing on outstanding results. Also, the order of the discussion in the following Sections are different from the presented in Fig. 5.7 and Tables 5.4 and 5.5 with the goal of bringing to light the most important results first, avoiding multiple similar considerations.

It is worth mentioning that, as seen in Fig. 5.9, $\left\langle f_{\text{burst}}^{\text{2Gyr}} \right\rangle$ is mainly null for most galaxies. Therefore, the correlations estimated for $\left\langle f_{\text{burst}}^{\text{2Gyr}} \right\rangle$ will not be discussed as probably little physical meaning can be extracted from it.

It is noticeable that the correlation tree (hierarchical clustering) among the different parameters is different for UV weak and upturn galaxies – as displayed by the upper and lateral dendrograms in Fig. 5.8 –, considering that such systems are very similar apart from their UV properties.

### 5.3.3.1   UV and UV-optical colours

One of the main results from the correlation maps is the difference between UV weak and upturn in terms of their UV and UV-optical colours against other parameters. Of course, these three colours have been used to classify these systems into UV weak or UV upturn (as in Fig. 2.3). By looking at the three last rows of tables 5.4 and 5.5, it is remarkable how such correlations change among both groups of galaxies. Nonetheless, FUV-NUV and FUV-$r$ are clustered together in both UV weak and upturn (Fig. 5.8), whereas NUV-$r$ behaves differently, being closer to other parameters in each type of UV bright class.

To further explore the trends for UV and UV-optical colours, this discussion is amplified in Sec. 5.3.4.

**FUV-NUV:**   for UV weak systems the overall trend is the very low level of correlations/anti-correlations against the other parameters (or the complete lack thereof – i.e. $|\rho| < 0.1$ for all of them except $\left\langle f_{\mathrm{burst}}^{\mathrm{2Gyr}} \right\rangle$). However, such correlations turn to mild when looking at the UV upturn population[6]. In other words, this means that the strength of the upturn is in fact linked to various parameters of these systems, notably $\langle \log t \rangle_r$, $\langle t_{\mathrm{form}} \rangle$, $\langle t_{\mathrm{last}} \rangle$, $\left\langle f_{\mathrm{burst}}^{\mathrm{2Gyr}} \right\rangle$, $\langle Z/Z_\odot \rangle$, and $\langle \mathrm{sSFR} \rangle$; this makes FUV-NUV an important parameter when looking for correlations for UV upturn galaxies.

**FUV-$r$:**   the changes for this colour are more subtle. UV upturn systems present an overall trend of anti-correlations with the other parameters, which is expected, since the magnitudes for the $r$-band tend to be lower (i.e. be brighter in the optical) when compared to FUV. Yet, there are some differences among their UV weak counterparts, with remarks on the values of $\rho$ estimated for $\langle \log t \rangle_r$, $\langle \mathrm{sSFR} \rangle$, $\langle t_{\mathrm{form}} \rangle$, and $\langle t_{\mathrm{last}} \rangle$. For UV upturn systems, $\rho$ calculated between FUV-$r$ and $\langle$sSFR$\rangle$ changes from no correlation ($\rho = -0.02$) to a mild anti-correlation for UV weak systems ($\rho = -0.21$); the other three timescale parameters, the values of $\rho$ change from a weak-to-mild anti-correlations to weak-to-mild correlations – i.e. their signal inverts among both types of systems –, with highlights on $\langle t_{\text{form}} \rangle$.

---

[6]With the exception of $\rho$ for $\left\langle f_{\mathrm{burst}}^{\mathrm{2Gyr}} \right\rangle$, which is understandable as $\left\langle f_{\mathrm{burst}}^{\mathrm{2Gyr}} \right\rangle$ is frequently null (see Fig. 5.9).

**NUV-$r$:** this colour may be interpreted as one of the main thermometers when looking for correlations changes among both subgroups of UV bright galaxies. The $r$-band tends to not significantly change between both types of systems, but the difference between the $r$-band and the 'pivoting' NUV-band can be an important marker of the different properties among both types of systems. In fact, among these three colours, NUV-$r$ is the only one that points to overall stronger correlations between the other physical properties of UV weak systems.

### 5.3.3.2 The various timescales: $\langle \log t \rangle_r$, $\langle t_{\text{form}} \rangle$, and $\langle t_{\text{last}} \rangle$

All the three timescale parameters appear clustered in Fig. 5.8 for both UV weak and UV upturn, which is an expected result as they are not linearly independent. For all the three parameters, $\rho > 0.5$ for both UV weak and upturn galaxies, except between $\langle t_{\text{last}} \rangle$ and $\langle t_{\text{form}} \rangle$ for UV weak, which is slightly smaller $\rho = 0.42$, still a mild-to-high correlation. Additionally, all three timescales strongly anti-correlate with $\langle$sSFR$\rangle$ and $\left\langle f_{\text{burst}}^{\text{2Gyr}} \right\rangle$ for both UV classes, which is also expected, specially for the following pairs: $\langle t_{\text{form}} \rangle$ and $\langle$sSFR$\rangle$, $\langle t_{\text{last}} \rangle$ and $\left\langle f_{\text{burst}}^{\text{2Gyr}} \right\rangle$, $\langle \log t \rangle_r$ and $\left\langle f_{\text{burst}}^{\text{2Gyr}} \right\rangle$.

### 5.3.3.3 $\langle Z/Z_{\odot} \rangle$

The $\rho$ between $\langle Z/Z_{\odot} \rangle$ and $\langle$sSFR$\rangle$, as well as $\langle t_{\text{last}} \rangle$, are very weak for the two UV classes. The differences of $\rho$ for UV weak and upturn appear for $\langle \log t \rangle_r$, $\langle t_{\text{last}} \rangle$, and $\left\langle f_{\text{burst}}^{\text{2Gyr}} \right\rangle$; in all cases $\rho \sim 0$ for UV upturn galaxies, but turn to mild for their weak counterparts ($0.2 < \rho < 0.4$).  Further changes for UV weak and upturn systems can be seen for the different timescales; $\rho$ does not show any significant correlation values for none of the three parameters for UV upturn systems.  On the other hand, for UV weak galaxies $\rho$ turns to mild for $\langle \log t \rangle_r$ and $\langle t_{\text{form}} \rangle$, while remaining null and robust for $\langle t_{\text{last}} \rangle$. These results are easily seen side-by-side in Fig. 5.7, and Tables 5.4 and 5.5.

### 5.3.3.4   $\langle$sSFR$\rangle$

The most important results are the anti-correlations between $\langle$sSFR$\rangle$ and the timescale parameters, which is detailed in Sec. 5.3.3.2. As discussed in Sec. 5.3.2, further exploration was made to assess the characteristics of $\langle$SFR$\rangle$ and $f_{\text{burst}}$ in several timescales.

## 5.3.4   Trends for the UV and UV-optical colours

It is worth exploring the trends shown by UV and UV-optical colours, as they act as direct measurements of the strength of the UV upturn, specially the FUV-NUV colour.

Fig. 5.9 explicitly depicts the colours FUV-NUV, FUV-$r$, and NUV-$r$ against six parameters among those explored by the previous heatmaps and clustermaps: $\langle \log t \rangle_r$, $\langle t_{\text{form}} \rangle$, $\langle t_{\text{last}} \rangle$, $\langle Z/Z_\odot \rangle$, $\langle$sSFR$\rangle$, and $\left\langle f_{\text{burst}}^{\text{2Gyr}} \right\rangle$. In this Fig., 2D-Gaussian kernel densities are also displayed in order to facilitate the distinction of the two groups, UV weak (in grey) and upturn (in red), except for the last row, as $\left\langle f_{\text{burst}}^{\text{2Gyr}} \right\rangle$ is mostly null.

The separations between the two groups of galaxies given FUV-NUV and FUV-$r$ are very clear in the first two columns, which is due to the criteria by Yi et al. (2011). On the other hand, for NUV-$r$ the distributions for both types of galaxies are mostly overlapped in all cases, except for $\langle$sSFR$\rangle$. In such case, the UV upturn systems seem to be concentrated in higher values, whereas UV weak galaxies reach lower values of $\langle$sSFR$\rangle$. This result is in agreement with the distributions of $\langle$sSFR$\rangle$ shown in Fig. 5.5.

Although such low values are consistent with ETGs, apparently UV upturn systems carry sightly higher values of $\langle$sSFR$\rangle$ and/or $\langle$SFR$\rangle$ at 0.1 Gyr when compared to UV weak counterparts in the same conditions. By further investigating the distributions of $\langle SFR \rangle$ and $f_{burst}$ over different timescales, as shown in Fig. 5.6, it is possible to see that $f_{burst}$ is consistently higher for UV weak galaxies (see mean values in Table 5.3), contradicting the first result given by $\langle SFR \rangle$ at 0.1 Gyr. Additionally, for $2 \times 10^9$ yr, UV weak systems had higher values of $\langle SFR \rangle$. According to this, UV weak systems have had a slightly higher star-formation activity over the timescales, specially at $2 \times 10^9$yr.

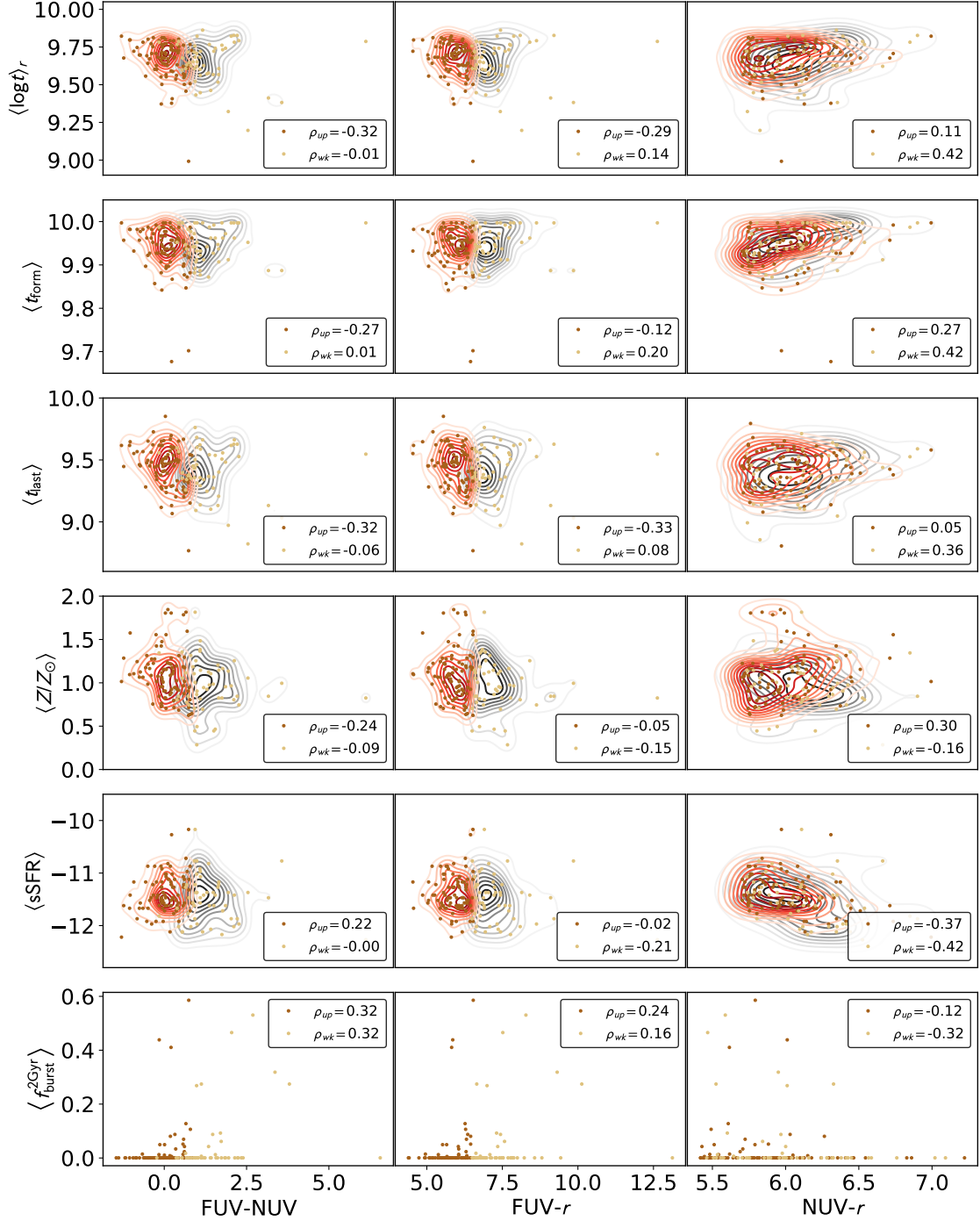

Figure 5.9: Sixteen plots featuring the UV and UV-optical (FUV-NUV, FUV-*r*, and NUV-*r* respectively) colours and the following physical parameters (from top to bottom): $\langle \log t \rangle_r$, $\langle t_{\text{form}} \rangle$, $\langle t_{\text{last}} \rangle$, $\langle Z/Z_\odot \rangle$, $\langle \text{sSFR} \rangle$, and $\left\langle f_{\text{burst}}^{2\text{Gyr}} \right\rangle$. UV weak and UV upturn systems are respectively depicted by light and dark brown round markers, as well as grey and red 2D-kernel density curves – with the exception of the last row, which has many values at zero, preventing the kernel density to be estimated. Additionally, the labels depict the corresponding values of $\rho$ for convenient visualisation ($\rho_{\text{wk}}$ for UV weak and $\rho_{\text{up}}$ for UV upturn).

The overall trend of the sub-plots for FUV-NUV depicted in Fig. 5.9 is that UV weak galaxies seem to present a larger dispersion. This reflects on the estimation of $\rho$, pointing to no correlations/anti-correlations for UV weak systems in this colour, against weak correlations for UV upturn galaxies.

These results (higher dispersion in FUV-NUV colour, broader range of $\langle sSFR \rangle$, higher values for $\langle SFR \rangle$ in the last 2 Gyr, and lower $\langle Z/Z_\odot \rangle$) indicate that UV weak systems probably host a wider range of stellar population types, when compared to their UV upturn counterparts. In other words, UV upturn galaxies seem to be evolving more passively than their UV weak counterparts. There are basically two hypotheses that may explain this:

i. for some reason, UV weak galaxies may have been subject to interactions such as accretion or/and minor mergers more frequently than UV upturn galaxies. The accretion of cold gas from the intergalactic medium would explain the lower overall metallicity and the gap of $\langle SFR \rangle$ in the last 2 Gyr;

ii. UV upturn systems may have evolved faster than their UV weak counterparts, i.e. in higher $z$.

These results point to the importance of measuring the FUV magnitude when investigating UV bright – and specially UV upturn – galaxies. The FUV-NUV colour is indeed an important marker for the study of the strength of the UV. Therefore this range of the electromagnetic spectrum should not be overlooked in the study of stellar populations through SED fitting.

### 5.3.5 Principal Component Analysis

As a final step in the analysis involving the differences and similarities of UV weak and upturn systems, I apply Principal Component Analysis (PCA) to both sets of data described in Sec. 5.3.3. The goal is to verify which parameters contribute to the total variance of such systems. This is an important step in order to validate the analyses made so far.

PCA aims at easing the analysis of high dimensional data, by reducing its dimensionality with the smallest lost of information (see, for instance, Abdi & Williams, 2010; Jolliffe & Cadima, 2016); it has been frequently used in Astrophysical research (see for instance Jeeson-Daniel et al., 2011; Chen et al., 2012; de Souza et al., 2014; Pace et al., 2019). Since PCA is sensitive to scale, one must standardise the parameters first (see Jolliffe & Cadima, 2016, Sec. 2C). The following equation shows the mathematical transformation to standardise a certain variable:

$$\overline{x} = \frac{x - \mu}{\sigma}, \tag{5.2}$$

in which $x$ is the original variable, $\mu$ is the mean, $\sigma$ is the standard deviation, and $\overline{x}$ is the standardised variable. PCA was estimated by making use of the SCIKIT-LEARN package in PYTHON (Pedregosa et al., 2011).

The results for 5 principal components (PCs) can be seen in Fig. 5.10. In this case, 5 PCs were enough to explain 95% of the variance between both groups (95.1% for UV upturn and 96% for UV weak). The contribution of each PC to the total variance is displayed in the title part of each sub-plot of Fig. 5.10.

It possible to see that no obvious differences appear in the first two PCs in Fig. 5.10,

which is a major indicator that these systems are quite similar among each other. In PC1 and PC2, the main contributions to the total variance come from the same variables for both UV weak and UV upturn galaxies; the differences appear in variables of secondary importance/contribution. For instance, $\langle \log t \rangle_r$, $\langle t_{\text{form}} \rangle$, $\langle t_{\text{last}} \rangle$, and $\langle \text{sSFR} \rangle$ are the four main variables for both UV classes in PC1. Nevertheless, differences start to appear in variables that contribute a little less to the total variance. For UV upturn systems, FUV-NUV and FUV-$r$ have an important role for the variance of PC1, whereas for UV weak systems NUV-$r$ and $\langle Z/Z_\odot \rangle$ are those that play a secondary role.

For PC2, FUV-NUV, FUV-$r$, and $\langle \text{sSFR} \rangle$ are the variables that most contribute to the variance for both systems as well. Yet, for secondary contributions, $\langle Z/Z_\odot \rangle$ seams to play a considerable role for UV upturn galaxies, whereas $\langle t_{\text{last}} \rangle$, $\left\langle f_{\text{burst}}^{\text{2Gyr}} \right\rangle$, and NUV-$r$ highly contribute to the variance of UV weak systems. It is worth mentioning that $\langle Z/Z_\odot \rangle$ contributes more for the variance of UV weak systems (as they are important for PC1) then for their UV upturn counterparts (becoming important for the variance only at PC2 and PC3);

PC3 depicts deeper differences in terms of contribution to the variance among the different variables. In this PC, NUV-$r$ is mostly important for UV upturn systems, whereas $\langle \text{sSFR} \rangle$ and $\langle t_{\text{form}} \rangle$ are the main contributors for the variance in UV weak galaxies. Then PC4, these similarities once again become stronger, being $\langle Z/Z_\odot \rangle$ important for the variance of both UV classes. Finally, the variance in PC5 is mostly explained by NUV-$r$ for UV weak systems, whereas $\left\langle f_{\text{burst}}^{\text{2Gyr}} \right\rangle$ seems more important for their upturn counterparts.

In other words, the main results show that:

- the main variables that contribute to the variance of PC1 and PC2 are roughly the same for both UV weak and upturn systems;

- differences between these two classes of galaxies appear in a secondary and more subtle level.

Figure 5.10: Principal components (PCs) resulting from the PCA technique. The results depict UV weak and UV upturn systems in light and dark brown respectively. The contributions to the total variance are available in the respective titles of the sub-plots (i.e. $v_{wk}$ for UV weak and $v_{up}$ for UV upturn galaxies).

### 5.3.6   Summary of results

In this Section I extend and summarise the discussion developed in this Chapter.  The order will not be same as throughout this Chapter, rather the discussion is developed in a 'hierarchical' logic, encompassing the main results herein presented.

The analysis performed by PCA has shown that the variance for both UV weak as well as UV upturn systems are dominated by the same variables.  This means that these systems, as expected, are not very different from each other – as a matter of fact, they were selected to be both massive red-sequence systems, with detectable UV emission, within the same range of $z$.

For PC1, the most important component, all timescales and $\langle sSFR \rangle$ are very important for the variance among both UV classes.  Nevertheless, the differences in the contribution of the variance among both groups can be detected in 'secondary' parameters.  Those parameters are $\langle Z/Z_\odot \rangle$ and UV and UV-optical colours; $\langle Z/Z_\odot \rangle$ and NUV-$r$ being important contributors for UV weak galaxies and FUV-NUV for UV upturn.

By combining these results with the previous analysis, in fact, the distributions of $\langle \log t \rangle_r$ and $\langle \log t \rangle_m$ are very similar for both UV classes, although the results show that UV upturn systems seem to harbour older stellar populations when compared to their weak counterparts.  This can been seen through the median values of $\langle \log t \rangle_r$ and $\langle \log t \rangle_m$, as well as the higher absolute skewness of UV upturn systems to the right.  These results are supported by the distributions of $D_n4000$ and $(g-r)$.  Yet, the age of the oldest stars ($\langle t_{form} \rangle$) in each type seem to be very similar among both systems.  In terms of $\langle t_{last} \rangle$, there seems to be a gap of $\sim 0.44$ Gyr between both systems (which needs to be confirmed, given the typical error bar of current age determinations in stellar population, as mentioned in Sec. 5.3.5).  This could support the idea that UV weak galaxies harbour slightly younger stellar populations.  This is backed by the extra analysis available in Tab. 5.3 and Fig. 5.6.

The metallicity also plays an important role, as seen in the analyses involving the direct comparison and correlations, which is supported by the PCA analysis. The distribution of $\langle Z/Z_\odot \rangle$ for UV weak galaxies has a longer tail towards lower values, is more skewed to the left, and its median is lower than those of UV upturn systems. This is also supported by the distribution of $D_n 4000$. Additionally, in the analysis of PCA, metallicity is more important for UV weak galaxies than for their upturn counterparts; this is an expected result, as $\langle Z/Z_\odot \rangle$ spreads through a higher range of values (i.e. reaches lower values) than in UV upturn systems.

UV and UV-optical colours are used to separate the galaxies into the UV classes herein used. Yet, FUV-NUV as well as FUV-$r$ are important contributors for the variance of UV upturn systems in PC1. When compared to UV weak systems, UV upturn galaxies show higher overall correlation scores of $\rho$ for FUV-NUV and NUV-$r$ against the other parameters. This is backed by Figs. 5.7, 5.8, and 5.9.

It is important to recall that the estimates for $\langle SFR \rangle$ and $\langle sSFR \rangle$ reflect the overfitting of young stellar populations present in Bruzual and Charlot (2003). This is caused by the lack of coverage in terms of the HR diagram regarding the post-main-sequence stellar evolutionary phases. Other results seem to be less affected by this due to the use of IR bands in the SED fitting procedure.

These results support the hypothesis that UV upturn may be evolving more passively than UV weak systems. Alternatively, UV upturn may have settled their stellar population at higher $z$. UV weak galaxies ones have lower median ages, more recent star burst events (see $\langle t_{last} \rangle$), higher mean values of $f_{burst}$ throughout several timescales, and lower metallicity, specially for 2 Gyr lookback time (see $\left\langle f_{burst}^{2Gyr} \right\rangle$). Two hypothesis could explain the difference in behaviour for UV weak systems:

i. for some reason, UV weak systems may have been more efficient in attracting gas from the intergalactic medium, which is known to be more metal-poor compared to the gas in the interior of galaxies (which have been recycled more often, therefore being more rich in metals);

ii. or they might have extended SFH, possibly caused by a larger number of interactions or gas-rich minor mergers throughout their evolution, which could also explain its characteristics, such as lower metallicities.

What we know is a drop, what we
don't know is an ocean.

Isaac Newton

If you can't fly, run. If you can't run,
walk. If you can't walk, crawl, but by
all means keep moving.

Martin Luther King Jr.

# 6
# Conclusions

This thesis is a result of approximately 4.5 years of very fruitful research work. The scientific goal was to explore the UV upturn in elliptical galaxies[1] from different angles:

1. the evolution of the fraction of UV upturn galaxies among all UV bright RSG group (Chap. 3) as a function of redshift and stellar mass;

2. the characterisation of the UV bright red-sequence galaxies in terms of emission lines, and its impacts on the evolution of the fraction of UV upturn (Chap. 4);

3. similarities and differences of the stellar populations in a de-biased sample of UV weak and upturn systems (Chap. 5).

With the goals refreshed, I present the main results of this study, as follows.

## 6.1   Main results

The main results of this thesis are presented in the following Sections in same order as analysed and discussed in the previous Chapters.

---

[1]In this thesis, I make use of a broader terms such as early-type and red-sequence galaxies, given that morphology is not directly taken into account.

### 6.1.1   Evolution of the UV upturn

It this study, I present a novel outlook on the evolution of the UV upturn. Differently from previous studies, it is the first time that the fraction of UV bright galaxies nesting the UV upturn phenomenon has been analysed. All previous studies have focused on the evolution regarding the strength of the upturn, usually with optical-UV and/or UV colours. Additionally, it is the first time that the UV upturn phenomenon has been studied by making use of Bayesian statistics. Therefore, by recalling the goals of this thesis (described in Sec. 1.5), it is clear that they have been successfully achieved.

This study has shown that the fraction of UV upturn galaxies among all UV bright RSGs, evolves in redshift; it increases up to redshift 0.2–0.25 and appears to decrease subsequently. However, given the thickness of credible intervals for $z > 0.25$, it is unclear if the trend in fact decreases, plateaus, or even continues to increase. Future investigations with larger sample of galaxies at $z > 0.25$ are needed to enlighten how this fraction evolves, as we look at earlier stages of the Universe.

Additionally, another important result is that the UV upturn in RSGs is more prevalent among more massive systems. This conclusion is supported both when the analysis is performed in terms of the cumulative distribution function (Fig. 2.4) or by using a logistic regression (Fig. 3.6).

### 6.1.2   Emission lines and the UV upturn phenomenon

The classification of UV upturn systems used in this research is based on the prescription proposed by Yi et al. (2011), which separates residual star formation (RSF), UV weak and UV upturn galaxies using colour cuts. An important finding of this thesis is that this method for galaxy classification is robust against potential AGN interlopers. However, it fails to properly clean the sample against systems with star-formation activity. In practice, approximately 23% of the galaxies classified as UV upturn in our sample are star-forming systems according to the WHAN diagnostic diagram. From the perspective of rare stellar evolutionary phases, the UV upturn hosts which are free from star formation are actually the most intriguing ones.

With this being said, by analysing the impact of each class of emission line, the main results contributing to the regression performed in Chap. 3, are those from the retired/passive class. It is noteworthy that these are the ones linked to evolved stellar populations. These results indicate that the evolution of the fraction of UV upturn is mostly supported by this class.

### 6.1.3 Similarities and differences in stellar population properties between UV weak and upturn systems

To grasp a feeling for what is happening in terms of stellar population properties, I have made use of the SED fitting results obtained with MAGPHYS code, made available by the GAMA collaboration. With these data at hand, PSM was applied to mitigate the effects of the confounding variables: $z$ and $\log M_\star$. The comparison was made in three levels:

a. by directly comparing the distributions of the stellar population properties in both UV classes;

b. by computing correlation ranks among the observed (colours) and stellar population properties, separately for each UV class;

c. by applying PCA to both UV weak and upturn sub-samples and evaluating differences and similarities among the variables that contribute to the total variance.

These analyses confirm that UV weak and UV upturn systems are, in general terms, very similar. This statement was one of the motivations for this part of the study, but it is also a finding from the analysis performed. This is not unexpected, as both types of galaxies are in the red-sequence, are bright enough in the UV to be detected by GALEX

Medium-depth Imaging Survey, and possess the same distributions of $\log M_\star$ and $z$ (as a result of the PSM).

The differences between both UV classes appear subtly throughout the analysis and they can be seen in their median ages, time since last burst of star formation, and metallicity. UV upturn systems present (median) higher ages and metallicities, which is also in accordance with the results for $(g - r)$ and $D_n4000$. Also there is a gap of approximately 0.44 Gyr between the times since last burst of star formation among both systems (which cannot be confirmed at the moment, given the current limitations of age estimations).

These results seem to indicate that either UV upturn galaxies have been evolving more passively than UV weak galaxies, or that they settled their stellar population at higher redshifts than their UV weak counterparts.

## 6.2   Perspectives

As is the case of many doctoral studies, the ideal research cannot be fulfilled in mere 4 or 5 years. Many of the investigations herein pursuit have the potential to keep moving forward. Some of the questions that still remain open and are great opportunities for research follow-ups are enumerated as follows.

1. As described in Sec. 6.1.1, it is important to study the evolution of the fraction of the UV upturn for $z > 0.25$ in order to probe its behaviour in earlier stages of the Universe. In order to accomplish that, it is necessary to make use of deeper observations, either ground-based (e.g. by making use of state-of-the-art telescopes such as LSST, Ivezić et al., 2019, or the Javalambre Physics of the Accelerated Universe Astrophysical Survey, J-PAS, Benitez et al., 2014; Bonoli et al., 2020) or in space (e.g. the Nancy Grace Roman Space Telescope – previously called WFIRST, Spergel et al., 2015);

2. Metallicity-related parameters such as Lick indices are very important for understanding the star formation and chemical enrichment histories of galaxies that host the UV upturn phenomenon, and those still remain to be analysed. Due to time limitations, this aspect could not be further explored and, therefore, it remains as a near future investigation goal.

3. Given the complexity of the stellar populations in UV upturn systems (as detailed in Sec. 1.3.1.1), actual understanding of the different components of such systems are yet to be unravelled. Ideally, it would be necessary to elaborate a new library of stellar populations that would be capable of encompassing all the rare populations linked to this phenomenon (i.e. binaries, EHB/HB stars, and so on). Only with such templates one could try to decipher quantitatively the contributions of each stellar population. Therefore, one of the long term follow-up goals of this thesis is to develop such library.

4. Environmental effects are also some of the issues that have been poorly explored in the past and still remain to be deeply investigated. This issue has the potential to be accomplished in the near-future by making use of the vast number of observations in archival databases.

5. With the rise of integral field spectroscopy, it is timely to study the spatial properties of UV upturn confirmed galaxies in the nearby Universe. This would enable to address several issues, including LINER-like emission sparse in their host galaxy. This is another opportunity that could be accomplished in the near future.

6. It is also timely to probe the UV emission of galaxies in semi-analytical models, such as the CMASS (e.g. Stoppacher et al., 2019). As discussed in Sec. 1.3, the UV emission is linked to hot components and it is appropriate to verify how semi-analytical simulations reproduce (or not) them.

> Well, I hope that if you are out there and read this and know that, yes, it's true I'm here, and I'm just as strange as you.

Frida Khalo

> Be happy while you're living, for you're a long time dead.

Scottish proverb

# 7
# Epilogue

During my doctoral studies (Dec. 2015 – Jul. 2020) herein presented to you, I have personally been through some of the toughest moments of my life. In 2016, during my first doctoral year, I lost my brother; in 2018 it was the time for my father, after many previous health complications; finally, this thesis is being finished during a pandemic of coronavirus (covid-19) – never before seen in our generation – and I have been writing in confinement. Unfortunately, covid-19 has also removed the lives of many people, including my aunt Francisca (whom we lovingly called Tia Dudu), and other acquaintances. As of the writing of this thesis, over 160,000 people have passed away in Brazil and 1,220,000 worldwide. The social, political, and economical impacts of this pandemic will last for several upcoming years.

With all this being said, I consider myself a fighter and a winner, as I have been able to develop my research as I dreamed before I even started my life as a researcher, even in the midst of these personal and collective events.

I end this thesis with a mix of feelings and sharing with you one of my favourite poems of all time, by Fernando Pessoa[1], a Portuguese poet and writer who lived from 1888 to 1935.

---

[1]https://en.wikipedia.org/wiki/Fernando_Pessoa

**Liberdade**

*Ai que prazer*

*Não cumprir um dever,*

*Ter um livro para ler*

*E não o fazer!*

*Ler é maçada,*

*Estudar é nada.*

*O sol doira*

*Sem literatura*

*O rio corre, bem ou mal,*

*Sem edição original.*

*E a brisa, essa,*

*De tão naturalmente matinal,*

*Como o tempo não tem pressa. . .*

*Livros são papéis pintados com tinta.*

*Estudar é uma coisa em que está indistinta*

*A distinção entre nada e coisa nenhuma.*

*Quanto é melhor, quanto há bruma,*

*Esperar por D.Sebastião,*

*Quer venha ou não!*

*Grande é a poesia, a bondade e as danças...*
*Mas o melhor do mundo são as crianças,*
*Flores, música, o luar, e o sol, que peca*
*Só quando, em vez de criar, seca.*

*Mais que isto*
*É Jesus Cristo,*
*Que não sabia nada de finanças*
*Nem consta que tivesse biblioteca. . .*

Fernando Pessoa (by his heteronym Alberto Caeiro)

# A
## Model

In what follows, I provide the STAN code used for the logistic model in place and the main code for running it with PYSTAN.

```
stan_code = """
    // DECLARATION OF VARIABLES
    data{
        int<lower=0> N;
        int<lower=0> K;
        int Y[N];
        matrix[N,K] X;
        real LogN;
    }

    // DEFINING THE PRIOR(S)
    parameters{
        vector[K] beta;
    }

    transformed parameters{
```

```
        vector[N] eta;
        eta = X * beta;
    }

    // MODEL: PROBABILITY, HYPERPRIORS, PRIORS, AND REGRESSION
    model{
        Y ~ bernoulli_logit(eta);
    }

    // DATA TO BE PLOTTED
    generated quantities{
        vector[N] etanew;
        real<lower=0, upper=1.0> pnew[N];
        etanew = X * beta;
        for (j in 1:N){
            pnew[j] = inv_logit(etanew[j]);
        }
    }
"""

iterations = 10000
chains      = 3        # HMC chains
warmup      = 3000     # How many of the first iterations we'll ignore
jobs        = -1       # Run code in parallel -- see pystan docs
seed        = 1

model = pystan.StanModel(model_code=stan_code)
```

```
fit = model.sampling(data=regression_data, seed=seed,
↪   iter=iterations, chains=chains, warmup=warmup, n_jobs=jobs,
↪   control=control)
```

# B

# Additional material for Chap. 5

In this Appendix, some additional material supporting the discussions in Chapter 5 is provided.

## B.1 Errors from SED fitting resulting parameters

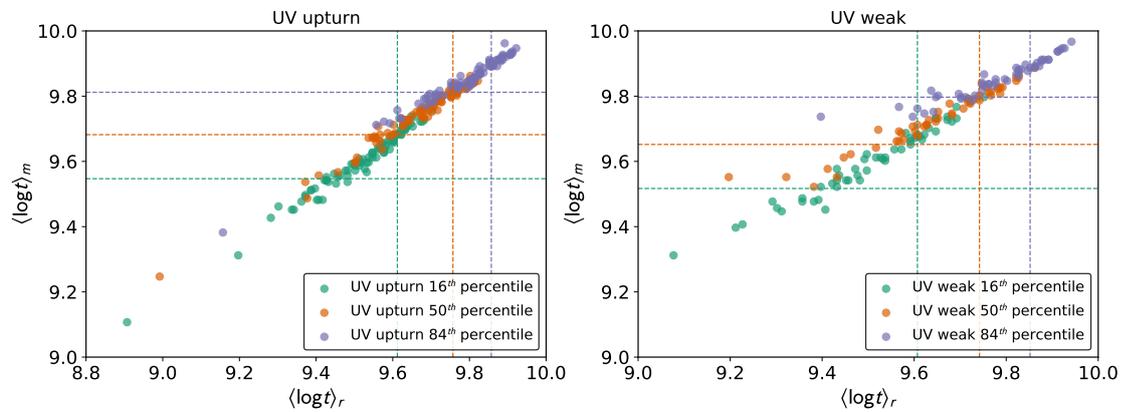

Figure B.1: Light-weighted *versus* mass-weighted ages 16th, 50th, and 84th quantiles for UV upturn (left panel) and UV weak (right panel) galaxies. The dashed lines represent the medians for each of the quantiles, according to the colour used. The quantiles used (16 and 84) represent the 1$\sigma$ distance from the median.